\documentclass[fleqn,usenatbib]{mnras}

\usepackage{newtxtext,newtxmath}

\usepackage[T1]{fontenc}

\DeclareRobustCommand{\VAN}[3]{#2}
\let\VANthebibliography\thebibliography
\def\thebibliography{\DeclareRobustCommand{\VAN}[3]{##3}\VANthebibliography}

\usepackage{graphicx}	
\usepackage{amsmath}	
\usepackage{hyperref}
\usepackage{xurl} 

\usepackage{orcidlink}

\usepackage[export]{adjustbox} 
\usepackage{subcaption}

\usepackage{booktabs} 
\usepackage{multirow} 

\newcommand{\Mjup}{M$_{\mathrm{Jup}}$}	
\newcommand{\Rhill}{R$_{\mathrm{Hill}}$}	

\title[Gap-carving planets in exoKuiper belts]{\textit{JWST}/MIRI coronagraphic search for planets in systems with gapped exoKuiper belts and proper motion anomalies}

\author[R. Bendahan-West et al.]{
R.~Bendahan-West~\orcidlink{0009-0000-0303-2145},$^{1}$\thanks{E-mail: rb941@exeter.ac.uk}
S.~Marino~\orcidlink{0000-0002-5352-2924},$^{1}$
A.~L.~Carter~\orcidlink{0000-0001-5365-4815},$^{2}$
V.~Squicciarini~\orcidlink{0000-0002-3122-6809},$^{1}$
A.~D.~James~\orcidlink{0009-0005-6943-6819},$^{1}$
\newauthor
A.~A.~Sefilian~\orcidlink{0000-0003-4623-1165},$^{3}$
T.~D.~Pearce~\orcidlink{0000-0001-5653-5635},$^{4}$
M.~F.~Friebe~\orcidlink{0009-0004-5747-7827},$^{5}$
C.~Lazzoni~\orcidlink{0000-0001-7819-9003},$^{1,6}$,
B.~Lakeland~\orcidlink{0000-0002-8122-2240},$^{7}$
S.~Ray~\orcidlink{0000-0003-2259-3911},$^{8}$
\newauthor
M.~C.~Wyatt~\orcidlink{0000-0001-9064-5598},$^{9}$
L.~Matr\`a~\orcidlink{0000-0003-4705-3188},$^{10}$
J.~Milli~\orcidlink{0000-0001-9325-2511},$^{11}$
V.~C.~Faramaz~\orcidlink{0000-0001-6403-841X},$^{3}$
Th.~Henning,$^{12}$,
S.~Hinkley~\orcidlink{0000-0001-8074-2562},$^{1}$
\newauthor
G.~M.~Kennedy~\orcidlink{0000-0001-6831-7547},$^{13,14}$
D.~Mesa~\orcidlink{0000-0001-8467-1933},$^{5}$
A.~Zurlo~\orcidlink{0000-0002-5903-8316}$^{15,16}$
\\
$^{1}$Department of Physics and Astronomy, University of Exeter, Stocker Road, Exeter EX4 4QL, UK\\
$^{2}$Space Telescope Science Institute (STScI), 3700 San Martin Drive, Baltimore, MD 21218, USA\\
$^{3}$Department of Astronomy and Steward Observatory, University of Arizona, 933 N Cherry Avenue, Tucson AZ 85721, USA\\
$^{4}$Department of Physics, University of Warwick, Gibbet Hill Road, Coventry CV4 7AL, UK\\
$^{5}$Astrophysikalisches Institut und Universit\"atssternwarte, Friedrich-Schiller-Universit\"at Jena, Schillerg\"aßchen 2-3, D-07745 Jena, Germany\\
$^{6}$INAF, Astronomical Observatory of Padua, Vicolo dell'Osservatorio 5, I - 35122, Padua, Italy\\
$^{7}$School of Physics and Astronomy, University of Birmingham, Edgbaston, Birmingham, B15 2TT\\
$^{8}$School of Mathematics and Physics, University of Queensland, St Lucia, QLD 4072, Australia\\
$^{9}$Institute of Astronomy, University of Cambridge, Madingley Road, Cambridge, CB3 0HA, UK\\
$^{10}$School of Physics, Trinity College Dublin, the University of Dublin, College Green, Dublin 2, Ireland\\
$^{11}$Univ. Grenoble Alpes, CNRS, IPAG, F-38000 Grenoble, France\\
$^{12}$Max-Planck-Insitut f\"ur Astronomie, K\"onigstuhl 17, 69117 Heidelberg, Germany\\
$^{13}$Malaghan Institute of Medical Research, Gate 7, Victoria University, Kelburn Parade, Wellington, New Zealand\\
$^{14}$Department of Physics, University of Warwick, Gibbet Hill Road, Coventry CV4 7AL, UK\\
$^{15}$Instituto de Estudios Astrofísicos, Facultad de Ingeniería y Ciencias, Universidad Diego Portales, Av. Ejército Libertador 441, Santiago, Chile\\
$^{16}$Millennium Nucleus on Young Exoplanets and their Moons (YEMS), Chile
}

\date{Accepted XXX. Received YYY; in original form ZZZ}

\pubyear{2025}

\begin{document}
\label{firstpage}
\pagerange{\pageref{firstpage}--\pageref{lastpage}}
\maketitle

\begin{abstract}
Over the past decade, ALMA has uncovered a range of substructures within exoKuiper belts, pointing to a population of undetected planets. With \textit{JWST}'s sensitivity, we now have the opportunity to identify these planets thought to be responsible for the observed substructures in debris discs. We present Cycle 1 \textit{JWST}/MIRI $11.4~\mu$m coronagraphic observations of three exoKuiper belts that exhibit gaps in their radial structures: HD\,92945, HD\,107146, and HD\,206893, to determine whether planets are responsible for carving these structures, as seen in our Solar System with the gas giants. We reduce the \textit{JWST}/MIRI data using \texttt{spaceKLIP}, and introduce new routines to mitigate the Brighter-Fatter effect and persistence. We do not detect any planet candidates, and all detected objects in the field-of-view are consistent with background stars or galaxies. However, by combining \textit{JWST} mass limits, archival observational constraints, and astrometric accelerations, we rule out a significant portion of planet parameter space, placing tight constraints on the planets possibly responsible for these gaps. To interpret these results, we explore multiple gap-carving scenarios in discs, either massless or with non-zero mass, including clearing by in-situ planet(s), as well as shaping by inner planets through mean-motion or secular apsidal resonances. Finally, we conclude that the planets causing the proper motion anomaly in these systems must reside within the inner 20~au.
\end{abstract}




\section{Introduction}
Although thousands of exoplanets have been discovered at small orbital radii through radial velocity and photometric transit surveys, only a limited number of planets more massive than Jupiter have been detected beyond ${\sim} 10$~au \citep[e.g.,][]{cloutier_2024}.  Over the past two decades, the number of wide-orbit planets have gradually increased, largely thanks to advances in high-contrast imaging instruments such as the Spectro-Polarimetric High-contrast Exoplanet REsearch \citep[SPHERE,][]{beuzit_2019} and the Gemini Planet Imager \citep[GPI,][]{macintosh_2014}. These facilities have enabled detections of increasingly lower-mass planets, with sensitivities now reaching down to ${\sim}2-3$~\Mjup~\citep[e.g.,][]{nielsen_2019, vigan_2021, squicciarini_2025}. Today, these ground-based capabilities are being surpassed by the unprecedented contrast and sensitivity offered by \textit{JWST}, now reaching down to Saturn masses and lower in some cases \citep{carter_2021, lagrange+2025}. Pushing these detection limits further is crucial for fully characterising the outer regions of planetary systems.

While high-contrast imaging has advanced our understanding of wide-orbit planets, an alternative and complementary approach to probing these outer regions involves studying exoKuiper belts, i.e., cold  debris discs at tens of au \citep{hughes_2018, marino_2022}. These dusty belts, composed of material ranging from observed $\mu$m-sized grains to inferred km-sized and larger planetesimals, are located at tens of au from their host stars and are considered a common feature of planetary systems, detected around $\sim$20\% of nearby AFGK--type stars \citep{sibthorpe_2018}. Through a collisional cascade, the large bodies in these cold belts grind down into smaller dust grains, which produce the infrared excess observed in these systems. This continuous replenishment of dust counteracts removal processes such as collisions, radiation pressure and Poynting-Robertson (PR) drag, allowing the discs to persist over $\sim$ Myr$-$Gyr timescales \citep{wyatt_2008}.

Observations of debris discs at different wavelengths can provide information regarding the distribution of different-sized grains. At shorter wavelengths (e.g., optical, near-infrared), smaller \micron-sized grains are probed as they scatter the light from the star, making the disc visible. Such scattered light observations have been carried out with different instruments; for instance, on board \textit{HST} \citep[e.g.,][]{golimowski2011, Schneider2014}, with ground based instruments like SPHERE \citep[e.g., HR~4796, TWA~7,][respectively]{milli_2017, ren_2021} or GPI \citep[e.g.,][]{esposito_2020, crotts_2024}, and more recently with \textit{JWST} \citep[e.g., Fomalhaut, Vega, $\beta$ Pic, Fomalhault~C, HD\,181327, $\epsilon$ Eridani,][respectively]{gaspar_2023, su_2024, rebollido_2024, lawson_2024, xie_2025, Wolff_2025}. At millimetre wavelengths, the thermal emission from larger mm-sized grains is detected using the high sensitivity and resolution of ALMA (e.g., REASONS; \citealt{matra_2025} and references therein). While the distribution of smaller grains is affected by non-gravitational forces, the larger grains remain largely unperturbed by these and are therefore thought to trace the parent planetesimal population more accurately. As a result, the structure and morphology of debris discs can be imaged and studied to infer valuable insight into the formation and dynamical evolution of planetary systems. In the Solar System, for example, the resonant populations within the Kuiper belt preserve evidence of Neptune's migration history, while the edges of both the asteroid and Kuiper belts encode information about the masses and orbits of giant planets \citep{Malhotra1995, ida_2000, morbidelli_2005, Morbidelli2020}.

The evidence for substructures and asymmetries in disc morphology is commonly interpreted as an indirect signature of unseen planetary companions. One of the most well-known examples is the warp observed in the $\beta$ Pic disc, which hinted at the presence of a massive planet \citep{mouillet_1997, augereau_2001} later confirmed through direct imaging \citep{Lagrange2009}. Similar warps have since been detected in other systems, suggesting the presence of undetected perturbers \citep[e.g., HD~110058, HD~111520,][respectively]{stasevic_2023, crotts_matthews_2024}. A minority of debris discs appear to be narrow and eccentric \citep[e.g., Fomalhaut, HR~4796, HD~202628,][respectively]{macgregor_2017,kennedy_2018, faramaz_2019}, which may indicate perturbations by eccentric planets \citep{wyatt_1999, Pearce_2014, kennedy_2020, rodet_2022}. Other discs show asymmetric clumps that may result from resonant trapping or recent giant collisions \citep[e.g. $\beta$ Pic; \citealt{telesco_2005, han_2023}, $\epsilon$ Eri;][]{booth_2023}. The sharpness of the inner edge can also provide insights into its origin, i.e., whether edges are sculpted by planets or consistent with collisional evolution alone \citep[e.g.,][]{marino_2021, imaz_blanco_2023, rafikov_2023, pearce2024}. As observational resolution continues to improve, substructures such as gaps are increasingly being detected in debris discs, akin to those commonly observed in protoplanetary discs \citep{andrews_2018}. Such observed gaps in debris discs are analogous to the gap between the asteroid and Kuiper belt in the Solar System, carved and populated by the gas giants. These gap structures provide compelling evidence for planetary companions that sculpt the morphology of these discs.

HD~92945, HD~107146, and HD~206893 are three well-characterised debris disc systems with observed gaps in their dust distribution \citep[][respectively]{marino_2019, marino_2018, marino_2020}. Direct imaging campaigns have searched for gap-carving planets in these systems, ruling out companions above 2-5~\Mjup~at the gap locations \citep{milli_2017, mesa_2021}, though inner planets have been detected in HD~206893 \citep{milli_2017, hinkley_2023}. Therefore, identifying the putative planets predicted to be embedded in these gaps remains an observational challenge.

Nevertheless, the origin of gaps in debris discs remains uncertain and can be explained by multiple different dynamical mechanisms involving the presence of unseen planets. The most straightforward scenario involves a single planet embedded within the gap, which gravitationally perturbs and scatters nearby debris within its chaotic zone. The amount of material being cleared at the gap location can also vary. In some scenarios, Trojans can be captured on the planet orbit and can create a detectable ring of material \citep[as seen around TWA~7b,][]{ren_2021, lagrange+2025, crotts_2025}. The mass required for an in-situ planet depends on the assumptions about the disc mass. In the simplest case, the disc is treated as massless, which does not capture planetesimal-driven planetary migration \citep{mustill_2012, marino_2018}. However, if the disc mass is taken into account, it can induce planet migration, allowing a less massive planet to carve a gap of comparable width as a single non-migrating planet would \citep{kirsh_2009, Morrison+2018, friebe_2022}. Alternatively, a chain of multiple lower-mass planets distributed across the gap could collectively clear the region in the disc, whether the latter is assumed to be massless \citep{Shannon+2016, lazzoni_2018} or massive \citep{Morrison+2018}. In some cases, a planet does not need to be located within the gap itself, and interactions from a planet located in the regions interior to the disc can carve a gap further out. For example, gaps could arise from the 2:1 mean motion resonance \citep{Tabeshian+2016, regaly_2018}, or through secular apsidal resonances induced by one precessing planet in a non-zero mass disc \citep{pearce_2015, Zheng+2017, sefilian_2021, sefilian_2023} or two precessing planets in a massless disc \citep{yelverton_2018}. These various scenarios imply different planet configurations and evolutionary histories, but all remain valid degenerate solutions until ruled out by observational constraints.

This paper aims to provide a comprehensive view of the planet constraints that can be placed for the three systems HD\,92945, HD\,107146 and HD\,206893, and to evaluate, based on these, the current understanding of possible gap-carving scenarios. In Section~\ref{sec: obs+reduction} we present new \textit{JWST}/MIRI coronagraphic observations at $11.4~\mu$m and describe the best practices for reducing \textit{JWST}/MIRI coronagraphic data using the \texttt{spaceKLIP} package. Section~\ref{sec: analysis} outlines the candidate vetting process and the construction of detection probability maps (DPMs), which are then used to rule out planetary parameters in each system. In Section~\ref{sec: planet constraints}, we use these DPMs to place limits on potential planets, focusing on companions near the inner edge of the exoKuiper belts and those responsible for the observed proper motion anomaly (PMa). Section~\ref{sec:gap-carving-scenarios} explores the various constraints on the planets responsible for carving the observed gaps in these discs, with each gap-carving scenario tailored to the system's specific characteristics. Finally, the main conclusions are summarised in Section~\ref{sec: conclusions}.

\section{Observations and Data Reduction}
\label{sec: obs+reduction}
This section presents \textit{JWST}/MIRI coronagraphy observations of three debris disc systems with known gaps: HD~92945, HD~107146, and HD~206893, as part of the Cycle 1 program GO~1668 (PI: S. Marino). Below, we outline the sample selection process for this program, detail the used MIRI coronagraphy observation strategy, and finally describe the data reduction process performed using \texttt{spaceKLIP}, incorporating its latest upgrades.

\subsection{Sample}
HD~92945, HD~107146, and HD~206893 originate from a larger sample containing five debris discs with gaps identified in ALMA data \citep{marino_2018,marino_2019,marino_2020,macgregor_2019,daley_2019, nederlander_2021}. Gaps have been detected in four additional systems using scattered light imaging alone \citep[HD~141569, HD~131835, HD~120326, and HD~141943,][respectively]{perrot_2016, bonnefoy_2017, feldt_2017, boccaletti_2019}. We consider discs resolved by ALMA because, contrary to scattered light detections, ALMA probes the mm-sized grains unaffected by non-gravitational forces that can distort the true extent of the planetesimal disc. Additionally, since the mm-sized dust grains are by-products of continuous planetesimal collisions, the distribution of planetesimals must enclose the presence of the observed gaps.

To image planets within the gaps of the discs, a face-on orientation is optimal. In face-on systems, the \textit{JWST} inner working angle obscures only the innermost region, allowing the rest of the disc and any potential planets to remain fully visible. Conversely, in edge-on systems, parts of the disc are obscured by the inner working angle, limiting the ability to detect planets embedded within the disc. Therefore, given the geometry of the systems, we excluded two of the five discs with gaps due to their edge-on orientation (HD~15115 and AU~Mic).

HD~92945, HD~107146, and HD~206893, share comparable ages, stellar masses, and distances, however, their specific disc characteristics differ slightly (e.g., gap location, number of gaps, gap symmetry, etc). A comprehensive summary of the stellar and disc parameters for these systems can be found in Table~\ref{tab:system_params}. All three targets show strong evidence for an inner companion with astrometric accelerations also known as proper motion anomaly \citep[PMa,][]{kervella+2019, kervella_2022}, measured from the comparison of proper motions between the Hipparcos and Gaia catalogues.

In particular, HD~92945 is a $100-300$~Myr old K1V star at 21.5~pc. This system's disc spans from $54-133$~au and contains a single 19~au gap centred at 72~au \citep{imaz_blanco_2023}. Previous ALMA and \textit{HST} observations hinted at asymmetry in the gap \citep[][respectively]{marino_2019, golimowski2011}, which has been confirmed by recent \textit{JWST}/NIRCam data \citep{Lazzoni_2025}. HD~107146 is a $100-250$~Myr old G2V star at 27.5~pc. Its disc spans from $44-144$~au and features a double-gapped radial profile with a narrow 7.7~au gap centred at 56~au followed by a wider 42~au gap centred at 78~au \citep{imaz_blanco_2023}. Both gaps do not exhibit strong signs of asymmetry \citep{marino_2018, imaz_blanco_2023}. HD~206893, a $140-170$~Myr old F5V star at 41~pc, has a disc spanning from $35-120$~au with a 40~au gap centred at 69~au \citep{imaz_blanco_2023}. Although ALMA observations indicate possible gap asymmetry \citep{marino_2020}, the gap remains consistent with being axisymmetric. This system also hosts two substellar companions: HD~206893~B, a $28.0\pm2.2~M_{\mathrm{Jup}}$ brown dwarf located at $9.6\pm0.3$~au \citep{milli_2017_hd206893,kammerer_2021, hinkley_2023}, and HD~206893~c, a $12.7\pm1.1~M_{\mathrm{Jup}}$ planet at $3.53\pm0.07$~au found to be responsible for the PMa in the system \citep{hinkley_2023}.

{
\setlength{\tabcolsep}{4pt}
\begin{table}
    \centering

\caption{Stellar and debris disc properties for HD~92945, HD~107146, and HD~206893. The disc inner and outer edges are denoted by $r_{\mathrm{in}}$ and $r_{\mathrm{out}}$, respectively, while the centre of the gap(s) is indicated by $r_{\mathrm{gap}}$. The gap widths ($w_{\mathrm{gap}}$) are calculated as the FWHM, based on the $1\sigma$ values given in \citet{imaz_blanco_2023}. Sources: $^{(1)}$ \citet{gaiadr3_collaboration_2023}, $^{(2)}$ \citet{torres_2006}, $^{(3)}$ \citet{harlan_1970}, $^{(4)}$ \citet{gray_2006}, $^{(5)}$ \citet{plavchan_2009}, $^{(6)}$ \citet{pearce2022}, $^{(7)}$ \citet{kervella_2004}, $^{(8)}$ \citet{Chen_2014}, $^{(9)}$ \citet{holland_2017}, $^{(10)}$ \citet{stanford-moore_2020}, $^{(11)}$ \citet{hinkley_2023}, $^{(12)}$ \citet{marino_2021}, $^{(13)}$ \citet{imaz_blanco_2023}.}
\label{tab:system_params}
\begin{tabular}{llcc}
    \hline
    \hline
    Parameters &  \textbf{HD\,92945}&\textbf{HD\,107146}& \textbf{HD~206893}\\
     \hline
   Distance [pc] &  $21.51 \pm 0.01$$^{(1)}$ & $27.47\pm0.02$$^{(1)}$ & $40.77\pm0.06$$^{(1)}$ \\
   Spectral Type &  K1V$^{(2)}$ &G2V$^{(3)}$ & F5V$^{(4)}$ \\
 Star mass [M$_{\odot}$]& $0.86\pm0.01$$^{(5)}$& $1.03^{+0.02}_{-0.04}$$^{(6)}$&$1.32^{+0.07}_{-0.06}$$^{(7)}$\\
   Age [Myr] &  $200\pm100$$^{(8,9,10)}$&$150^{+100}_{-50}$$^{(8,9,10)}$& $155\pm15$$^{(11)}$\\
   Inclination [$^\circ$] &  $65.4\pm0.6$$^{(12)}$&$19.9\pm0.6$$^{(12)}$& $40\pm3$$^{(12)}$\\
   PA [$^\circ$] &  $100.0\pm0.6$$^{(12)}$&$153.3\pm1.5$$^{(12)}$& $62\pm4$$^{(12)}$\\
   $r_{\mathrm{in}}$ [au] &  $54\pm2$$^{(13)}$&$44\pm2$$^{(13)}$& $35\pm8$$^{(13)}$\\
   $r_{\mathrm{out}}$ [au] &  $133\pm6$$^{(13)}$&$144.3\pm1.0$$^{(13)}$& $120\pm20$$^{(13)}$\\
   $r_{\mathrm{gap,1}}$ [au] &  $72.0\pm1.5$$^{(13)}$&$56.0\pm0.7$$^{(13)}$& $69\pm3$$^{(13)}$\\
   $w_{\mathrm{gap,1}}$ [au] &  $19\pm9$$^{(13)}$&$7.7\pm1.4$$^{(13)}$& $40\pm9$$^{(13)}$\\
   $r_{\mathrm{gap,2}}$ [au] &  ... &$78.3\pm1.2$$^{(13)}$& ... \\
   $w_{\mathrm{gap,2}}$ [au] &  ... &$42\pm6$$^{(13)}$& ... \\
   \hline
\end{tabular}
\end{table}
}

\subsection{Observations}

We present \textit{JWST}/MIRI observations using the 4-quadrant phase-mask \citep[4QPM,][]{rouan2000, lajoie_2014, Boccaletti_2015} and the F1140C coronagraphic filter as part of the Cycle 1 program GO~1668 (PI: S. Marino). The program was designed to obtain an optimal data reduction, allowing for different point-spread function (PSF) subtraction methods to be performed: Angular Differential Imaging \citep[ADI,][]{muller_1987, Marois_2006}, Reference Differential Imaging \citep[RDI,][]{Ruane_2019}, and a combination of ADI and RDI (ADI+RDI). All the data reduction was conducted using \texttt{spaceKLIP} \citep{Kammerer_2022, carter_jwst_2023}.

\begin{table*}
    \centering
\caption{Observing parameters for the GO~1668 \textit{JWST} program, with science targets in bold and corresponding PSF references in italics. Spectral types and Kmag are obtained from Simbad. Exposure times represent total durations per star; individual exposures require division by the number of dithers and rolls. Note that background observations were performed for every science roll and following the dithered reference observations, all using the same settings as their counterpart observations.}
\label{tab:obs_params}
    \begin{tabular}{lcccccccccl}
        \hline
        \hline
         Star&  SpT&  Kmag&  Readout&  $N_{\mathrm{groups}}$&  $N_{\mathrm{ints}}$&  $t_{\mathrm{exp}}$ (s)&  $N_{\mathrm{dithers}}$&  $N_{\mathrm{rolls}}$&  Roll angle ($^\circ$)&$t_{\mathrm{total}}$ (s)\\
         \hline
         \textbf{HD~92945} &  K1V&  5.660&  FASTR1&  1251&  6&  1800.236&  1&  2&  7&3600.472\\ 
         \textit{HD~95234}&  M1III&  1.537&  FASTR1&  145&  2&  69.747&  9&  1&  --&627.722\\ \hline
         \textbf{HD~107146}&  G2V&  5.540&  FASTR1&  1251&  6&  1800.236&  1&  2&  7&3600.472\\ 
         \textit{HD~111067}&  K4III&  1.917&  FASTR1&  275&  2&  132.064&  9&  1&  --&1188.573\\ \hline
         \textbf{HD~206893}&  F5V&  5.593&  FASTR1&  1251&  6&  1800.236&  1&  2&  7&3600.472\\ 
         \textit{HD~208445}&  M4III&  2.061&  FASTR1&  240&  2&  115.286&  9&  1&  --&1037.575\\ 
         \hline
    \end{tabular}
\end{table*}

Based on pre-launch estimates \citep[from \texttt{PanCAKE} simulations;][]{carter_pancake_2021}, the MIRI F1140C filter provided the best planet mass sensitivity at the location of the gaps for the three systems. The instrument settings were optimised using the \texttt{PanCAKE} tool and are summarised in Table~\ref{tab:obs_params}.

Following pre-launch performance predictions, we opted for an observing strategy allowing for two roll angles on the science targets in addition to PSF reference stars. Each roll for the science observations is shifted by $7^\circ$ with 30-minute exposure times. To improve the spatial diversity for the PSF subtraction, we performed 9-POINT-SMALL-GRID dithers \citep{Lajoie_2016} for the reference stars. The exposure time of the PSF stars is scaled such that the signal-to-noise achieved per dither observation is comparable to the signal-to-noise of the science observations.

The different reference stars have been selected using the SearchCal\footnote{\url{https://www.jmmc.fr/english/tools/proposal-preparation/search-cal/}} tool. We looked for the brightest star within $20^\circ$ of the science target and that would not saturate in less than 5 groups in the target acquisition. Note that spectral type matching between science and reference targets is not problematic for observations past 5~$\mu$m, as the emission from the science targets and reference stars is in the Rayleigh-Jeans regime\footnote{\url{https://jwst-docs.stsci.edu/methods-and-roadmaps/jwst-high-contrast-imaging/jwst-high-contrast-imaging-proposal-planning/hci-psf-reference-stars}}.
We also verified that these stars lack any stellar companions within ${\sim}100$~au, as indicated by the Gaia Renormalized Unit Weight Error (RUWE)~$<1.4$ \citep[][and more details in \S\ref{subsec: ruwe}]{gaiadr3_collaboration_2023} and astrometric accelerations larger than 3$\sigma$ \citep{kervella+2019}. Additionally, we ensured there was no infrared excess indicative of a discernible disc that could interfere with the chosen reference stars.

For MIRI coronagraphy, background observations are required to counter an inherent stray light artefact known as the "glow stick" \citep{boccaletti_2022}, which is seen to dominate any science or reference observations. The recommended solution is to observe a nearby area of the sky without bright sources, using the same integration parameters (i.e., same number of groups, integrations, exposure time) as the corresponding science or reference observations. This background observation will identically reproduce the "glow stick", allowing for an optimal subtraction of this artefact. The program was divided into three uninterrupted sequences, each dedicated to one of the science stars. Each sequence included the two science rolls, reference dithers, and background observations. This approach minimises potential wavefront drifts between observations, which could otherwise compromise the PSF subtractions.

\subsection{Data reduction with \texttt{spaceKLIP}}
\label{sec: data-reduction}
The data reduction process follows the example and guidance given by the ERS-01386 coronagraphic program on HIP~65426~b \citep{hinkley_ers_2022a, carter_jwst_2023}. The entire data processing uses the python package \texttt{spaceKLIP}\footnote{\url{https://github.com/spacetelescope/spaceKLIP}}, a custom pipeline combining coronagraphic tools such as the official \texttt{jwst} pipeline\footnote{\url{https://jwst-pipeline.readthedocs.io}} \citep{jwst_package_2022} and \texttt{pyKLIP} subtraction techniques \citep{wang_pyklip_2015}. The reduction used the \texttt{spaceKLIP} version \verb|2.2.1.dev15+gf97eb19|, the \texttt{jwst} pipeline version 1.18.1, the \texttt{pyKLIP} version 2.8.2, the CAlibration REference Data System (CRDS) version 12.1.10, and the CRDS context file \verb|jwst_1364.pmap|.

We downloaded the uncalibrated data files (version 2023\_2a) from the MAST archive\footnote{The data described here may be obtained from the MAST archive at \href{https://doi.org/10.17909/gsf1-2q34}{doi:10.17909/gsf1-2q34}} and followed through the different \texttt{spaceKLIP} reduction stages, as described below. Note that here we focus on the key \texttt{spaceKLIP} parameters and highlight any deviations from the default pipeline. For a more comprehensive description of the \texttt{spaceKLIP} steps, refer to \citet{Kammerer_2022}, \citet{carter_jwst_2023} and the \textit{readthedocs} page\footnote{\url{https://spaceklip.readthedocs.io/en/latest/index.html}}. The next steps are presented in the order they were applied during data reduction.

\begin{figure*}
    \centering
    \includegraphics[width=\linewidth]{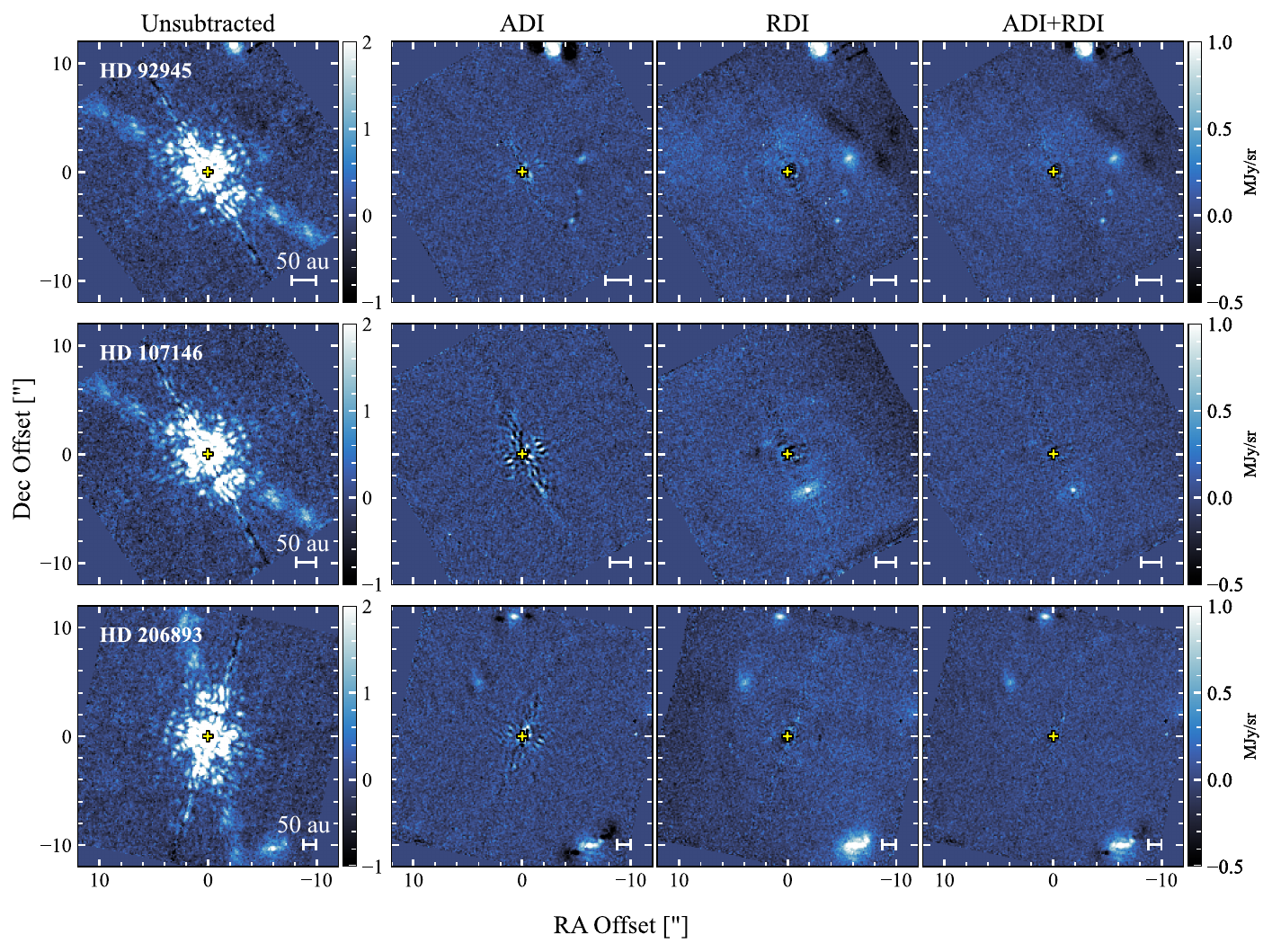}
    \vspace{-5mm}
    \caption{Grid displaying the unsubtracted images and the results of the different PSF subtraction techniques (ADI, RDI, and ADI+RDI) used for HD\,92945, HD\,107146, and HD\,206893. All unsubtracted data share the same colour scale, while all the PSF subtracted data share a different colour scale, both with units of MJy/sr. The data reduction was done using \texttt{spaceKLIP} as described in \S\ref{sec: data-reduction}, with the images shown using the maximum KL modes. The full MIRI field-of-view ($24\arcsec\times24\arcsec$) is shown with all images rotated to North-East. The scale bar at the bottom right corner represents a projected distance of 50~au.}
    \label{fig:subtractions}
\end{figure*}

\textbf{I.~ Stage 1:} Most of the Stage 1 parameters have not been changed and followed the default values used in the MIRI coronagraphy tutorial available online on the \texttt{spaceKLIP} \textit{readthedocs} page$^6$ (accessed on 20/08/2025). The only changes are the following:

\textit{a) LIKELY ramp fitting:} The ramp fitting used in this paper differs from the default ramp fitting algorithm \verb|OLS_C| (Ordinary Least Squares). While \verb|OLS_C| fits straight-line segments to the ramp assuming uniform Gaussian noise, the \verb|LIKELY| algorithm we employed instead maximises a likelihood function that models the detector’s true noise properties, including both photon and read noise. More details about this method are found in \citet{brandt2024a, brandt2024b}. This approach provides more accurate slope estimates and improves the identification of cosmic ray events, leading to more reliable flux measurements.

We found that in all the uncalibrated files, the first ${\approx}10$ groups and the last 2 groups of each integration behave differently from the remaining groups (see Figure~\ref{fig: ramp_fitting_visualisation}). To account for this, we excluded these 12 groups during ramp fitting, which results in only a ${\approx}1$\% loss of usage groups.

\textit{b) Brighter-Fatter effect correction:} The Brighter-Fatter effect (BFE) appears when the detector response becomes non-linear due to strong flux contrasts between neighbouring pixels, resulting in distorted and broadened PSFs near bright pixels \citep{argyriou_2023}. In our case, the reference stars have a higher detector count than the science targets, producing noticeable BFE and causing the reference PSF to be different and broader near those brighter pixels. For the ERS program observing HIP~65426 \citep{carter_jwst_2023}, the BFE was negligible as the detector count levels between the reference and science observations were better matched. This discrepancy was due to the observing parameter optimisation in \texttt{PanCAKE} used for program GO\,1668, which preferred long ramps over matching BFE. A new functionality to match the detector counts in \texttt{PanCAKE} is under development.

To correct for this effect, we use the \verb|mask_groups| option implemented in the Stage 1 pipeline of \texttt{spaceKLIP}. This function trims the necessary amount of groups during the ramp fitting of the brighter images, in our case the reference and corresponding background observations, to match the BFE between the reference and science PSFs. Three masking methods are available: `custom', `basic' and `advanced'. The `custom' method uniformly trims a fixed number of groups across all pixels and integrations. The `basic' method optimises the number of groups to trim over a central subset of the image, where the PSF lobes and BFE are strongest. We adopt the `advanced' method, which performs the optimisation on a pixel-by-pixel basis and yields the most accurate correction for this dataset. More details about the `advanced' method and the differences between the methods are provided in Appendix~\ref{app: bfe}.

\textbf{II.~ Stage 2:} The main goal of this step is to calibrate each integration of the images from counts/s to MJy/sr. We kept the default parameters.

\textbf{III.~ Image processing:} The following steps describe the steps performed before the PSF subtraction. We build on the online MIRI coronagraphy tutorial$^6$, with parameters empirically adjusted to best match this dataset. Here, we only focus on the parameters that were modified from their default values. In addition, we introduce two new functions: background observation cleaning and persistence trimming.

\textit{a) Bad pixel cleaning:} First, the pixels flagged in the Data Quality array during the ramp fitting stage were filled in with the \verb|interp2d| method with a kernel size of 7 pixels. We found that increasing the kernel size from the default 3 pixels to 7 while using \verb|interp2d| produced less residuals in the final PSF-subtracted reductions. We then used sigma-clipping (\verb|sigclip|) with a sigma of 3 to find additional bad pixels and cleaned them using the \verb|interp2d| method with a kernel size of 7 pixels. We then identified bad pixels from temporal variations across integrations using \verb|timeints| and a sigma of 4, which we then cleaned with \verb|timemed|. Finally, the remaining ${\sim}5-10$ bad pixels were cleaned manually with the \verb|custom| method, using an \verb|interp2d| kernel of 7 pixels. It is not always possible to apply an accurate correction to all bad pixels (e.g., those in PSF lobes), which can leave some visible residuals (see discussion in \S\ref{source_vetting}).

\textit{b) Background observation cleaning:} Although background observations are designed to sample an empty region of the sky, contaminating sources can still fall within the background field-of-view. This was observed in the science backgrounds of the HD\,92945 observations, which resulted in negative sources contaminating the final reductions for this star (Figure~\ref{fig:bg_cleaning}). Since background observations must be repeated and centred on diagonally opposed quadrant, we can compare the two backgrounds to identify and remove such contaminants. This is done automatically by flagging $\pm3\sigma$ outliers between the two background images, yielding cleaner final PSF-subtracted reductions (see Appendix~\ref{app: bg_cleaning}). The method is implemented in \texttt{spaceKLIP} as \verb|clean_backgrounds|.

\textit{c) Background subtraction:} 
We apply the \verb|subtract_background_godoy| function from \texttt{spaceKLIP}, described in \citet{godoy2024}. Consequently, unlike \citet{carter_jwst_2023}, we do not discard the first integration of the science or reference observations, which was previously done because of an apparent increase in noise explained by Reset Switch Charge Decay (RSCD). This method therefore prevents from losing 16\% of the science observations and 50\% of the reference observations.

\textit{d) Image alignment:} Following \citet{carter_jwst_2023}, we did not realign the MIRI images due to the challenging alignment of the complex MIRI 4QPM PSF structure.

\textit{e) Persistence trimming:} Bright sources can leave a lasting imprint on the \textit{JWST} detectors through a phenomenon known as persistence, in which the signal from earlier exposures is not completely cleared before the following ones \citep{dicken_2024}. Persistence can introduce spurious structure into the data that may be misinterpreted as real sources. We detect this effect in all the MIRI observations from this program (Figure~\ref{fig:persistence}), as well as in other programs \citep[e.g., James et al., in prep.; see also the source identified as a speckle in][]{godoy2024}. 

During MIRI coronagraphic observations, the star is first placed in two different target acquisition (TA) regions\footnote{\url{https://jwst-docs.stsci.edu/jwst-mid-infrared-instrument/miri-operations/miri-target-acquisition/miri-coronagraphic-imaging-target-acquisition##MIRICoronagraphicImagingTargetAcquisition-4QPMtargetacquisition}}, which coincide with the locations where persistence is observed. To mitigate this effect, we identify the stellar positions in both TA images and mask the corresponding pixels in any integrations affected by persistence. The full process is described in Appendix~\ref{app: persistence_trimming} and is implemented in \texttt{spaceKLIP} as the \verb|persistence_trimming| function.

\textbf{IV.~ PSF subtraction:} The PSF subtraction techniques used follow three different principal component analysis (PCA) based methods implemented in \texttt{spaceKLIP} through \texttt{pyKLIP}: ADI, RDI and a combination of ADI+RDI. More details are provided in \citet[][\S2.6]{carter_jwst_2023}. Figure~\ref{fig:subtractions} shows an overview of the resulting images using all PSF subtraction strategies for HD\,92945, HD\,107146, and HD\,206893.

There are two main additional parameters to include during the PSF subtraction: the number of KLIP PCA modes (also known as KL~modes) used to determine how aggressive the subtraction is, and the number of annuli/subsections \citep{carter_jwst_2023}. In our case, we are limited to 6~KL~modes for ADI (6 integrations in a single roll), 18~KL~modes for RDI (9 dithers of 2 integrations each), and 24~KL~modes for ADI+RDI (sum of ADI and RDI). The data reduction was done on the full range of KL modes [1, max], and we scan all the KL modes to look for significant point sources. The ability to find point sources depends on the number of KL modes used (e.g., level of residuals and over-subtraction) and ultimately affects the contrast (see \S\ref{contrast_curves}).

PSF subtraction can be performed on images as a whole or images partitioned into different annuli and subsections. Splitting the image into different annuli/subsections allows the possibility to target PSF subtraction in specific regions of the image. While the image gets clearer (i.e., fewer residuals) when increasing the number of annuli and subsections, the resulting contrast deteriorates through a decrease in the algorithmic throughput of the subtraction process (i.e., any signal suffers from stronger over-subtraction). We derived the deepest contrast when using the image as a whole i.e., a single annulus and a single subsection.

\section{Analysis}
\label{sec: analysis}
This section details the analysis of the PSF-subtracted images taken to extract any planet candidate detection. We first examine in \S\ref{source_vetting} whether the sources observed in the processed images are planet candidates or background contaminants based on multi-wavelength and multi-epoch archival observations of the three systems. After the source vetting, we quantify in \S\ref{contrast_curves} the contrast sensitivity of the observations and constrain the potential mass and semi-major axis of planets in the systems using detection probability maps (DPMs) and PMa in \S\ref{sec: DPM}.

\begin{table*}
    \centering
    \caption{Properties of the sources detected within the field-of-view of the observations. Sources are numbered from North to South, as shown in Figure~\ref{fig:bkg_vetting}. Columns 4-6 list the coordinates derived from the PSF-fitting, with the corresponding observation epoch given in column 2. The source morphology (extended or point-like) is determined from the PSF-fitting results shown in Figure~\ref{fig: psf-fitting}. For extended sources, fluxes are measured via aperture photometry, whereas the flux of HD\,92945 C4 is derived from point-source fitting. We note that the flux of HD\,107146 C1 may be affected by its location between two quadrants of the 4QPM.}
    \begin{tabular}{lcccccccc}
        \hline
        \hline
         Target & Obs. date & Source&  $\Delta$RA, $\Delta$Dec&  RA &  Dec &  Flux 11~$\mu$m &  Extended?&  Archival detection?\\
         & & & [$\arcsec$]& [s]& [$\arcsec$]& [$\mu$Jy] & & \\
         \hline
         HD\,92945& 13 June 2023 & C1&  -2.6, 11.4&  10:43:27.69&  -29.03.41.21&  103.5&  Y&  ALMA + \textit{HST}\\ 
         & & C2&  -5.3, 1.2&  10:43:27.48&  -29.03.51.41&  19.8&  Y&  ALMA + \textit{HST}\\ 
         & & C3&  -4.9, -1.9&  10:43:27.51&  -29.03.54.51&  2.9&  Y&  ALMA + \textit{HST}\\ 
         & & C4&  -4.3, -4.5&  10:43:27.56&  -29.03.57.11&  6.8& N &  Gaia\\ \hline
         HD\,107146& 15 June 2023 & C1&  -1.7, -3.2&  12:19:06.10&  16.32.47.17&  35.5&  Y&  \textit{HST}\\ \hline
         HD\,206893& 19 June 2023 & C1&  1.0, 10.9&  21:45:22.12&  -12.46.49.17&  8.7&  Y &  N\\ 
         & & C2& 4.2, 4.9& 21:45:22.34& -12.46.55.17& 8.5& Y&N\\
         & & C3& -5.9, -10.1& 21:45:21.65& -12.47.10.17& 92.3& Y&\textit{HST}\\
        \hline
    \end{tabular}
    \label{tab:source_vetting}
\end{table*}

\begin{figure*}
    \centering
    \includegraphics[width=\linewidth]{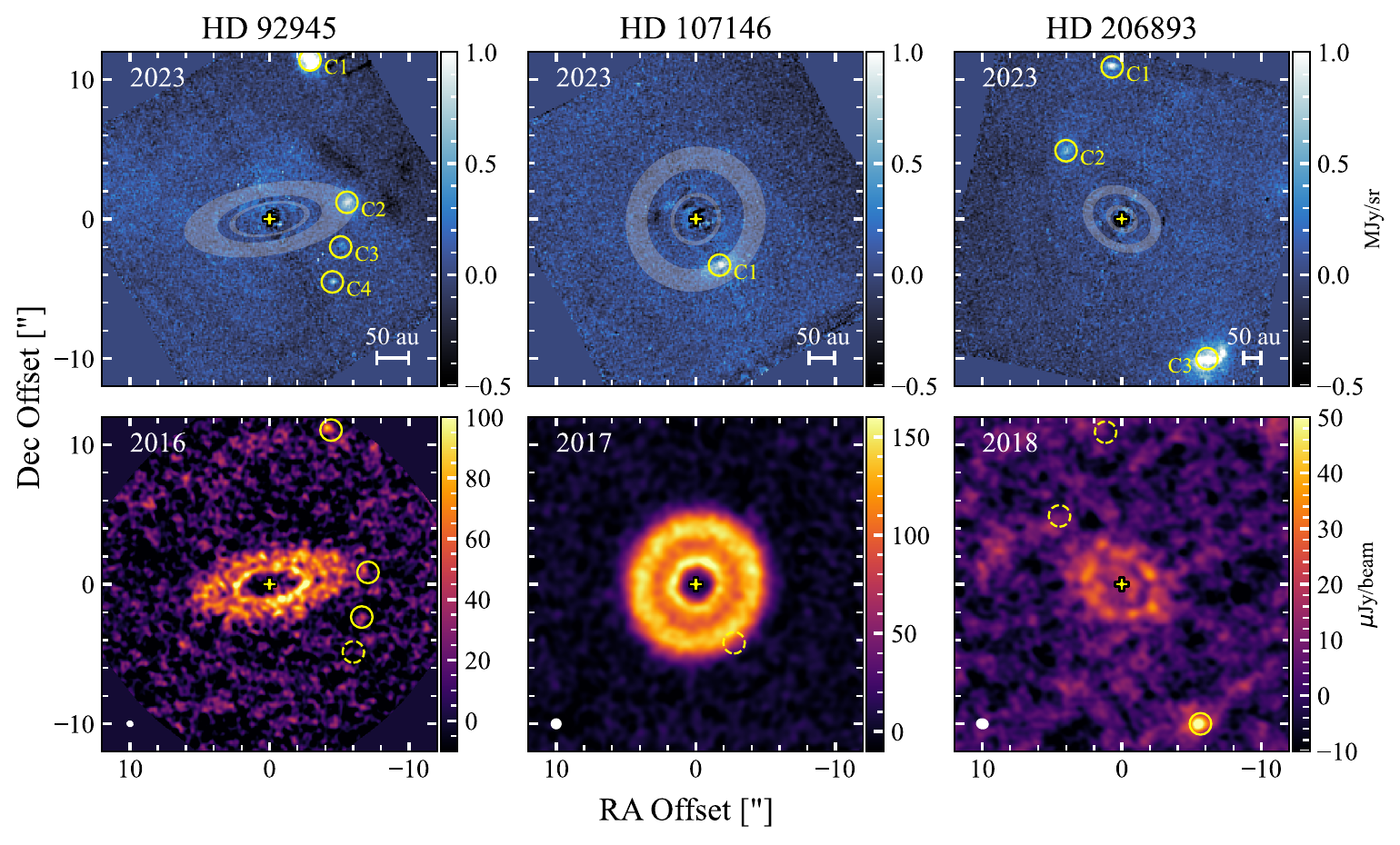}
    \caption{Comparison between the $11.4~\mu$m \textit{JWST}/MIRI data (top row) and archival ALMA observations (bottom row) for HD\,92945, HD\,107146, and HD\,206893 obtained from \citet[][respectively]{marino_2019, marino_2018, marino_2020}. The radial extent of the ALMA disc is over-plotted on the \textit{JWST} data to guide the eye. The positions of the sources in the \textit{JWST} observations are traced back to account for the proper motion of these stars to determine whether they are co-moving or background objects. Solid yellow circles indicate sources observed in both \textit{JWST} and ALMA (C1, C2, and C3 in HD\,92945, and C3 in HD\,206893), while dashed yellow circles denote sources without ALMA counterparts (C4 in HD\,92945, C1 in HD\,107146, and C1 and C2 in HD\,206893). Only C4 in HD\,92945 is consistent with being a star, while the others are consistent with background galaxies.The scale bar in the bottom right corner represents a projected distance of 50~au. The white circle in the bottom left corner of the ALMA plots denotes the beam size used in the observations.}
    \label{fig:bkg_vetting}
\end{figure*}

\subsection{Source vetting}
\label{source_vetting}

As illustrated in Figure~\ref{fig:subtractions}, the processed images contain several candidate sources. However, we found that no sources are consistent with a companion planet. All the observed candidates are found to be background objects. From the PSF-fitting, the majority of the sources are extended, indicating that they are likely background galaxies. The following section describes the vetting process characterising each source.

A prominent residual at ${\approx}1\arcsec$ from the image centre is identified in the RDI panels of HD\,92945 and HD\,107146 (Figure~\ref{fig:subtractions}). This residual is caused by a bad pixel within a PSF lobe, which makes the lower-left lobe (in the de-rotated frame) consistently the brightest across all observations. This asymmetry in the PSF is difficult to correct during the BFE correction and PSF subtraction steps (Figure~\ref{fig: BFE correction}). We do not consider this feature as a candidate source in our analysis.

For the vetting process, we examined multi-wavelength and multi-epoch archival data for all three systems to verify if the same sources were detected previously. For data collected at different epochs, we considered the star’s proper motion to determine whether an observed source is co-moving with the star or is a background object. Most sources detected in the \textit{JWST}/MIRI 11.4~$\mu$m observations are background contaminants identified in archival ALMA (mm \& sub-mm) data, archival high-contrast imaging data with \textit{HST} (optical \& IR), and the Gaia archive. Additionally, we forward modelled each source with a point-like PSF using the \verb|extract_companions| function in \texttt{spaceKLIP} (see resulting fits in Figure~\ref{fig: psf-fitting}). Planets and background stars should be well fitted with a point-like model whereas background galaxies should be better fitted with an extended PSF model. Table~\ref{tab:source_vetting} summarises the properties of the sources observed within the field-of-view of the observations, while Figure~\ref{fig:bkg_vetting} compares the \textit{JWST}/MIRI observations with archival ALMA observations.

Four sources are detected in the HD\,92945 observations (left panels of Figure~\ref{fig:bkg_vetting}). By tracing back the stellar proper motion, C1, C2, and C3 align with counterparts in archival \textit{HST}/ACS data \citep[F606W and F814W,][]{golimowski2011} and in ALMA data \citep[0.86 mm,][]{marino_2019}. Their expected positions as background objects are marked by yellow circles in the lower panel of Figure~\ref{fig:bkg_vetting}, consistent with ALMA detections at $>3\sigma$ significance (5.5$\sigma$, 4.2$\sigma$, 3.2$\sigma$, for C1, C2, C3, respectively). PSF-fitting further shows that all three sources are extended (Figure~\ref{PSF-fitting-hd92945}), supporting their interpretation as background galaxies. Within a $15\arcsec$~radius of HD\,92945, the Gaia archive reports only one star, Gaia DR3 5455707157212258048, which, after accounting for HD\,92945's proper motion, lies $\sim$0.6\arcsec from the position of C4. The offset could reflect the absence of a proper motion measurement for the Gaia star. PSF-fitting shows that C4 is consistent with a point source and is therefore most likely a background star.

For HD\,107146, only one source is detected within the field-of-view (middle panels of Figure~\ref{fig:bkg_vetting}). Although it is not seen in archival ALMA observations \citep[1.14 mm,][]{marino_2018}, the source is clearly visible in archival \textit{HST}/ACS imaging as a background galaxy \citep[F606W and F814W,][]{Ardila2004, Schneider2014}. Its predicted location within the disc around 2020 \citep{Schneider2014}, is consistent with current observations. PSF fitting (Figure~\ref{PSF-fitting-hd107146}) confirms that the source is extended, further supporting its classification as a background galaxy. We note that its measured flux may be strongly attenuated by its position at the boundary between two quadrants of the 4QPM.

Three sources are visible in the HD\,206893 observations (right panels of Figure~\ref{fig:bkg_vetting}). None of the sources are consistent with the two previously known companions in the system, HD\,206893~B and c \citep[at separations of 0.2\arcsec and 0.1\arcsec, respectively,][]{hinkley_2023}, which lie too close to the star within the MIRI inner working angle \citep[0.36\arcsec,][]{boccaletti_2022}. By tracing back the proper motion of HD\,206893, we find that C3 aligns with counterparts in archival ALMA data \citep[0.89 mm,][]{marino_2020} and raw \textit{HST}/NICMOS data \citep[F110W and F160W,][]{milli_2017}, confirming it as a background object. PSF-fitting (Figure~\ref{PSF-fitting-hd206893}) shows that C3 is extended, consistent with being classified as a background galaxy. The other two sources, C1 and C2, have no counterparts in previous observations. PSF-fitting indicates that both are extended, suggesting they are also likely background galaxies.

\subsection{Contrast curves}
\label{contrast_curves}
\begin{figure}
    \centering
    \includegraphics[width=\linewidth]{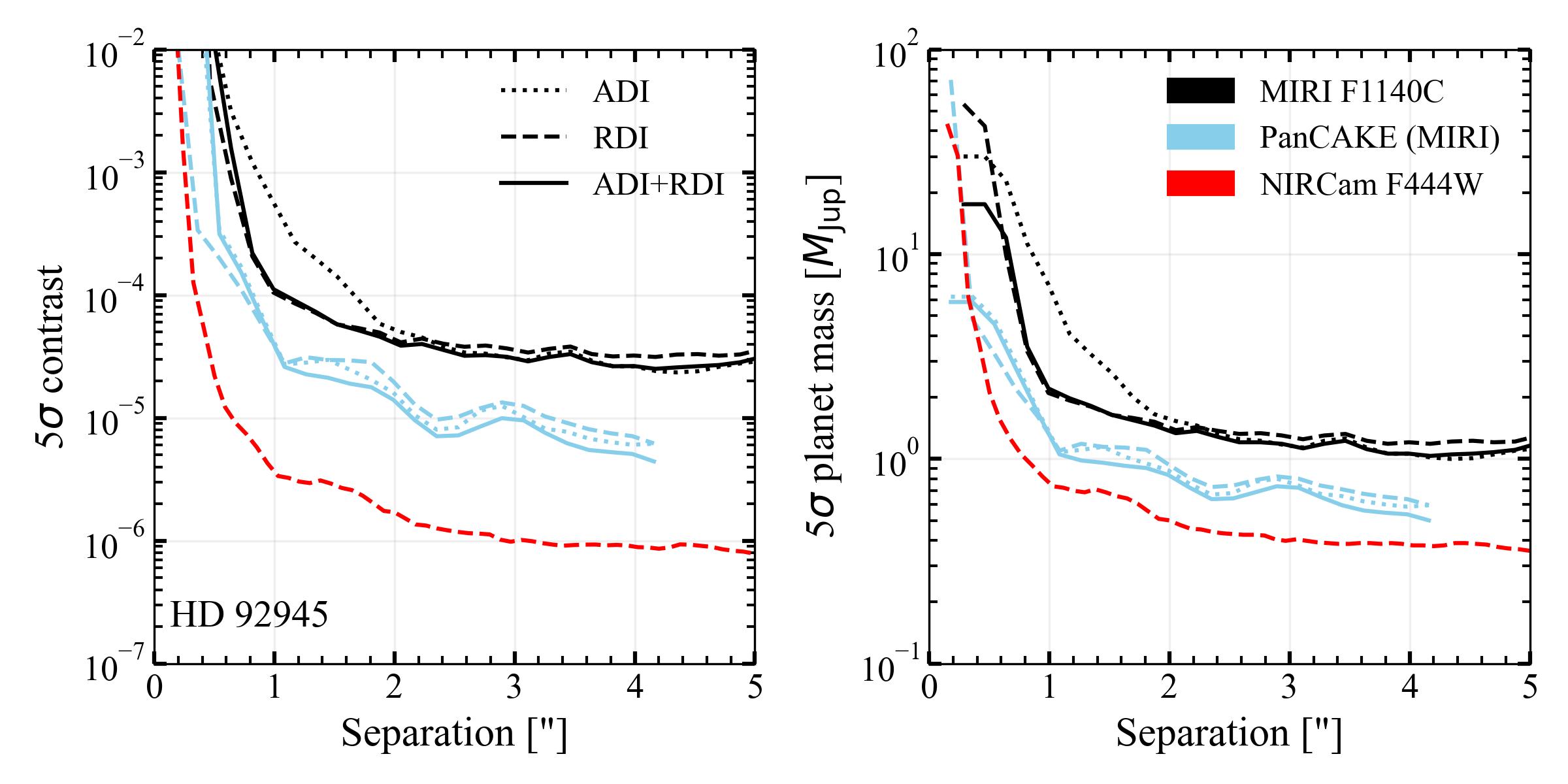}
    \includegraphics[width=\linewidth]{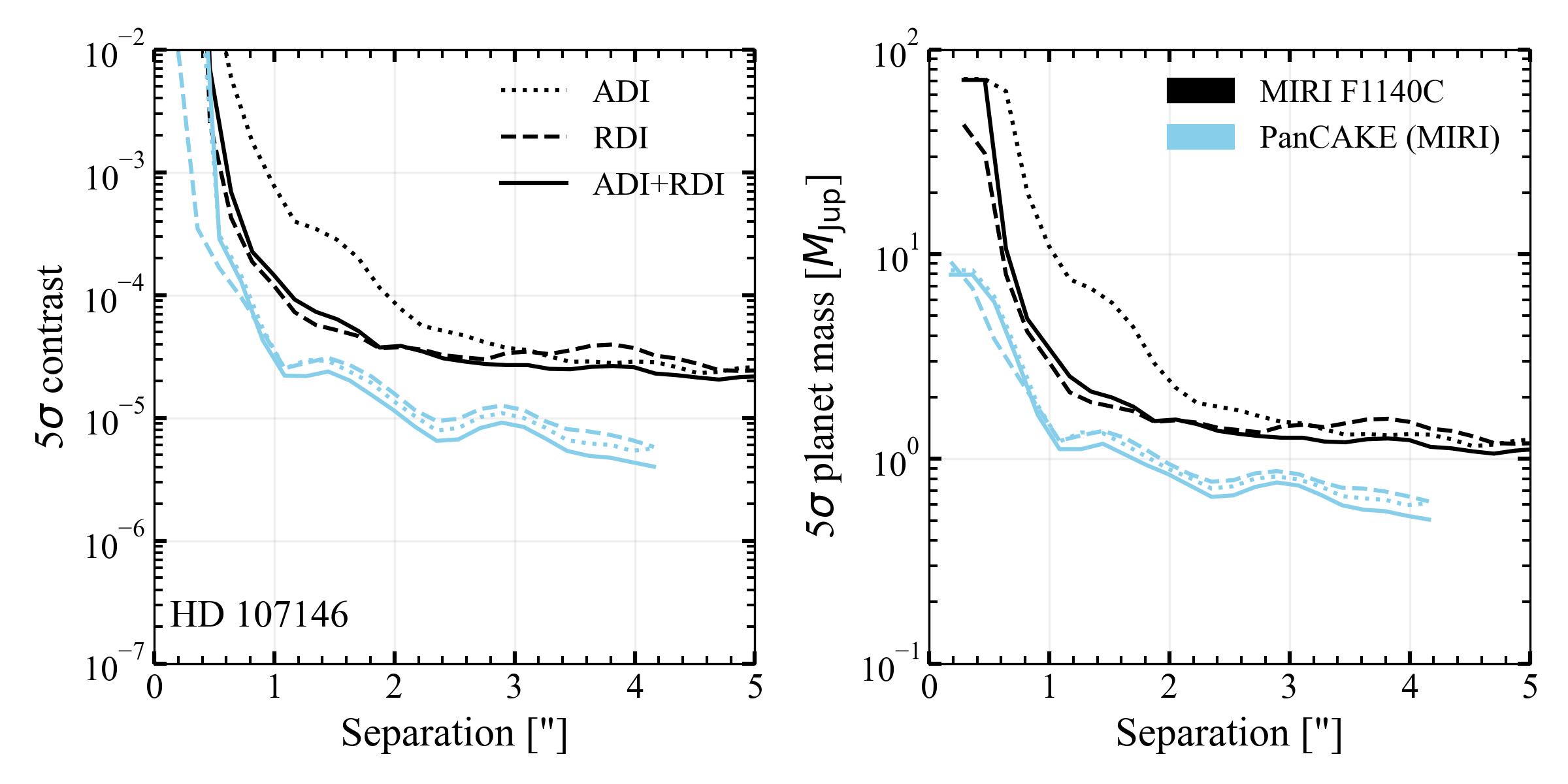}
    \includegraphics[width=\linewidth]{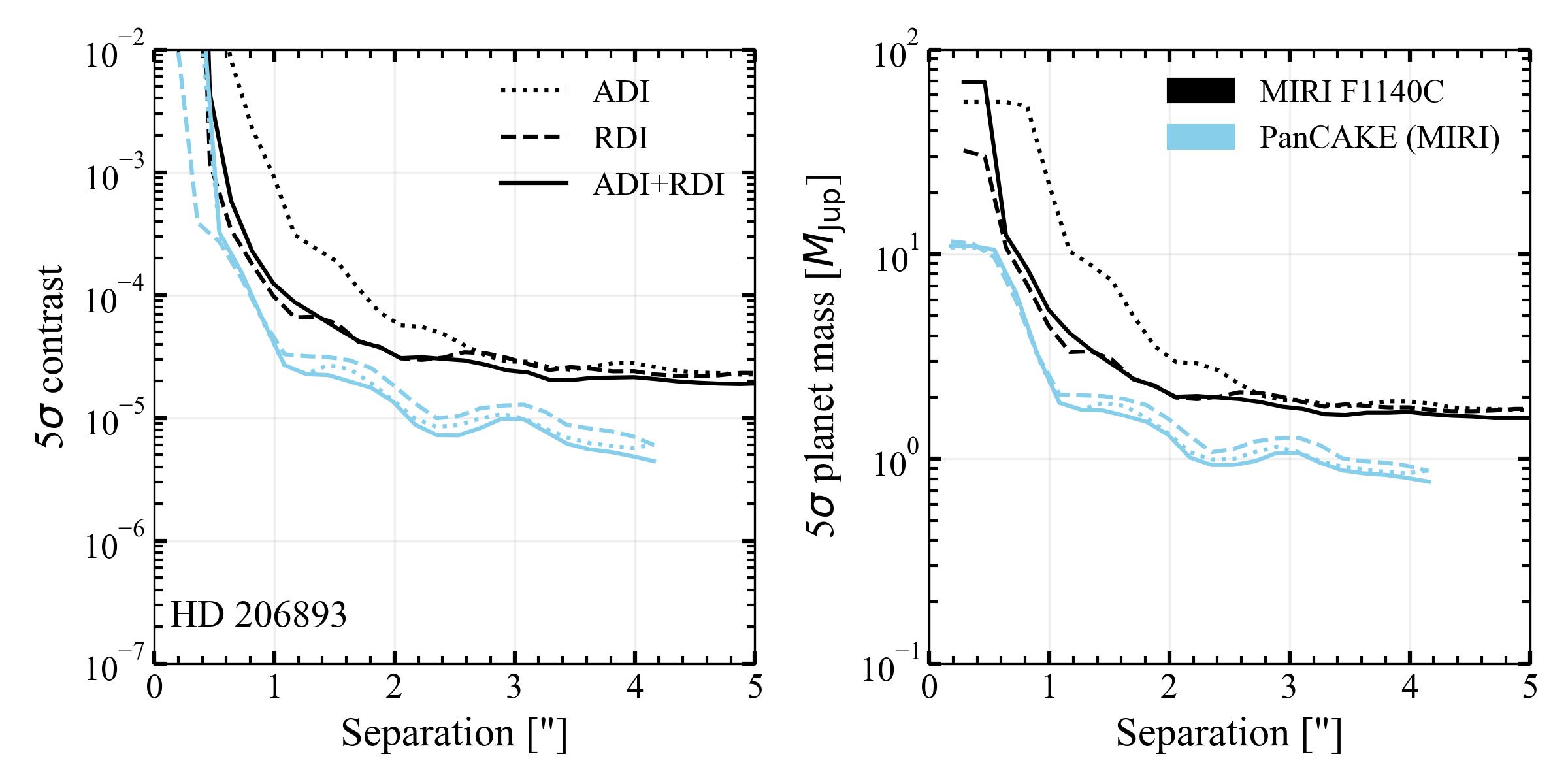}
    \vspace{-5mm}
    \caption{Calibrated $5\sigma$ contrast curves (left) for the MIRI 1140C observations with the respective mass sensitivities in units of \Mjup~(right), as a function of projected separation. Black curves correspond to MIRI F1140C contrasts, blue curves correspond to \texttt{PanCAKE} contrast predictions, and for HD\,92945 only, the red curve corresponds to the NIRCam F444W contrast \citep[from GO\,3989,][]{Lazzoni_2025}. The different line styles represent contrast obtained using different PSF subtraction techniques (i.e., ADI, RDI, and ADI+RDI). All contrasts were calculated with \texttt{spaceKLIP} using the maximum number of KL modes and with 1 annulus/1 subsection.}
    \label{fig:contrast curves}
\end{figure}

Even though we did not detect any companions, non-detections can be used to set upper limits on the presence of planets in these systems. For this, we use calibrated contrast curves, produced by the \texttt{spaceKLIP} package, which also take into account the transmission of the coronagraphic mask. These contrast curves correspond to the flux level of a companion sufficient to produce a $5\sigma$ detection. The contrast curve is calibrated by injecting fake companions at different angular separations and performing the PSF subtraction. As the flux of the injected planets is known, the flux that is being lost while performing the PSF subtraction can be quantified and a correction applied to the contrast derived. Figure~\ref{fig:contrast curves} demonstrates the contrast capabilities for the obtained MIRI observations. 

The contrast depth depends on the number of KL modes used for PSF subtraction. The deepest contrast for each subtraction method is illustrated in Figure~\ref{fig:contrast curves}, corresponding to the maximum number of KL modes: 6 for ADI, 18 for RDI, and 24 for ADI+RDI. Beyond 1 mode for ADI, 6 modes for RDI, and 6 modes for ADI+RDI, increasing the number of KL modes produces little change in the achieved contrast, although the deepest values are still obtained at the maximum. We also note that RDI provides the best performance at separations $<1\arcsec$, while ADI+RDI yields deeper contrast in the background-dominated regime at larger separations.

Additionally, we compare the measured contrasts to literature results and predicted contrast performance. We reach a sensitivity corresponding to contrast ratios of ${\approx}5\times10^{-5}$ for separations ${\gtrsim} 1.5$\arcsec, comparable to typical \textit{JWST}/MIRI performance \citep[e.g.,][]{boccaletti_2022, matthews+2024, Malin_2025, Sanghi_2025}. However, we measure a discrepancy between the observed contrast and predicted sensitivities from \texttt{PanCAKE} simulations (Figure~\ref{fig:contrast curves}), which is more significant in the background-dominated regime ($>1\arcsec$). This mismatch is expected for MIRI coronagraphic observations, as \texttt{PanCAKE} simulations currently over-predict contrast capabilities. This over-prediction is due to several MIRI-specific artefacts that are not included in pancake simulations (including the "glow stick" subtraction and BFE correction). Although the \texttt{PanCAKE} MIRI predictions were empirically scaled to match the performance from the ERS-01386 program on HIP~65426~b \citep{carter_jwst_2023}, the impact of background subtraction and BFE appears to be more pronounced in the presented GO~1668 observations than in the ERS program. This therefore highlights the need for further refinement of MIRI \texttt{PanCAKE} predictions to accurately reflect current observational limits and using empirical MIRI contrast curves may provide more reliable performance estimates.

To convert contrast into mass sensitivity, as shown in the right panels of Figure~\ref{fig:contrast curves}, we use a combination of ATMO-ceq \citep{phillips_2020} and BEX \citep{linder_2019} planetary evolution models, joined together following the approach in \citet{carter_2021}. This conversion is done using the \texttt{MADYS} package\footnote{\url{https://github.com/vsquicciarini/madys}} \citep{Squicciarini2022}, using the \verb|bex-atmo2023-ceq| evolution model. For Figure~\ref{fig:contrast curves}, the system ages are fixed to 200~Myr, 150~Myr and 155~Myr for HD\,92945, HD\,107146, HD\,206893, respectively (see Table~\ref{tab:system_params}), and the contrast and masses are shown as a function of projected separation (age uncertainty and de-projection will be taken into account in \S\ref{sec: DPM}).

For HD\,92945, we compare the contrast and mass sensitivities between \textit{JWST}/MIRI at $11.4~\mu$m and \textit{JWST}/NIRCam at $4.4~\mu$m \citep{Lazzoni_2025}. While NIRCam delivers significantly deeper contrast than MIRI, the corresponding improvement in mass sensitivity is less pronounced. There is also a significant caveat here, where NIRCam wavelengths are much more sensitive to potential effects of disequilibrium chemistry and clouds \citep[e.g.,][]{crotts_2025, Bardalez-Gagliuffi_2025}, so the chemical equilibrium model (ceq) used is likely the most optimistic case for NIRCam. We also strengthen that the calculated sensitivities are expressed as a function of projected separation, and de-projection is required for robust constraints on planet mass and location. For example, the difference in mass sensitivity between NIRCam and MIRI appears more prominent in Figure~\ref{fig:contrast curves}, whereas in Figure~\ref{fig:detectability} the discrepancy is reduced once we take into account inclination, stellar age uncertainty and planet eccentricity (see \S\ref{sec: DPM}).

\subsection{Detection probability maps (DPM)}
\label{sec: DPM}

In this section, we use non-detections and contrast curves to assess which types of planets can be ruled out as a function of their mass and semi-major axis. Since contrast curves are defined in terms of projected separation, we must de-project them by accounting for the full range of orbital configurations that could place a planet at a given separation \citep[e.g.,][]{2020ExoDMC}. To do this, we build detection probability maps (DPMs), calculated by combining the \texttt{MADYS} tool and an improved version of \texttt{ExoDMC}\footnote{\url{https://github.com/mbonav/Exo_DMC}}, which is now a dependency of \texttt{MADYS}. This tool generates DPMs and indicates the probability of detecting a planet of a given mass and semi-major axis if it exists, given the achieved contrast. Given non-detections, these DPMs allow to rule out planets with a high detection probability.

To build such maps, we define a grid of planet masses (0.01 -- 100\,\Mjup) and semi-major axes (0.1 -- 12\arcsec) uniformly spaced in log-space. For every point in this grid, we first convert the planet mass to a predicted magnitude using planetary evolution models, with the stellar age and uncertainties sampled uniformly. Specifically, we use the same \verb|bex-atmo2023-ceq| model in \texttt{MADYS}, as done previously for Figure~\ref{fig:contrast curves}. While we define a broad parameter grid, the achieved contrast limits only probe planet masses where the models are fully tabulated, requiring no extrapolation for the detectable mass range in any of the three systems. However, such models remain only partially validated for effects such as cloud coverage and disequilibrium chemistry, which could affect the estimated mass sensitivities.

For each point in the grid, we also convert the semi-major axis into a distribution of projected separations. We assume that any planet in the outer regions is co-planar with the disc and compute a set of separations by drawing random orbital configurations. Inclinations are sampled from a normal distribution constrained by ALMA observations, while eccentricities are drawn from a half-Gaussian distribution\footnote{Negative values are discarded} centred at 0 with a standard deviation of 0.1. 

These predicted magnitudes and projected separations are then compared to the contrast limits of the observations. A grid point is considered detectable if the planet's expected magnitude is brighter than the $5\sigma$ contrast limits at the corresponding separation. This process is repeated across the entire grid to compute the $5\sigma$ detection probability, incorporating random orbital configurations and sampling over the system's age range to account for age uncertainties.

\begin{figure}
    \centering
    \includegraphics[trim=0.3cm 0.1cm 0.3cm 1.2cm, clip=true, width=\linewidth]{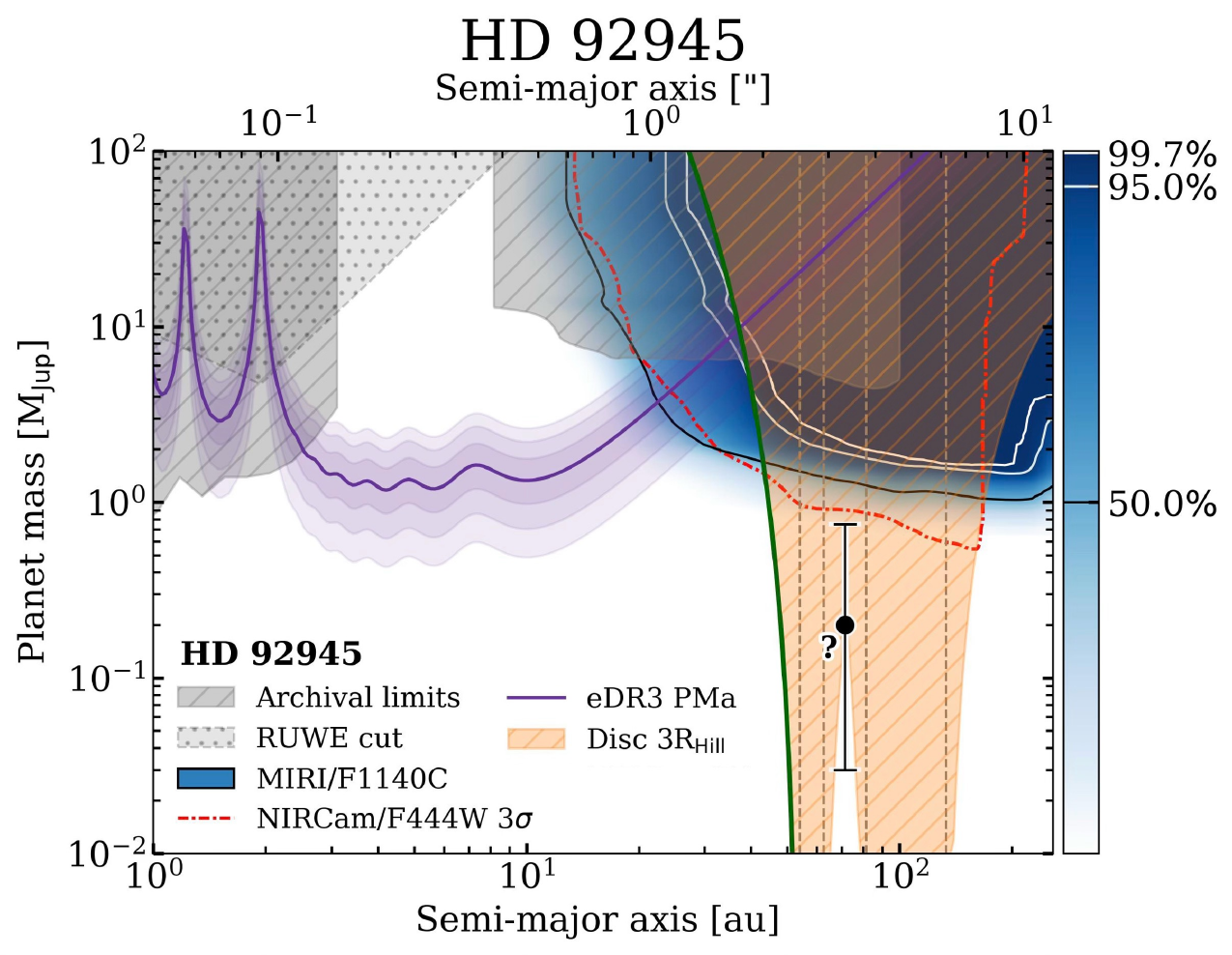}
    \includegraphics[trim=0.2cm 0.1cm 0.3cm 0.97cm, clip=true, width=\linewidth]{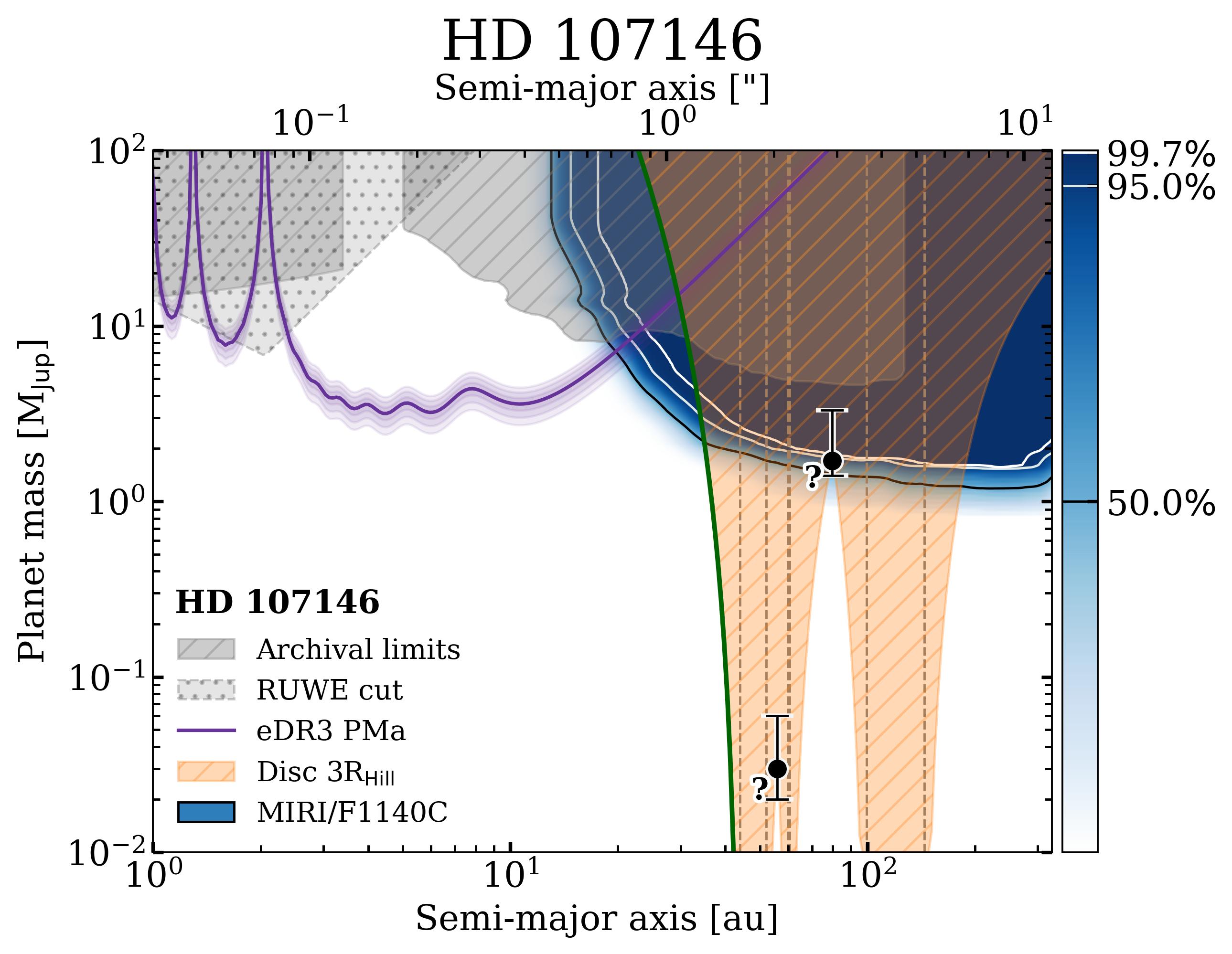}
    \includegraphics[trim=0.2cm 0.1cm 0.3cm 0.97cm, clip=true, width=\linewidth]{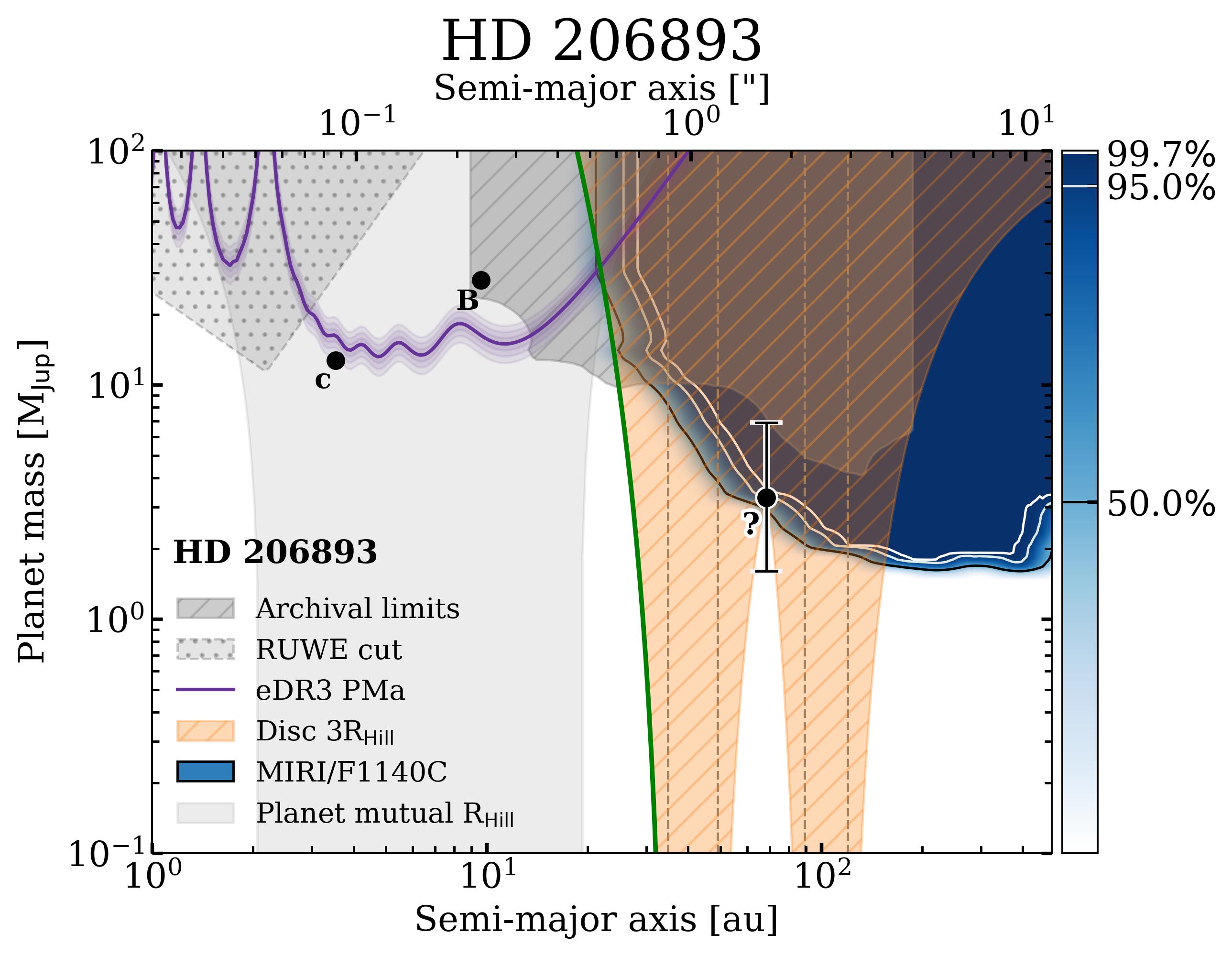}
    \vspace{-5mm}
    \caption{DPMs for HD\,92945, HD\,107146, and HD\,206893 (top to bottom). The blue shading shows the probability of a 5$\sigma$ planet detection with MIRI at 11.4~\micron, with contours marking the 50, 95 and 99.7\% confidence levels. For HD\,92945, the red dash-dotted line shows the 99.7\% detection limit from NIRCam at 4.4~\micron. The darker grey shaded regions denote constraints from archival direct imaging and RV data, and grey dotted regions from Gaia astrometry (RUWE). The orange hatching marks limits imposed by the disc morphology, where planets within 3~\Rhill~of the disc edges would disrupt it (\S\ref{subsec: disc stability}). The purple curve describes the planet mass and separation combinations required to explain the observed PMa signal. The green solid line highlights the orbital parameters for a planet shaping the disc inner edge (\S\ref{constraints inner edge}), and the black dot with a question mark marks the possible parameters for a planet located at the centre of the gaps in a massless disc (\S\ref{sec: single-planet-massless-disc}). For HD\,206893, the light-grey region rules out planet parameters to ensure planet stability based on mutual \Rhill. All DPM components are detailed in \S\ref{sec: DPM}.}
    \label{fig:detectability}
\end{figure}

When calculating the DPMs, we use the deepest available contrast curves to exclude the widest possible range of planet configurations in each system. For these MIRI observations, we have decided to combine the contrasts obtained from the three PSF subtraction techniques illustrated in Figure~\ref{fig:contrast curves}. Since no planets were detected across all subtraction methods and KL modes, we select, at each separation, the best (i.e., lowest) contrast achieved across all techniques and KL modes. While this approach only yields a marginal improvement over the contrast from ADI+RDI with maximum KL modes, it results in a contrast that maximises the sensitivity to potential planets. The resulting DPMs are shown in Figure~\ref{fig:detectability}, where the shade of blue indicates the probability of a $5\sigma$ planet detection.

For HD\,92945, we have added the $99.7$\% contour from the DPM calculated using \textit{JWST}/NIRCam coronagraphic data at $4.4~\mu$m (F444W) from \citet{Lazzoni_2025} for comparison (red line in Figure~\ref{fig:detectability}). Compared to MIRI/F1140C, NIRCam/F444W achieves deeper mass sensitivity for this system, with an improvement of roughly a factor of 3 and particularly enhanced sensitivity at smaller separations. However, this observed mass improvement assumes that the planet has a cloudless atmosphere in chemical equilibrium. Planet evolution models at these shorter wavelengths \citep[from the NIR and up to $\approx$8~\micron,][]{matthews+2024, crotts_2025} are strongly affected by uncertainties in atmospheric properties, such as composition, cloud coverage, but also vertical mixing, and disequilibrium chemistry, which in turn significantly influence the reliability of predicted fluxes (e.g., seen for Eps Ind Ab; \cite{matthews+2024}, TWA~7b; \cite{lagrange+2025, crotts_2025}, and more details in \citealt{bowens-rubin_2025}). These effects are much less pronounced at longer wavelengths like MIRI's $11.4~\mu$m, making MIRI limits more robust, despite being less constraining.

Beyond the \textit{JWST} constraints on the DPMs, additional shaded regions seen in Figure~\ref{fig:detectability} further exclude possible planet configurations in these systems. The following subsections describe how these regions are calculated. The functions used to generate the curves based on disc extent (\S\ref{subsec: disc stability}), proper motion anomaly (\S\ref{subsec: pma}), and Gaia astrometry (\S\ref{subsec: ruwe}) have been made available online\footnote{\url{https://github.com/raphhbw/ExePMa}}.

\subsubsection{Planet limits from archival direct imaging and RV data}
The grey-hatched regions in Figure~\ref{fig:detectability} rule out the presence of additional companions based on previous upper limits from ground-based direct imaging (VLT/SPHERE) and radial velocity (RV) monitoring (HARPS, SOPHIE, and HIRES). These archival data provide valuable complementary constraints to \textit{JWST}/MIRI observations, particularly at smaller separations (for RV) and in overlapping regions (for SPHERE). Note that HD\,206893~B was detected with SPHERE, however, the presence of additional companions was ruled out \citep{milli_2017_hd206893}. 

SPHERE contrast curves were retrieved from the SPHERE High-Contrast Data Center \citep[HC-DC,][]{delorme_2017}. SPHERE contrast limits were converted into mass constraints using the same approach as for \textit{JWST} (i.e., DPM using ATMO+BEX models via \texttt{MADYS}). The $99.7$\% contour from these DPMs is used as the archival direct imaging upper limit.

RV data were obtained from the HARPS, SOPHIE, and HIRES spectrographs, with SOPHIE observations retrieved from the OHP archive\footnote{\url{http://atlas.obs-hp.fr/sophie/}}, and HIRES and HARPS data taken from \citet{tal-or_2019} and \citet{trifonov_2020}, respectively, as those works correct the standard pipeline RVs for known small systematics. Detection limits were derived following the local power analysis method outlined by \citet{meunier_2012}, where planetary detectability is assessed on a semi-major axis--mass grid. For each point on the grid, we calculate a Keplerian signal at the observation timestamps, with the relevant period, at 100 random phases and produce the relevant Keplerian-only Lomb-Scargle periodogram. A signal is considered ruled out if, within a narrow window around the injected period, the periodogram peak exceeds all peaks in the periodogram of the RV data, meaning a planet that should have been detected is absent. The detection limit at a given semi-major axis is then defined as the lowest planet mass excluded across all 100 random phases.

\subsubsection{Planet limits from the disc extent}
\label{subsec: disc stability}
In addition to planet detection limits, we include dynamical constraints based on the observed disc extent and gap locations. The vertical dashed lines in Figure~\ref{fig:detectability} mark the inner and outer edges of the discs, as well as the boundaries of the observed gaps. The orange-hatched regions represent dynamically unstable zones, where planets on a circular orbit located within 3 Hill radii (\Rhill) of a disc edge would significantly disturb the surrounding material if $m_{\mathrm{plt}} \ll m_{\mathrm{star}}$ \citep{Gladman1993, ida_2000, depater_2001, kirsh_2009, friebe_2022, pearce2024}. The boundaries of the orange region in Figure~\ref{fig:detectability} follow \citet{Pearce_2014} and are defined by:
\begin{equation}
    m_{\mathrm{plt}} = 3m_{\mathrm{star}} \left|\frac{r_\mathrm{edge}}{a_{\mathrm{plt}}} - 1\right|^3 \left( \frac{1}{N_{\mathrm{R_{Hill}}}}\right)^3,
\label{eq:inner_edge_in}
\end{equation}
where $a_{\mathrm{plt}}$ and $m_{\mathrm{plt}}$ are the semi-major axis and mass of the planet, $m_{\mathrm{star}}$ the stellar mass, and $N_{\mathrm{R_{Hill}}}$ (set to 3) specifies the distance from the disc edges ($r_\mathrm{edge}$) in \Rhill. The absolute value accounts for both cases where the disc edge lies inside or outside of the planet's orbit.

Any planet within this 3~\Rhill~zone would likely clear material from the disc and shift the disc edges to locations inconsistent with current observations. Therefore, any planet sculpting the disc edges cannot lie in these excluded regions. Similar constraints can be applied to planets located within the gaps, which we discuss further in \S\ref{sec: single-planet-massless-disc}.

\subsubsection{Planet limits from mutual interactions}
\label{subsec: mutual hill radius}
For systems with confirmed planets, as in HD\,206893, the positions of the known companions are marked in Figure~\ref{fig:detectability} by black dots. In such cases, additional dynamical constraints can be applied by considering the potential for mutual close encounters between the planets. To avoid such interactions that can lead to dynamical instability on short timescales, planetary orbits must be sufficiently spaced, typically by a minimum number of mutual \Rhill. 

Following \citet{chambers1996}, we consider two planets to be dynamically unstable if their orbital separation is less than a critical $2\sqrt{3} \approx 3.5$ times their mutual \Rhill, where the mutual \Rhill~is defined as Equation (1) in \citet{chambers1996}. In our specific case, HD\,206893~B and c are spaced out by 4.3 mutual \Rhill, which satisfies the stability criterion we set. 

To identify regions where no additional planets could exist without violating this stability criterion, we compute the $2\sqrt{3}$ mutual \Rhill~zones around each known companions. These excluded regions are shown as the light grey area in Figure~\ref{fig:detectability} and are combined into a single region, since the known planets do not lie within each other's unstable zones. Note that the stability criterion depends on factors we are not exploring here, including mean motion resonances between planets which can maintain stability, or planet eccentricity and additional planets in the system \citep[see details in][]{chambers1996, Andrew+2009}. Nevertheless, we use the excluded region to set upper limits on where dynamically stable planets could reside.

\subsubsection{Planet limits from proper motion anomaly (PMa)}
\label{subsec: pma}
All three systems exhibit strong evidence for an inner companion, as indicated by astrometric accelerations, also known as proper motion anomalies (PMa). PMa is defined as the difference between the proper motion measured by Gaia eDR3, interpreted as the instantaneous proper motion vector at J2016.0, and a long-term proper motion baseline (24.75 yr) computed from the variation of sky coordinates between Hipparcos (J1991.25) and Gaia eDR3 itself \citep{kervella+2019}. Starting from PMa components from the \citet{kervella_2022} catalogue, we looked for orbits from unseen companions which could explain the observed PMa. As done in \citet{marino_2020}, for each system we assumed a single companion on a circular orbit seen coplanar to the disc. The PMa information is not sufficient to fully solve the orbit, however, it provides valuable constraints, resulting in a degenerate curve in the mass -- semi-major axis space.

The purple line and shaded purple regions (68, 95, and 99.7\% confidence level) in Figure~\ref{fig:detectability} represent the mass of the planet needed in order to explain the observed PMa. While this method assumes a single planet on a circular orbit, the disc morphology does not indicate the presence of highly eccentric massive planets that would force a disc eccentricity \citep{wyatt_1999, wyatt_2005, Pearce_2014, faramaz_2014}. In the case of HD\,206893, where two massive substellar objects are present, the derived purple curve remains consistent with the PMa in the system attributed to planet c, aligning with \citet{hinkley_2023}.

\subsubsection{Planet limits from Gaia astrometry (RUWE)}
\label{subsec: ruwe}
Finally, we use Gaia astrometry and the Renormalised Unit Weight Error (RUWE) parameter to construct the grey dotted regions shown in Figure~\ref{fig:detectability}. The RUWE quantifies the excess noise in Gaia's astrometric data, with RUWE~$>1.4$ typically indicating a strong signal, often attributed to unresolved astrometric perturbations from an unseen companion \citep{lindegren_2021}. 

In all three systems studied here RUWE~$<1.4$, which allows us to set upper limits on any possible astrometric signal and exclude regions of the planet mass--semi-major axis parameter space where a companion would have induced a detectable signal. For short-period companions, i.e., orbital periods shorter than Gaia DR3's 1038-day baseline, we can exclude planet parameters following the approach in \citet[][see their \S3.3.1]{limbach2024}, under the main assumption of RUWE~$<1.4$. For periods longer than the Gaia baseline, we apply the mass--separation relation using Equation (32) of \citet{Kiefer+2024}, to further rule out planet parameters. The combined excluded region is shown as the grey dotted area in the top left corners of the DPMs in Figure~\ref{fig:detectability}.

\begin{table*}
    \centering
    \caption{Observational sensitivity to planets in HD\,92945, HD\,107146 and HD\,206893, derived from the DPMs shown in Figure~\ref{fig:detectability}. Mass upper limits are given as a function of semi-major axis for three scenarios: a planet sculpting the inner edge of the disc, a planet located at the observed gap(s), and a planet responsible for the PMa signal. These values (and footnotes) represent the deepest observational limits currently attainable at these locations, though dynamical arguments may impose stricter constraints. Limits are quoted at the 99.7\% confidence level, with those in brackets indicating the 50\% level, derived from the contours in the DPMs. For HD\,206893, dashes (--) indicate that MIRI observations lack the sensitivity to detect a planet sculpting the inner edge of the disc. The crosses ($\times$) mark systems where no second gap has been observed. For HD\,92945, we also show the upper limits from \textit{JWST}/NIRCam at $4.4$~\micron~ presented in \citet{Lazzoni_2025}.}
    \renewcommand{\arraystretch}{1.2} 
    \begin{tabular}{lcc|cc|cc|cc}
        \hline
        \hline
        \multirow{2}{*}{Star} & \multicolumn{2}{c|}{inner edge} & \multicolumn{2}{c|}{gap$_1$} & \multicolumn{2}{c|}{gap$_2$} & \multicolumn{2}{c}{PMa companion} \\
        \cmidrule(lr){2-3} \cmidrule(lr){4-5} \cmidrule(lr){6-7} \cmidrule(lr){8-9}
        & a$_\mathrm{plt}$ [au] & m$_\mathrm{plt}$ [\Mjup] & a$_\mathrm{p}$ [au] & m$_\mathrm{p}$ [\Mjup] & a$_\mathrm{p}$ [au] & m$_\mathrm{p}$ [\Mjup] & a$_\mathrm{p}$ [au] & m$_\mathrm{p}$ [\Mjup] \\
        \midrule
        HD\,92945 (MIRI)& $36$ ($43$) $^{\mathrm{a}}$&  $<12.4$ ($<1.7$) $^{\mathrm{a}}$& $72$ & $<2.1$ ($1.3$)& $\times$ & $\times$ & $2.5 - 30$& $0.4 - 5$\\
 HD\,92945 (NIRCam)& $44$ ($46$)& $<1.3$ ($0.6$)& $72$ & $<0.9$ ($0.4$)& $\times$& $\times$& $2.5 - 25$&$0.4 - 5$\\
        HD\,107146& $33$ ($35$)& $<4.1$ ($2.1$)& $56$ & $<2.1$ ($1.6$)& $78$ & $<1.8$ ($1.4$)& $2.5 - 20$& $2.5 - 8$\\
        HD\,206893& -- (--) $^{\mathrm{b}}$& -- (--) $^{\mathrm{b}}$& $69$ & $<4.0$ ($2.8$)& $\times$& $\times$& $3.53\pm0.07$ $^{(1)}$& $12.7\pm1.1$ $^{(1)}$\\
        \bottomrule

    \end{tabular}
  \\
    \begin{flushleft}
    $^{\mathrm{a}}$ SPHERE observations more constraining at the inner edge, $6.5$~\Mjup~at $40$~au, given at the 3$\sigma$ confidence level.\\
    $^{\mathrm{b}}$ SPHERE observations more constraining at the inner edge, $10$~\Mjup~at $22.5$~au, given at the 3$\sigma$ confidence level.\\
    Planet reference: $^{(1)}$ HD\,206893~c ; \citet{hinkley_2023}
    \end{flushleft}
    \label{tab:planet-constraints}
\end{table*}

\section{Planet constraints using the detection probability maps}
\label{sec: planet constraints}
Using the mass sensitivity of \textit{JWST}/MIRI observations, along with the additional constraints shown in the DPMs (Figure~\ref{fig:detectability}), we assess where potential planets could reside in these systems. Certain locations are of particular interest, such as planets near the inner edge of the disc, which may be responsible for truncating the inner edge (\S\ref{constraints inner edge}). The DPMs also help constrain the properties of companions responsible for the strong PMa signals (\S\ref{planet causing pma}). Later in \S\ref{sec:gap-carving-scenarios}, we explore the possibility that the gaps are carved by planets. 

Table~\ref{tab:planet-constraints} summarises the updated planet limits derived from the DPM contours at both the 99.7\% and 50\% confidence levels for the location of interest. These values reflect the observational sensitivities, although dynamical arguments could impose stricter limits. 

\subsection{Planet truncating the disc inner edge}
\label{constraints inner edge}
All three systems show inner edges that are consistent with being steep \citep{imaz_blanco_2023} and shaped by planets \citep{pearce2024}. The planet's location is defined by its proximity to the inner edge without destabilising the disc, defined as 3~\Rhill~interior to the disc's inner edge (green curve in Figure~\ref{fig:detectability}). The corresponding minimum planet mass is set by requiring that clearing the disc over 10 diffusion timescales occurs within the system's age, ensuring sufficient time for the planet to carve the observed edge \citep{pearce2022, pearce2024}. Adopting only 1 diffusion timescale would yield a higher minimum mass, so using 10 provides a more conservative estimate \citep{costa_2024}. For consistency, we adopt the minimum planet masses derived by \citet[][see their \S4.3]{imaz_blanco_2023}, as we use the same stellar and disc parameters. Two scenarios are explored: a single planet truncating the inner edge, which requires a relatively massive planet (tens of Earth masses), and a chain of equal-mass planets, which can achieve the same effect with significantly lower-mass planets (a few Earth masses each).

Although the minimum planet masses required for disc truncation lie below the detection limits of the \textit{JWST}/MIRI observations, the planet masses ruled out by the DPMs (see Table~\ref{tab:planet-constraints}) provide tighter upper limits on the masses of these potential single sculptors. These observational constraints are obtained from the intersection of the probability contours with the green curve in Figure~\ref{fig:detectability}.

Based on MIRI observations alone, a sculpting planet in HD\,92945 would need to be located at $40-47$~au with a mass between $0.26 - 6.5$~\Mjup. For HD\,107146, the corresponding sculpting planet would be at $33-38$~au with a mass ranging between $0.4 - 4.1$~\Mjup, and for HD\,206893, the planet would need to be at $22-30$~au with a mass between $0.35 - 10$~\Mjup.

NIRCam 4.4~\micron~observations of HD\,92945 \citep{Lazzoni_2025} provide tighter constraints, narrowing the allowed parameters for a single inner edge sculptor to $44-47$~au with a mass between $0.26 - 1.3$~\Mjup. However, NIRCam wavelengths are more model-dependent than those from MIRI.

While we have assumed that the disc is sculpted by a planet at 3~\Rhill~from its inner edge, several caveats should be considered. First, this interpretation assumes that the inner edge of the disc is actively shaped by a planet that remains near this location. However, it is possible that a planet initially sculpted the edge earlier in the system's history and has since migrated inward, decoupling from the current position of the disc edge. In such cases, the planets responsible for the truncation may no longer reside near the inner edge of these systems. Additionally, the disc edge could be shaped by mechanisms that do not require a planet at all, and could be set by planetesimal formation alone or collisional evolution within the disc \citep{kennedy_2010}. These scenarios highlight that while our interpretation provides one possible explanation, it is not the only one. Further discussion of these caveats and alternative scenarios can be found in \citet[][see their \S6]{pearce2024}.

\subsection{Planet causing the PMa}
\label{planet causing pma}
While the PMa indicates the presence of a perturber, it does not constrain its orbital radius, as it depends on a degenerate combination of planet mass and separation. To break this degeneracy, we assess how the constraints from the DPMs rule out different planet parameters. At small separations (typically less than a few au), radial velocity monitoring and the Gaia RUWE parameter can exclude massive, close-in planets as the source of the PMa, provided no significant astrometric or spectroscopic signal is detected. At wide separations (tens of au), direct imaging limits from ground-based facilities or \textit{JWST} rule out companions massive enough to induce the observed PMa. 

The DPMs in Figure~\ref{fig:detectability} show that, for all three systems, the planets responsible for the PMa signal cannot reside near the disc's inner edge, and therefore likely does not play a role in actively shaping it. Instead, the observational constraints indicate that the perturber in all cases is confined to the inner regions of the planetary system. The allowed range of semi-major axis and masses for such companion are summarised in Table~\ref{tab:planet-constraints}, and are typically around 2.5 -- 30~au and 0.4 -- 8~\Mjup. In the case of HD\,206893, this interpretation is consistent with the detection of planet c at 3.5~au, which has been shown to explain the observed PMa signal \citep{hinkley_2023}.

In recent years, a broader trend has emerged in which PMa signals tend to trace planets located in the inner regions of planetary systems. This pattern is not only seen in the three systems presented here, but also in AF Lep \citep[AF Lep b,][]{DeRosa2023, Mesa2023, Franson2023}, $\beta$ Pic \citep[$\beta$ Pic c,][]{brandt_2021_betapic}, and HR~8799 \citep[HR~8799~e,][]{brandt_2021_hr8799}. However, the presence of a PMa perturber at small separations does not rule out the presence of additional, further-out planets. For example, HD~206893~B, $\beta$ Pic b, and HR~8799 b,c and d all orbit at wider separations than their corresponding PMa planets, with HR~8799~b located as far out as $\approx$70 au. Another compelling example for planet formation at wider separations is TWA 7 b, a planet candidate detected within the gap of a debris disc around 50~au \citep{lagrange+2025, crotts_2025}. These cases highlight that while PMa signals typically arise from inner planets, other planets may form at larger separations and play a key role in shaping the observed disc morphology. A similar configuration may therefore apply to the systems studied in this work.

\section{Gap carving scenarios}
\label{sec:gap-carving-scenarios}
There are multiple gap carving scenarios to consider, and while hypothesising a planet embedded within the gap itself is the most straightforward solution, it is not the only one. Different mechanisms ultimately predict different planetary parameters, including their location relative to the gap, as well as evolutionary histories.

In the following subsections, we assess four distinct gap-carving scenarios for each of the systems. The first scenario involves a single planet embedded within the disc, carving a gap around its orbit through scattering (\S\ref{sec: embedded planet}). The second scenario (\S\ref{sec: 2:1MMR}) examines whether a planet located near the disc's inner edge could generate a gap via the 2:1 mean motion resonance (MMR). The remaining two scenarios focus on secular apsidal resonances: one involving two inner planets and a massless disc (\S\ref{sec:two_planets_sec_res_YK18}), and the other with a single inner planet and a massive (non-zero mass) disc (\S\ref{sec:sec_res_massive_disc}). Overall, these four gap-carving scenarios are adapted to each system individually, with slight modifications where necessary, e.g., to account for the presence of two gaps in HD\,107146 or the two known companions in HD\,206893. Table~\ref{tab:summary-gap-formation} provides a summary of the validity of each gap-carving scenario with corresponding planet predictions.

{
\setlength{\tabcolsep}{1pt}
\begin{table*}
\centering
\caption{Summary of the gap-carving scenarios for HD\,92945, HD\,107146, and HD\,206893 discussed in §\ref{sec:gap-carving-scenarios}. Some theoretical predictions remain unaffected by the new \textit{JWST} observational constraints, and references to the original studies with still valid predictions are provided. However, note that these values might need updating with new disc parameters. For HD\,92945's secular resonance in a massless disc + 2 planets scenario, the \textit{JWST}/NIRCam constraints are given in brackets.}
\begin{tabular}{l|c|cc|c}
    \hline
    \hline
     \multirow{2}{*}{Gap-carving scenarios} & \multirow{2}{*}{HD\,92945} & \multicolumn{2}{c}{HD\,107146} & \multirow{2}{*}{HD\,206893} \\
     \cmidrule(lr){3-4}
      & & gap 1 & gap 2 & \\
    \hline
    \textbf{$\bullet$~Embedded planet:} & & & & \\
    \multirow{2}{*}{Single planet + massless disc (\S\ref{sec: single-planet-massless-disc})} & a$_{\mathrm{plt}}$: $72$~au & a$_{\mathrm{plt}}$: $56$~au & a$_{\mathrm{plt}}$: $78$~au & a$_{\mathrm{plt}}$: $69$~au \\
   \vspace{5pt}
    & m$_{\mathrm{plt}}$: $0.03-0.75$~\Mjup & m$_{\mathrm{plt}}$: $0.02-0.06$~\Mjup & m$_{\mathrm{plt}}$: $1.4-1.8$~\Mjup & m$_{\mathrm{plt}}$: $1.6-4.0$~\Mjup \\
   Chain of planets + massless disc (\S\ref{chain of planets}) & $\sim$2 planets with ${\approx}0.02$~\Mjup$^\ddagger$ & \multicolumn{2}{c}{If single wide gap:} & $\sim$2-3 planets with ${\approx}0.04$~\Mjup$^\ddagger$ \\
   \vspace{5pt}
    & & \multicolumn{2}{c}{$\sim$2-3 planets with ${\approx}0.04$~\Mjup$^\ddagger$} & \\
    Single planet + massive disc (\S\ref{subsec: single planet in massive disc}) & m$_{\mathrm{plt}}\lesssim0.26$~\Mjup$^{(1)}$ & \multicolumn{2}{c}{If single wide gap:} & m$_{\mathrm{plt}}\lesssim2.5$~\Mjup$^{(1)}$ \\
    & M$_{\mathrm{disc}}\lesssim 75$~M$_{\oplus}$$^{(1)}$ & \multicolumn{2}{c}{m$_{\mathrm{plt}}\lesssim1.5$~\Mjup$^{(1)}$} & M$_{\mathrm{disc}}\lesssim 190$~M$_{\oplus}$$^{(1)}$ \\
    &  & \multicolumn{2}{c}{M$_{\mathrm{disc}}\lesssim 50$~M$_{\oplus}$$^{(1)}$} &  \\
    \hline
    \textbf{$\bullet$~2:1 MMR} (\S\ref{sec: 2:1MMR}): & Ruled out & Ruled out & Ruled out & Ruled out \\
    \hline
    \textbf{$\bullet$~Sec. res. in massless disc + 2 planets} (\S\ref{sec:two_planets_sec_res_YK18}): & a$_1$: $11-20$~au ($11-15$~au) & \multicolumn{2}{c}{\multirow{4}{*}{Ruled out*$^{(2)}$}} & \multirow{4}{*}{Ruled out}\\
    & m$_1$: $1.4-3.1$~\Mjup~($1.4-1.9$~\Mjup) & & \\
    & a$_2$: $40-47$~au ($44-47$~au) & & \\
    & m$_2$: $0.3-7.8$~\Mjup~($0.3-1.5$~\Mjup) & & & \\
    \hline
    \textbf{$\bullet$~Sec. res. in massive disc + 1 planet} (\S\ref{sec:sec_res_massive_disc}): & & \multicolumn{2}{c}{If single wide gap:} & \\
     & a$_{\mathrm{plt}}$: $10-25$~au & \multicolumn{2}{c}{a$_{\mathrm{plt}}$: $8-15$~au} & With HD\,206893~B only$^\dagger$:\\
     & m$_{\mathrm{plt}}$: $0.5-3$~\Mjup & \multicolumn{2}{c}{m$_{\mathrm{plt}}$: $2-5$~\Mjup} & M$_{\mathrm{disc}}$: $\approx400$~M$_{\oplus}$ \\
     & M$_{\mathrm{disc}}$: $10-100$~M$_{\oplus}$ & \multicolumn{2}{c}{M$_{\mathrm{disc}}$: $30-100$~M$_{\oplus}$} & \\
     \vspace{-7pt}
    \\
     \hline
\end{tabular}
\\
\begin{flushleft}
$^\ddagger$ Assuming an inter-planet spacing of 16~\Rhill. More compact configurations would lead to lower planet masses.\\
* Two gaps could be opened in specific planet configurations (Friebe \& Sefilian, in prep.)\\
$^\dagger$ Friebe \& Sefilian (in prep.) explore the addition of HD\,206893~c, but find it difficult to reproduce the observed gap.\\
References: $^{(1)}$ \citet{friebe_2022}, $^{(2)}$ \citet{yelverton_2018}
\end{flushleft}
\label{tab:summary-gap-formation}
\end{table*}
}

\subsection{General caveats}
Throughout this section, we consider the gap location, the gap width and the level of gap asymmetry to evaluate the different scenarios. Using the gap depth could be a useful tool to exclude some gap-carving mechanisms; however, current constraints allow for a wide range of possible depths, from 50\% full to more than 90\% depleted, and therefore are not constraining \citep{imaz_blanco_2023}. The gap depth is therefore excluded as a criterion from the following discussion.

When interpreting the planet predictions in this work, it is important to keep in mind the assumptions on which they are based. The constraints on gap-carving planets are based on analytical arguments. More detailed $N$-body simulations could refine these results, however, such tailored modelling lies beyond the scope of this paper. Additionally, the predictions depend strongly on both the disc parameters and the observational upper limits obtained from the DPMs. Disc parameters can vary depending on the wavelength used to observe the disc as well as the models used to fit the data. Small changes in disc parameters could lead to large differences in predicted planet properties due to the scaling laws that we use. Similarly, improvements in observational sensitivity would impact the validity of predicted planet properties. Despite these caveats, the constraints presented here represent the most up-to-date estimates for the planets that could be responsible for carving the observed gaps, given current theoretical understanding.

\subsection{Planet(s) embedded in the disc}
\label{sec: embedded planet}
The most commonly used mechanism for carving a gap in a debris disc involves a planet on a circular orbit embedded within the disc, scattering nearby planetesimals. Non-resonant planetesimals close to the planet are expected to be cleared, resulting in an azimuthally symmetric gap. For a more asymmetric gap, some level of planet eccentricity would be required. In this first scenario, we investigate three configurations: a massive single planet located at the centre of the gap in a massless disc (\S\ref{sec: single-planet-massless-disc}), a chain of multiple equal-mass planets populating the gap (\S\ref{chain of planets}), and a single planet clearing a gap through migration within a massive disc (\S\ref{subsec: single planet in massive disc}). Note that this list is not exhaustive, and variants or combinations of these scenarios could be possible. 

\subsubsection{Single planet in a massless disc}
\label{sec: single-planet-massless-disc}
We start by assuming a massless disc ($M_{\mathrm{disc}} \ll m_{\mathrm{plt}}$) and a single planet on a circular orbit located at the centre of the gap. In this scenario, the planet is expected to scatter material found within 3~\Rhill\footnote{We use the Hill radius criterion rather than resonance overlap \citep[chaotic zones;][]{Wisdom1980}, as it is more robust regarding planet eccentricity \citep{Pearce_wyatt_2014}. Approximating the gap width as the chaotic zone has been done in the past and produces results consistent with $N$-body simulations \citep[e.g., Figure 5 in][]{marino_2018}.} on either side of its orbit \citep{Gladman1993, ida_2000, kirsh_2009, friebe_2022}, resulting in a gap width of 6~\Rhill.

The mass of a planet ($m_{\mathrm{plt}}$) required to carve such a gap, assuming $m_{\mathrm{plt}} \ll m_{\mathrm{star}}$, is given by:
\begin{equation}
    m_{\mathrm{plt}} = 3m_{\mathrm{star}} \left(\frac{w_{\mathrm{gap}}}{6a_{\mathrm{plt}}}\right)^{3},
    \label{eq:gap-carving-planet}
\end{equation}
where $a_{\mathrm{plt}}$ is the planet's semi-major axis, which in this scenario would be equal to the central radius of the gap ($r_{\mathrm{gap}}$), and $w_{\mathrm{gap}}$ is the observed width of the gap. Using the disc parameters listed in Table\,\ref{tab:system_params}, we estimate the mass of the gap-carving planet for each system. These predicted planet masses and locations are visualised as the black dots with a question mark in Figure\,\ref{fig:detectability}. 

Using Equation~(\ref{eq:gap-carving-planet}), for HD\,92945, the estimated planet mass required to carve the observed 19~au gap at 72~au, ranges from 0.03 to 0.75~\Mjup. The range of planet mass is visible by the black error bars in Figure~\ref{fig:detectability} and is calculated from the uncertainty on $w_{\mathrm{gap}}$, $r_{\mathrm{gap}}$ and $m_{\mathrm{star}}$. \textit{JWST}/MIRI data excludes planets more massive than 2~\Mjup~at this location, but this sensitivity does not reach the predicted mass range. The same can be said about the \textit{JWST}/NIRCam data that excludes planets more massive than 0.9~\Mjup~at this location \citep{Lazzoni_2025}. Since both observational limits remain above the predicted mass range, the single planet scenario remains a possible explanation for the observed gap in HD\,92945, if the gap is truly axisymmetric.

Applying the same analysis to HD\,206893, a planet at $69$~au with a mass between 1.6 and 6.9~\Mjup~would be required to explain the observed $40$~au gap. In this case, the \textit{JWST}/MIRI observations can rule out planets more massive than 4~\Mjup, which in turn refines the predicted mass range to $1.6-4.0$~\Mjup. Therefore, this scenario remains a possible cause for the observed gap, assuming an axisymmetric gap.

For HD\,107146, we consider two embedded planets in a massless disc, each located at the centre of one of the two observed gaps. We would therefore require a first planet at 56~au with a mass ranging from 0.02 to 0.06~\Mjup~and a second planet at 78~au with a mass between 1.4 and 3.3~\Mjup. \textit{JWST}/MIRI constraints can rule out planets more massive than 1.8~\Mjup~at the outer gap location, tightening the predicted mass for that second planet to $1.4-1.8$~\Mjup. Conversely, the predicted mass range at the first gap location is significantly lower than the observational sensitivity. We note that a configuration with both planets centred within the gaps would be considered dynamically stable following the planet stability argument made in \S\ref{subsec: mutual hill radius}, although the planet separation (3.7 mutual \Rhill) lies close to the critical threshold of $2\sqrt{3} \approx 3.5$ mutual \Rhill. A more detailed evaluation of the dynamical stability of these predictions would require dedicated $N$-body simulations.

When evaluating this gap-carving scenario, it is important to consider the timescale necessary to open a gap in these discs. Previous $N$-body simulations have shown that planets with masses $\geq 10$~M$_{\oplus}$ can carve gaps by depleting at least 50\% of the material in less than 10~Myr for a ${\sim}1M_{\odot}$ at separations of ${\sim}75$~au \citep[e.g., Figure~5 in][]{marino_2018}. These timescale estimates are found to be shorter but comparable to those inferred from scattering diffusion and clearing timescale \citep{Tremaine1993, morrison_2015}. Even allowing for uncertainties in stellar ages, these results indicate that there is enough time for the predicted planets to have sculpted the gaps seen in the observed discs.

While the gap is assumed to be cleared of non-resonant material, resonant material could still remain trapped within the gap and make it appear less depleted than predicted. This is seen in the TWA~7 system, where a planet is found embedded in the gap alongside co-orbital material \citep{ren_2021, lagrange+2025, crotts_2025}. Although the gap depths can be estimated from observations and disc modelling, the measurements remain uncertain. Current uncertainties allow for both shallow ($<50$\%) and deep ($>90$\%) gaps \citep{imaz_blanco_2023}. Consequently, gap depth alone does not yet provide sufficient reliability to place strong constraints on planet masses. Tighter constraints would therefore require higher sensitivity observations and dedicated $N$-body simulations accounting for the clearing of both the resonant and non-resonant material \citep[see, e.g., Figure~6 \& 7 in][]{marino_2018}.

\subsubsection{Chain of multiple planets in a massless disc}
\label{chain of planets}
A chain of lower-mass planets distributed across the gap could produce comparable gap widths if the disc mass is much smaller than the planet mass. In this case, the system's age and gap width set a lower limit on the planet masses required to effectively clear material via scattering, while dynamical stability places a limit on the number of planets that can coexist within the gap \citep{Shannon+2016}.

To evaluate this scenario, we use Equation (4) from \citet{Shannon+2016}, which defines the minimum mass of the planets required to carve the gap for a given inter-planet spacing. Whereas their calculation assumed a typical spacing of 20~\Rhill~\citep{fang_2013}, we adopt 16~\Rhill, which reduces the estimated planet masses by a factor of two \citep{Shannon+2016} and allows for more planets to populate the gap. Since no results are available for more compact planet configurations, 16~\Rhill~provides the most conservative estimate.

For HD\,92945, given the inter-planet spacing of 16~\Rhill, the minimum mass of the planets in the gap would be ${\sim}7$~M$_\oplus$ ($\sim$0.02~\Mjup). A 16~\Rhill~spacing corresponds to $\sim$23~au, meaning only $\sim$2 planets could occupy the $\sim$20~au wide gap, one at each edge. For HD\,107146, combining both depleted regions into a single wide gap of ${\sim}50$~au yields a minimum planet mass of ${\sim}14$~M$_\oplus$ ($\sim$0.04~\Mjup). The 16~\Rhill~spacing of ${\sim}30$~au would then allow $\sim$2-3 planets to populate the gap. For HD\,206893, the required minimum planet mass is ${\sim}12$~M$_\oplus$ ($\sim$0.04~\Mjup), with an inter-planet spacing of ${\sim}23$~au permitting $\sim$2-3 planets to populate the gap width. 

In all three systems, the predicted mass limits are significantly lower than the detection sensitivity of the observations, meaning that such planets cannot be ruled out, and this scenarios remains valid. We note, however, that calculations are restricted to an inter-planet spacing of 16~\Rhill, which for a more compact configuration, would result in lower planet mass predictions, but exploring such cases requires tailored $N$-body simulations.

\subsubsection{Single planet in a massive disc}
\label{subsec: single planet in massive disc}
A debris disc with a mass comparable to the planet mass can significantly influence the planet embedded in the disc. As a single planet interacts with surrounding material, it predominantly scatters debris outward and exchanges angular momentum, causing the planet to lose energy and gradually migrate inward\footnote{The effect will vary if multiple planets are considered \citep{wyatt2003,Bonsor2014}} \citep{kirsh_2009}. The extent of this migration depends on both the disc mass and the planet mass.

\citet{friebe_2022} demonstrated that the same gap width in a debris disc can be reproduced by either a massive, nearly stationary planet (akin to the massless disc scenario) or a lower-mass planet that carves the gap as it migrates via planetesimal scattering. This degeneracy, illustrated in Figure~5 of \citet{friebe_2022}, highlights that different combinations of planet and disc mass can lead to the same gap width. In contrast, the depth would vary, as lower-mass planets would not eject material, leaving remaining scattered material to populate a shallower gap. As a result, the gap depth may inform about the planet's migration history and migration rate \citep{wyatt2003}.

With the current \textit{JWST observations}, we cannot exclude any of the planet and disc mass combinations predicted from the gap widths in \citet{friebe_2022}. Consequently, assuming axisymmetric gaps\footnote{In the case of a more asymmetric gap, the planet would likely need to follow a more eccentric orbit.}, the results of \citet{friebe_2022} remain consistent with a migrating planet as the origin of the gap. While this would not qualitatively change the conclusions, the associated mass estimates should be updated using revised disc parameters \citep{imaz_blanco_2023}.

In addition to scattering material, a migrating planet can also trap surrounding debris as co-orbital material (e.g. Jupiter's Trojan population), leading to observable substructures in the discs. In the case of TWA~7, the planet candidate observed within the disc gap is accompanied by co-orbital material, strongly suggesting resonant trapping and giving the appearance of a double-gap disc structure \citep{ren_2021, lagrange+2025}. A similar mechanism could explain the double gap morphology seen in HD\,107146, where a single planet migrating inwards might have accumulated a Trojan population, producing the two apparent gaps \citep[e.g., \S3.2.1 in][]{friebe_2022}.

There are several extensions to this scenario. For example, there could be two or more planets embedded in a massive disc. In such cases, planets may experience divergent migration \citep{Morrison+2018}, clearing broad gaps, or convergent migration if, for example, they are locked in resonance \citep{walsh_2011}. Another extension is the case of a single planet embedded in a disc with an evolving mass. A planet born embedded in a massive and dispersing protoplanetary disc, may carve a gap much wider than its chaotic zone due to sweeping secular resonances \citep{Zheng+2017}. Similar to the one planet case, testing these scenarios requires dedicated $N$-body simulations, which are beyond the scope of this paper.

\subsection{2:1 mean motion resonance}
\label{sec: 2:1MMR}
A planet orbiting near the inner edge of the disc, rather than within the gap itself, could open a gap at its external 2:1 MMR \citep[e.g.,][]{Tabeshian+2016, Tabeshian_2017}. In this scenario, the planet's semi-major axis is set by requiring its 2:1 MMR to coincide with the observed gap location, while its mass depends on the width of the gap. Rearranging Equation (8) from \citet{Tabeshian+2016} and expressing it in terms of width and not fractional width, the planet mass is given by:
\begin{equation}
    m_{\mathrm{plt}} = \frac{\mathrm{M}_{\mathrm{Jup}}}{0.009}\left(\frac{w_{\mathrm{gap}}}{2\sqrt{2\ln{(2)}}\times r_{\mathrm{gap}}}-0.009\right),
\label{eq:2:1MMR}
\end{equation}
assuming the planet is on a circular orbit. We use this relation to predict the planet mass and location, and compare these to the planet parameters excluded by the DPMs (Figure~\ref{fig:detectability}) to evaluate whether this MMR-driven scenario can explain the observed gaps.

In the case of HD\,92945, if the observed gap were shaped by the 2:1 MMR, the responsible planet would need a mass of $5.2-17.5$~\Mjup~at 45~au. A planet with these predicted orbital parameters would have been detected by \textit{JWST}/MIRI observations, therefore ruling out the 2:1 MMR mechanism as the origin of the gap in HD\,92945.

For HD\,107146, we first test whether the innermost gap could result from a 2:1 MMR, while assuming that a different mechanism is shaping the second gap \citep[e.g., 3:1 MMR,][]{Tabeshian_2017}. By setting the 2:1 MMR at the centre of the inner gap, we find that a planet with a mass between 4.1 and 6.9~\Mjup~at 35~au would be capable of carving the width of the first gap. In this configuration, the 3:1 MMR would be at ${\sim}73$~au, coinciding with the second gap, which can also be found to carve observable gaps when planet eccentricity becomes important \citep{Tabeshian_2017}. However, a planet with these predicted parameters would have been detected with the MIRI observations, therefore ruling out this scenario.

Alternatively, if the second gap in HD\,107146 were shaped by a 2:1 MMR, the corresponding planet would have a mass ranging from 20.3 to 28.8~\Mjup~at 49~au. Such a planet would have been detected with MIRI and can therefore rule out this scenario as the origin of the second gap in HD\,107146.

Finally, applying the same analysis to HD\,206893, the 2:1 MMR scenario would require a planet with a mass between 20 and 33~\Mjup~located at 43~au to produce the observed gap. However, a planet of this mass would have been observed with MIRI. Additionally, a planet at 43~au would lie within an observed ring of material between the inner edge (35~au) and the gap's inner edge (49~au), and would therefore truncate the disc in a way that contradicts the observed structure. The 2:1 MMR scenario is therefore ruled out as the origin of the gap in HD\,206893.

Overall, the 2:1 MMR scenario fails to explain the observed gaps in the three systems studied. Even when accounting for uncertainties in gap locations and widths, the required planet masses to carve such wide gaps are generally too high and already excluded by observational limits. While we have focused on cases where the MMR was exterior to the planet, we could also consider resonances interior to the planet \citep{Tabeshian+2016, Tabeshian_2017}, therefore placing planets near the outer edge of the discs. However, these configurations lead to the same conclusions, where the required planets are too massive and are therefore ruled out by MIRI observations. Finally, as with the embedded planet in a massive disc scenario, a lower-mass migrating planet could in principle open wider gaps through the sweeping of material into specific MMRs \citep[e.g.,][]{wyatt2003, friebe_2022}. Evaluating such scenario is outside the scope of this paper.

\subsection{Secular resonances in a massless disc due to two planets} 
\label{sec:two_planets_sec_res_YK18}
The observed gaps could also be carved by secular apsidal resonances induced by two inner planets on slightly eccentric orbits, assuming a massless disc for simplicity. In this scenario, the gap is not shaped by direct scattering, but through long-term gravitational interactions that excite planetesimal eccentricities. These secular resonances occur when a planetesimal's pericentre precession rate matches one of the system's eigenfrequencies, which governs the planet's orbital evolution \citep[e.g.][]{murray_1999}. At these resonance locations, eccentricities grow over time and material is gradually depleted locally, producing observable asymmetric gaps \citep{pearce_2015, yelverton_2018}. A two-planet system would generate two resonances, each dominated by one planet, but the resonances differ in width. In \citet{yelverton_2018}, one resonance is typically too narrow to generate a detectable gap, while the resonance dominated by the outer planet is generally broad enough to carve an observable gap \citep[see \S3 of][]{yelverton_2018}. The location and width of the gap depend on the planet's semi-major axis and mass ratios, as well as the planet's eccentricities.

Following the approach used in \citet{yelverton_2018}, we only focus on the most dominant secular resonance and treat the second resonance as negligible. This simplification allows us to use the analytical expression \citep[rearranged from Equation (24) in][]{yelverton_2018} for the location of the gap ($r_{\mathrm{gap}}$) due to the secular resonance:
\begin{equation}
    r_{\mathrm{gap}} = \left(\frac{m_{2}}{m_{1}}\right)^{2/7}a_{2}^{11/7}a_{1}^{-4/7},
    \label{eq:2planets+massless_sec_res}
\end{equation}
where $a_1$ and $a_2$ are the semi-major axes of the inner (1) and outer planet (2), and $m_1$ and $m_2$ are their respective masses. 

In addition to predicting the gap location, the width of the resonance depends on which resonance dominates. Table~3 in \citet{yelverton_2018} provides the corresponding expressions for the resonance width. Here we approximate the width of the resonance to the width of the gap as in \cite{yelverton_2018}, but note that this is only a first-order approximation. In the simpler cases, the width of the resonance only depends on the outer planet's semi-major axis and eccentricity, which can be approximated as:
\begin{equation}
     w_{\mathrm{gap}} \approx \frac{e_2a_2}{0.07},
    \label{eq:2planets+massless_sec_res_width}
\end{equation}
where $w_{\mathrm{gap}}$ is the width of the gap (or resonance in this approximation) and $e_2$ is the eccentricity of the outer planet. Using both $r_{\mathrm{gap}} $ and $w_{\mathrm{gap}}$, we can assess whether a given two-planet configuration is capable of producing the observed gap.

\begin{figure}
    \centering
    \begin{subfigure}{\linewidth}
        \includegraphics[trim=0.5cm 0.3cm 0.3cm 0.2cm, clip=true, width=\linewidth]{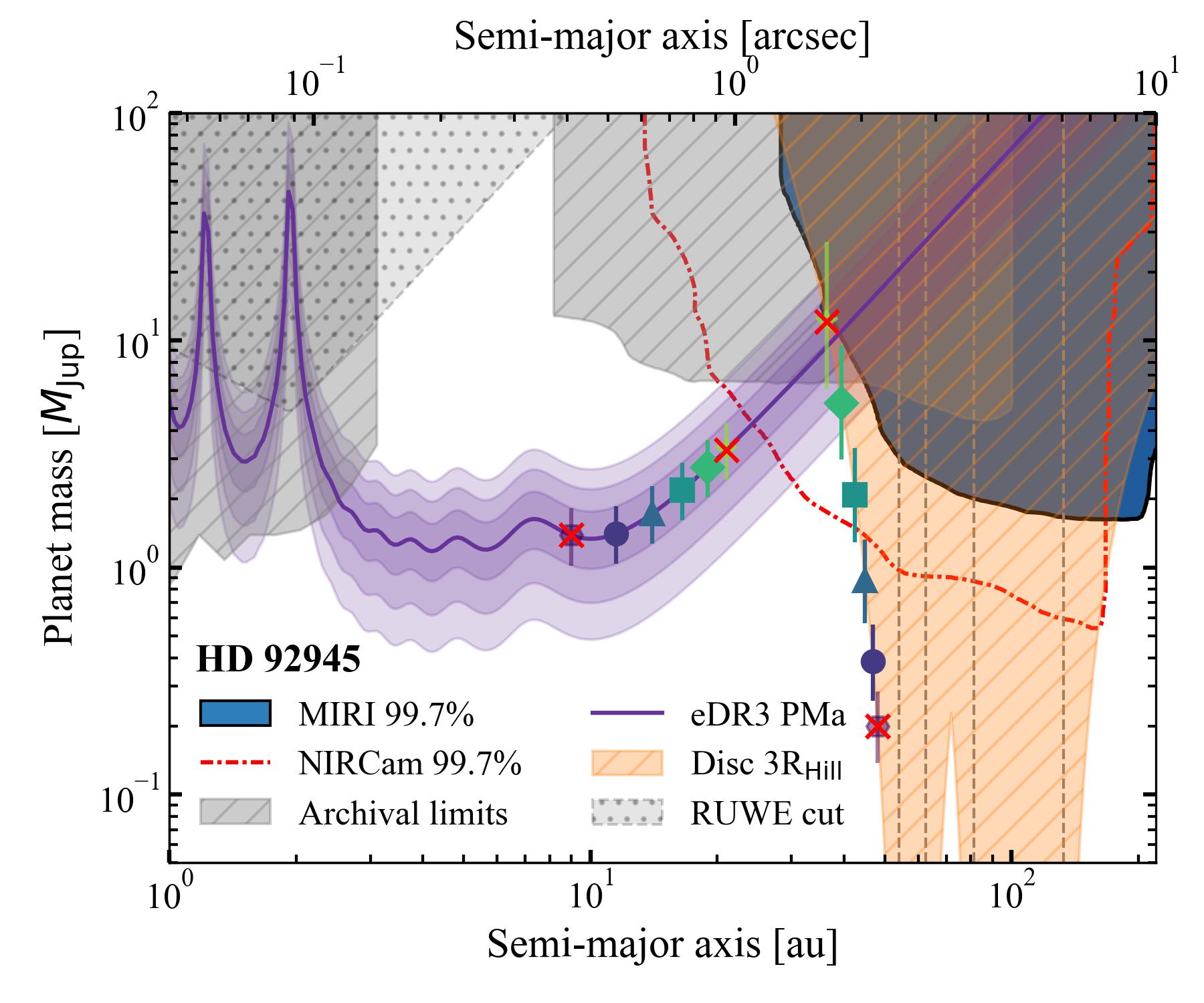}
        \vspace{-5mm}
        \caption{Secular resonance in a massless disc due to two planets in HD\,92945. Each point with a different shape and shade of colour represents a pair of planets that would create a gap through secular resonance at the observed gap location. Points with a red cross denote a combination of planets that is not possible, due to either one of the planets being in the region ruled out by observational limits or the clearing timescale of the inner edge of the disc being too long. The \textit{JWST}/MIRI and \textit{JWST}/NIRCam $99.7\%$ mass upper limits are shown as the blue region and the red dash-dotted line, respectively.}
        \label{fig:2planets_sec_res_hd92945}
    \end{subfigure}
    \begin{subfigure}{\linewidth}
        \includegraphics[trim=0.1cm 0.1cm 0.2cm 0.97cm, clip=true, width=\linewidth]{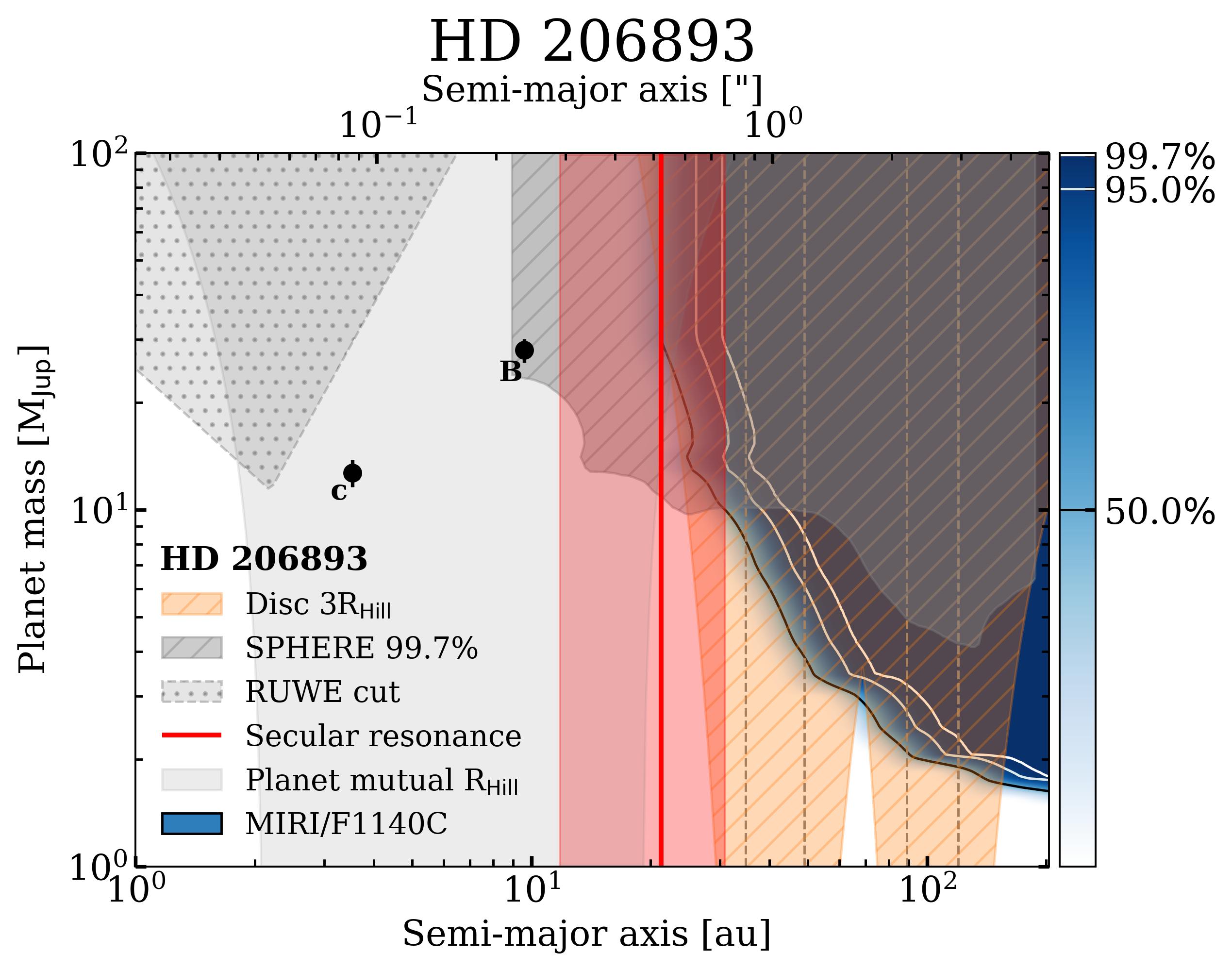}
        \caption{Secular resonance in a massless disc due to the two companions in HD\,206893. HD\,206893~B and c set the location of the secular resonance at 21~au, denoted by the vertical red line, with the red shaded region representing the resonance width, calculated with Equation~(\ref{eq:2planets+massless_sec_res_width}). This scenario is ruled out as the secular resonance location does not match the gap location.}
        \label{fig:2planets_sec_res_hd206893}
    \end{subfigure}
    \caption{Plots evaluating the secular resonance in a massless disc due to two planets scenario for HD\,92945 and HD\,206893. The general structure of both plots follows the DPM shown in Figure\,\ref{fig:detectability}.}
    \label{fig:2planets+massless}
\end{figure}

\subsubsection{Caveats}
This scenario assumes that only a single gap is formed via secular resonances, even though two resonant locations exist in principle. As shown in \citet{yelverton_2018}, only the dominant (widest) resonance is retained while the other is considered negligible, resulting in the formation of just one gap. However, when the planet semi-major axes are comparable and the mass of the outer planet is smaller or comparable to the mass of the inner planet, both resonances have a comparable width, and only considering the widest resonance is no longer applicable \citep[see Figures 4 and 5 in][]{yelverton_2018}. In this case, both resonances could be important enough to carve distinct gaps in the disc (see Friebe \& Sefilian, in prep.).

In this work, we use the simple approximation to find the gap location rather than numerically solving for the two resonance locations and widths. Despite this approximation, the predicted secular resonance location remains consistent with the numerical approach \citep[e.g., blue lines in Figure 9 of][comparing both methods for a similar planet configuration to HD\,206893]{marino_2020}.

\subsubsection{HD 92945}
Equation~(\ref{eq:2planets+massless_sec_res}) has four free parameters and only one constraint ($r_{\mathrm{gap}}$), making this a degenerate problem. However, we can use additional restrictions based on the PMa, disc extent, and direct imaging limits, to restrict the degeneracy in planet parameters. We place the inner planet on the PMa curve, as a planet is required in the inner regions of the system to explain the observed PMa in the system. This choice consequently fixes $m_1$ as a function of $a_1$. By rearranging Equation~(\ref{eq:2planets+massless_sec_res}) for $m_2$, we can determine the outer planet's mass as a function of $a_2$ by fixing the secular resonance location at the gap location (72\,au). Finally, the observed sharp inner edge \citep{imaz_blanco_2023}, which suggests the presence of a sculpting planet \citep{pearce2024}, allows us to place the outer planet as close as possible to the disc's inner edge (at 3~\Rhill), therefore lifting the degeneracy on $m_2$ and $a_2$. While this entire process yields a set of planet pairs as a function of $a_1$ that are capable of carving a gap via secular resonance at the correct location, it represents only one possible simplification to an otherwise highly degenerate problem. The evaluation process is illustrated in Figure~\ref{fig:2planets_sec_res_hd92945}.

By shifting the inner planet along the PMa curve, we assess the validity of each planet pair against the constrained parameter space from DPMs, determining the range of planet parameters that could explain the observed gap. One of the main limiting factors for the possible planet configurations is the direct imaging sensitivity, which excludes a planet pair when at least one of them falls within these excluded regions (seen as a crossed out configuration in Figure~\ref{fig:2planets_sec_res_hd206893}). In addition, we also consider the inner edge clearing timescale \citep[used in][]{pearce2022, pearce2024} to limit the valid planet configurations. If the mass of the outer planet is too low, the clearing timescale would exceed the system's age, allowing us to rule out the planet pair. Note that we have tested additional criteria, such as the mutual Hill radius and gap clearing timescales, though neither limits the possible planet configurations for HD\,92945.

We apply the method outlined above to quantify the population of planet pairs capable of producing the observed gap in HD\,92945. We incorporate the $1\sigma$ uncertainty on the PMa companion's mass (i.e., $m_1$), to capture the resulting range in the outer planet's location and mass (visible as the error bars in Figure~\ref{fig:2planets_sec_res_hd92945}). Using the \textit{JWST}/MIRI and SPHERE detection limits, the inner planet is expected to lie between $\sim$11 and 20~au with a mass of 1.4 to 3.1~\Mjup, while the outer planet would be located between 40 and 47~au with a mass of 0.3 to 6.5~\Mjup. Alternatively, using the \textit{JWST}/NIRCam red line in Figure~\ref{fig:2planets_sec_res_hd92945} would restrict the inner planet to $11-15$~au with a mass of $1.4-1.9$~\Mjup, and the outer planet to $44-47$~au with a mass ranging from 0.3 to 1.5~\Mjup. However, for a more conservative assessment and for reasons discussed in \S\ref{sec: DPM}, we adopt the constraints based on the \textit{JWST}/MIRI data.

It should be noted that secular resonances in this scenario require the planets to have some eccentricity. Our analysis adopts complementary dynamical arguments that assume circular orbits (from PMa and inner-edge sculpting). While strictly circular orbits would remove the quoted mass predictions, even slight eccentricities \citep[e.g., $e=0.05$,][]{yelverton_2018} are sufficient to generate gaps via secular resonances, meaning results remain applicable under realistic orbital conditions.

\subsubsection{HD 107146}
For HD\,107146, the two-planet secular resonance scenario in a massless disc faces the key limitation that the observed gaps in this system do not exhibit the strong asymmetry expected from this mechanism. $N$-body simulations by \citet{yelverton_2018}, which explored a wide range of two planet configurations, show that carving a gap without introducing significant azimuthal asymmetry is challenging. Since the gaps generated by this scenario are inherently asymmetric, the results are inconsistent with the observations, ruling out secular resonance in a massless disc due to two planets as the origin of the gaps in HD\,107146.

\subsubsection{HD 206893}
Evaluating this secular resonance scenario for HD\,206893 is more straightforward than for the previous two systems, as two companions (HD\,206893~B and c) have already been detected in this system. Using Equation~(\ref{eq:2planets+massless_sec_res}) with the orbital parameters of these planets \citep[from Table 3 in][]{hinkley_2023}, we find the secular resonance location to be at 21~au, which lies interior to the disc's inner edge at 35~au (see the vertical red line in Figure\,\ref{fig:2planets_sec_res_hd206893}). Given the observed planet configuration, the outer resonance dominates over the inner resonance, which following Table 3 in \citet{yelverton_2018} allows us to estimate the resonance width using Equation~(\ref{eq:2planets+massless_sec_res_width}). The depleted region produced by this resonance is then $21\pm6$\,au, as indicated by the red shaded area in Figure~\ref{fig:2planets_sec_res_hd206893}. Since the calculated secular resonance location does not align with the observed gap, we confidently rule out this secular resonance in a massless disc scenario as the origin of the gap in HD\,206893. 

From HD\,206893's DPM (Figure~\ref{fig:detectability}), we identify a narrow region in planet parameter space between the light-grey region excluded by mutual \Rhill~and the disc's inner edge, where a third planet could, in principle, exist. However, this region is seen to overlap with the predicted secular resonance at 21~au (Figure~\ref {fig:2planets_sec_res_hd206893}), which could affect the location's dynamical stability. This overlap suggests that the disc's inner edge may instead be carved by the secular resonance induced by the two known companions, making the presence of an additional planet near the inner edge unlikely. Further assessing such scenario would require dedicated dynamical analysis.

\subsection{Secular resonances in a massive disc due to a single planet}
\label{sec:sec_res_massive_disc}
We explore a gap-carving mechanism driven by secular apsidal resonances in discs with non-zero mass located exterior to a single planet. In this case the planetary precession is driven by the disc's back-reaction, with the disc's self-gravity modulating the evolution of planetesimals in addition to the planet \citep{sefilian_2021, sefilian_2023}.\footnote{For a somewhat similar result involving a disc that back-reacts on the planet but does not perturb itself, see \citet{pearce_2015}. In that case, the occurrence of a secular resonance requires a disc-to-planet mass ratio of $M_{\mathrm{disc}}/m_{\mathrm{plt}} \sim 1$, in contrast to the scenario in \citet{sefilian_2021, sefilian_2023}, where resonances occur when $M_{\mathrm{disc}}/m_{\mathrm{plt}} \lesssim 1$ (see also \S6.5 in \citealt{sefilian_2023}).} A key difference introduced by including disc mass is the evolution of the gap morphology. While secular resonances in both massless and massive discs initially produce highly asymmetric gaps, a planet interacting with a massive disc undergoes secular resonant friction (if $M_{\mathrm{disc}}/m_{\mathrm{plt}} \lesssim 1$), which gradually circularises the planet's orbit \citep{Tremaine1998}, and consequently results in a more symmetric gap over time \citep{sefilian_2023}.

To evaluate the valid planet parameters, Equation~(19) in \citet{sefilian_2021} relates a planet's semi-major axis and mass to the location of the secular resonance (i.e., gap location), given a fixed disc mass. Their Figure~7 and 12 illustrate how different combinations of planetary parameters and disc masses can produce a resonance capable of explaining the gap locations in HD 107146 and HD 92945, respectively. Using the constraints derived from DPMs, we can refine the previous viable planet parameter range. If we assume the planet responsible for this secular resonance also drives the PMa signal, we can overlay the PMa curve from the DPMs onto Figure~7 and 12 of \citet{sefilian_2021} to further narrow down the possible planet-disc configurations consistent with the observed gap. Note that these configurations correspond to disc-to-planet mass ratios of $M_{\mathrm{disc}}/m_{\mathrm{plt}} \lesssim 1$; for higher ratios, the system would not feature secular resonances and the disc may not be sufficiently stirred to produce observable signatures \citep{Sefilian2024}.

\subsubsection{HD 92945}
For HD\,92945, we identify the overlap between the valid planet-disc parameter space that satisfies gap opening timescales and maximum disc mass requirements \citep[white region in Figure 12 of][]{sefilian_2021} and the PMa-derived planet constraints (given in Table~\ref{tab:planet-constraints}).
This results in a planet located at ${\sim}10-25$~au with a mass between ${\sim}0.5-3$~\Mjup. Finally, by requiring that a secular resonance falls at the gap location, this constrains the disc mass to lie between ${\sim}10-100$ M$_{\oplus}$. Using the limits derived from the \textit{JWST}/NIRCam observations \citep{Lazzoni_2025} does not provide any tighter constraints. This scenario therefore remains valid to explain the observed gap in HD\,92945.

\subsubsection{HD 107146}
Figure 7 in \citet{sefilian_2021} illustrates the range of planet and disc parameters in which this scenario is capable of producing a gap in HD\,107146, assuming the disc has a single wide gap. Assuming the planet responsible for this secular resonance also produces the PMa signal, we can constrain the viable planet parameter space in the same way as for HD\,92945. Under this assumption, the planet is expected to be located between ${\sim}8-15$~au with a mass ranging from ${\sim}2-5$~\Mjup. To place the secular resonance at the observed gap location, the corresponding disc mass must lie between ${\sim}30-100$~M$_{\oplus}$.

However, a caveat in this analysis is that HD\,107146 features two distinct gaps in its disc, and not one single wide gap. This double-gap profile complicates the interpretation as this mechanism has only been shown to produce single gaps. In principle, this complication can be addressed if the system hosts two planets interior to the disc. Through their secular interactions with a sufficiently massive disc, the planets can sculpt two distinct axisymmetric gaps, each located at a corresponding secular resonance. Friebe \& Sefilian (in prep.) demonstrate that a broad range of planetary configurations, that remain below current detection limits, can sculpt two gaps at the observed locations, provided the disc mass does not exceed ${\sim}100 M_{\oplus}$.

\subsubsection{HD 206893}
Evaluating this scenario for HD\,206893 is challenging due to the presence of two already detected massive companions in the system. \citet{sefilian_2021} considered the influence of only HD\,206893~B, constraining the disc mass to ${\approx}170$~M$_{\oplus}$ (with updated orbital parameters for HD\,206893~B, this estimate increases to ${\approx}400$~M$_{\oplus}$). However, the presence of a second massive planet significantly affects the system's dynamics, making an analysis based solely on HD\,206893~B incomplete. The impact of an additional planet in this secular resonance scenario with a massive disc is explored in Friebe \& Sefilian (in prep.). They find that, while considering any additional inner planet will lower the disc mass required to maintain the resonance location at the observed location, accounting for a non-zero disc mass shifts the secular resonances inward relative to the results of \citet{yelverton_2018}, making it difficult to explain the observed gap.

\subsection{Alternative scenarios}

\subsubsection{Gap carving planet(s) orbiting exterior to the disc}
An additional scenario worth considering is the influence of planets on orbits exterior to the disc. Although gap formation by such companions has not been studied in great detail, evidence from systems such as HD\,106906 suggests that wide-orbit planets can strongly perturb debris structures through secular perturbations \citep{nesvold_2017, farhat_2023}. Exterior companions could provide an alternative pathway to explain some of the observed disc substructures, highlighting a mechanism worth exploring in future dynamical studies. For the three systems studied in this work, MIRI observations rule out companions ${>}1.5-2$~\Mjup~at $3\sigma$ confidence beyond the disc outer edges.

\subsubsection{Planet-less scenario}
So far, we have explored four gap-carving scenarios driven by single or multiple planets, but gaps could also form from planet-less mechanisms. As discussed in \citet{marino_2019, marino_2020}, several planet-less processes may explain the observed gaps across the different systems, particularly since all three discs exhibit gaps near 70~au. Possible mechanisms include: hydrodynamical processes that produce dust gaps in protoplanetary discs, which are then inherited by planetesimals formed from that dust and the collisionally generated (secondary) dust; a different size distribution in the gap, reducing the mass in dust, but not necessarily the local solid mass (e.g. by having larger planetesimals that dominate the mass budget); a change in the strength of solids or a higher dynamical excitation at the gap location making their collisional evolution faster \citep[see discussion in][for alternative gap origins]{marino_2019, marino_2020}.

\section{Conclusions}
\label{sec: conclusions}
In this paper, we present Cycle 1 \textit{JWST}/MIRI coronagraphic observations at $11.4~\mu$m (F1140C) targeting three exoKuiper belt systems: HD\,92945, HD\,107146, and HD\,206893, all of which show radial gaps in the discs, perhaps suggestive of planet-disc interactions. Each system also shows a significant PMa, further supporting the presence of unseen companions. The main goal of these \textit{JWST} observations was to detect the planets thought to be responsible for the observed gaps.

The MIRI observations were processed with \texttt{spaceKLIP}, using the ADI, RDI, and ADI+RDI PSF subtraction techniques. The reductions incorporate key improvements tailored to MIRI-specific artefacts, including corrections for the Brighter-Fatter effect and persistence trimming. We achieved the highest mass sensitivity with the ADI+RDI approach, converting the magnitudes to mass using a planet evolution model combining both ATMO-ceq and BEX.

No planet candidates were detected in the data. All detected sources were confirmed to be either background galaxies (identified through multi-wavelength, multi-epoch data) or a background star. Additionally, PSF-fitting was done to all the sources to catalogue the fluxes and positions of these background contaminants.

To quantify the observational sensitivity, we constructed detection probability maps (DPMs) for each system, translating contrast limits into de-projected separation and planet mass space. For HD\,92945, we also incorporated \textit{JWST}/NIRCam coronagraphic data at $4.4~\mu$m (F444W) from \citet{Lazzoni_2025} for comparison. These DPMs include not only the \textit{JWST} limits but also constraints from archival observations (ground-based direct imaging and radial velocity), disc morphology, Gaia astrometric noise (RUWE), and the PMa signal. While we did not reach the predicted sensitivity due to unaccounted MIRI artefacts in the \texttt{PanCAKE} simulations, the DPMs provide valuable constraints on the population of potential planets in each system.

We used these constraints to assess which planet configurations could explain the sculpting of the inner edge of the discs, the measured PMa signals and the observed gaps in the discs. We show that the planets responsible for the PMa are not the same as the ones shaping the inner edge of the discs. Instead, the PMa planets are likely located within the inner $20$~au of the system, consistent with trends observed in other systems. Only the next generation of instruments, from the ELTs to future Gaia releases, will expand the search for the planets responsible for the observed PMa with increased angular resolution and refined astrometric constraints.

Regarding planets responsible for the observed gaps, we evaluate four possible gap-carving scenarios for each system, using the constraints provided by the DPMs. Each mechanism was assessed based on its ability to reproduce the observed gap location, gap width, and observed gap symmetry. These \textit{JWST} observations allow us to rule out or place tighter constraints on the possible gap-carving planets. These results are summarised in Table~\ref{tab:summary-gap-formation}.

\noindent $\bullet$ For HD\,92945, three gap-carving scenarios remain possible: embedded planet(s) carving the gap, or planets interior to the disc inner edge producing gaps through secular resonances.
    
\noindent $\bullet$ For HD\,107146, the gaps could be explained by embedded planets in each gap or by migrating planet(s) capturing Trojan populations. The presence of two gaps and their level of symmetry make it currently difficult to explain them with secular resonances, though more detailed work including $N$-body simulations would be required to assess these scenarios further (Friebe \& Sefilian, in prep.).

\noindent $\bullet$ For HD\,206893, most gap-carving scenarios can be ruled out. Evaluation of the secular resonance in a massless disc due to the two known companions shows that the secular resonance location does not coincide with the observed gap, though it may be responsible for the sculpting of the disc's inner edge. Instead, the observed gap in this system is best explained by the presence of embedded planet(s).

Across all three systems, while some gap-forming mechanisms can be ruled out, multiple viable pathways remain for producing the observed disc substructures. Removing this degeneracy will require either detecting the inner planets causing the PMa's or using future direct imaging instruments with increased sensitivity to detect planets at the gap locations themselves. Disentangling the underlying gap-carving mechanisms will then require combining such future planet searches with dedicated $N$-body simulations to directly compare theoretical predictions with the observed disc morphologies.

\section*{Acknowledgements}
We would like to thank the referee for their constructive feedback that helped us improved the quality of this paper. RBW is supported by a Royal Society grant (RF-ERE-221025). SM acknowledges funding by the Royal Society through a Royal Society University Research Fellowship (URF-R1-221669) and the European Union through the FEED ERC project (grant number 101162711). A.A.S. is supported by the Heising-Simons Foundation through a 51 Pegasi b Fellowship. TDP is supported by a UKRI Stephen Hawking Fellowship and a Warwick Prize Fellowship, the latter made possible by a generous philanthropic donation. LM acknowledges funding by the European Union through the E-BEANS ERC project (grant number 100117693). Views and opinions expressed are however those of the author(s) only and do not necessarily reflect those of the European Union or the European Research Council Executive Agency. Neither the European Union nor the granting authority can be held responsible for them.

This work is based on observations made with the NASA/ESA/CSA James Webb Space Telescope. The data were obtained from the Mikulski Archive for Space Telescopes at the Space Telescope Science Institute, which is operated by the Association of Universities for Research in Astronomy, Inc., under NASA contract NAS 5-03127 for JWST. These observations are associated with program \#1668. This work has made use of the SPHERE Data Centre, jointly operated by OSUG/IPAG (Grenoble), PYTHEAS/LAM/CeSAM (Marseille), OCA/Lagrange (Nice) and Observatoire de Paris/LESIA (Paris) and supported by a grant from Labex OSUG@2020 (Investissements d’avenir – ANR10 LABX56).

RBW thanks W. O. Balmer, E. Bogat, R. Ferrer-Chavez, R. Kane and K. Franson for helpful discussions.

\section*{Data Availability}
All the JWST data used in this paper can be found in MAST at \href{https://doi.org/10.17909/gsf1-2q34}{doi:10.17909/gsf1-2q34}.



\bibliographystyle{mnras}
\bibliography{references} 

@ARTICLE{carter_jwst_2023,
       author = {{Carter}, Aarynn L. and {Hinkley}, Sasha and {Kammerer}, Jens and {Skemer}, Andrew and {Biller}, Beth A. and {Leisenring}, Jarron M. and {Millar-Blanchaer}, Maxwell A. and {Petrus}, Simon and {Stone}, Jordan M. and {Ward-Duong}, Kimberly and {Wang}, Jason J. and {Girard}, Julien H. and {Hines}, Dean C. and {Perrin}, Marshall D. and {Pueyo}, Laurent and {Balmer}, William O. and {Bonavita}, Mariangela and {Bonnefoy}, Mickael and {Chauvin}, Gael and {Choquet}, Elodie and {Christiaens}, Valentin and {Danielski}, Camilla and {Kennedy}, Grant M. and {Matthews}, Elisabeth C. and {Miles}, Brittany E. and {Patapis}, Polychronis and {Ray}, Shrishmoy and {Rickman}, Emily and {Sallum}, Steph and {Stapelfeldt}, Karl R. and {Whiteford}, Niall and {Zhou}, Yifan and {Absil}, Olivier and {Boccaletti}, Anthony and {Booth}, Mark and {Bowler}, Brendan P. and {Chen}, Christine H. and {Currie}, Thayne and {Fortney}, Jonathan J. and {Grady}, Carol A. and {Greebaum}, Alexandra Z. and {Henning}, Thomas and {Hoch}, Kielan K.~W. and {Janson}, Markus and {Kalas}, Paul and {Kenworthy}, Matthew A. and {Kervella}, Pierre and {Kraus}, Adam L. and {Lagage}, Pierre-Olivier and {Liu}, Michael C. and {Macintosh}, Bruce and {Marino}, Sebastian and {Marley}, Mark S. and {Marois}, Christian and {Matthews}, Brenda C. and {Mawet}, Dimitri and {McElwain}, Michael W. and {Metchev}, Stanimir and {Meyer}, Michael R. and {Molliere}, Paul and {Moran}, Sarah E. and {Morley}, Caroline V. and {Mukherjee}, Sagnick and {Pantin}, Eric and {Quirrenbach}, Andreas and {Rebollido}, Isabel and {Ren}, Bin B. and {Schneider}, Glenn and {Vasist}, Malavika and {Worthen}, Kadin and {Wyatt}, Mark C. and {Briesemeister}, Zackery W. and {Bryan}, Marta L. and {Calissendorff}, Per and {Cantalloube}, Faustine and {Cugno}, Gabriele and {De Furio}, Matthew and {Dupuy}, Trent J. and {Factor}, Samuel M. and {Faherty}, Jacqueline K. and {Fitzgerald}, Michael P. and {Franson}, Kyle and {Gonzales}, Eileen C. and {Hood}, Callie E. and {Howe}, Alex R. and {Kuzuhara}, Masayuki and {Lagrange}, Anne-Marie and {Lawson}, Kellen and {Lazzoni}, Cecilia and {Lew}, Ben W.~P. and {Liu}, Pengyu and {Llop-Sayson}, Jorge and {Lloyd}, James P. and {Martinez}, Raquel A. and {Mazoyer}, Johan and {Palma-Bifani}, Paulina and {Quanz}, Sascha P. and {Redai}, Jea Adams and {Samland}, Matthias and {Schlieder}, Joshua E. and {Tamura}, Motohide and {Tan}, Xianyu and {Uyama}, Taichi and {Vigan}, Arthur and {Vos}, Johanna M. and {Wagner}, Kevin and {Wolff}, Schuyler G. and {Ygouf}, Marie and {Zhang}, Xi and {Zhang}, Keming and {Zhang}, Zhoujian},
        title = "{The JWST Early Release Science Program for Direct Observations of Exoplanetary Systems I: High-contrast Imaging of the Exoplanet HIP 65426 b from 2 to 16 {\ensuremath{\mu}}m}",
      journal = {\apjl},
         year = 2023,
        month = jul,
       volume = {951},
       number = {1},
          eid = {L20},
        pages = {L20},
          doi = {10.3847/2041-8213/acd93e},
archivePrefix = {arXiv},
       eprint = {2208.14990},
 primaryClass = {astro-ph.EP},
       adsurl = {https://ui.adsabs.harvard.edu/abs/2023ApJ...951L..20C}
}

@ARTICLE{Zheng+2017,
       author = {{Zheng}, Xiaochen and {Lin}, Douglas N.~C. and {Kouwenhoven}, M.~B.~N. and {Mao}, Shude and {Zhang}, Xiaojia},
        title = "{Clearing Residual Planetesimals by Sweeping Secular Resonances in Transitional Disks: A Lone-planet Scenario for the Wide Gaps in Debris Disks around Vega and Fomalhaut}",
      journal = {\apj},
         year = 2017,
        month = nov,
       volume = {849},
       number = {2},
          eid = {98},
        pages = {98},
          doi = {10.3847/1538-4357/aa8ef3},
archivePrefix = {arXiv},
       eprint = {1709.07382},
 primaryClass = {astro-ph.EP},
       adsurl = {https://ui.adsabs.harvard.edu/abs/2017ApJ...849...98Z}
}

@ARTICLE{meunier_2012,
       author = {{Meunier}, N. and {Lagrange}, A. -M. and {De Bondt}, K.},
        title = "{Comparison of different exoplanet mass detection limit methods using a sample of main-sequence intermediate-type stars}",
      journal = {\aap},
         year = 2012,
        month = sep,
       volume = {545},
          eid = {A87},
        pages = {A87},
          doi = {10.1051/0004-6361/201219163},
archivePrefix = {arXiv},
       eprint = {1207.4329},
 primaryClass = {astro-ph.EP},
       adsurl = {https://ui.adsabs.harvard.edu/abs/2012A&A...545A..87M}
}

@ARTICLE{Lazzoni_2025,
       author = {{Lazzoni}, C. and {Bendahan-West}, R. and {Marino}, S. and {Lawson}, K.~D. and {Carter}, A. and {Squicciarini}, V. and {Strampelli}, G. and {Hinkley}, S. and {Kennedy}, G. and {James}, A.~D. and {Milli}, J. and {Ray}, S.},
        title = "{JWST/NIRCam observations of HD\raisebox{-0.5ex}\textasciitilde92945 debris disk: An asymmetric disk with a gap}",
      journal = {arXiv e-prints},
         year = 2025,
        month = nov,
          eid = {arXiv:2511.07561},
        pages = {arXiv:2511.07561},
          doi = {10.48550/arXiv.2511.07561},
archivePrefix = {arXiv},
       eprint = {2511.07561},
 primaryClass = {astro-ph.EP},
       adsurl = {https://ui.adsabs.harvard.edu/abs/2025arXiv251107561L}
}

@ARTICLE{Sanghi_2025,
       author = {{Sanghi}, Aniket and {Beichman}, Charles and {Mawet}, Dimitri and {Balmer}, William O. and {Godoy}, Nicolas and {Pueyo}, Laurent and {Boccaletti}, Anthony and {Sommer}, Max and {Bidot}, Alexis and {Choquet}, Elodie and {Kervella}, Pierre and {Lagage}, Pierre-Olivier and {Leisenring}, Jarron and {Llop-Sayson}, Jorge and {Ressler}, Michael and {Wagner}, Kevin and {Wyatt}, Mark},
        title = "{Worlds Next Door: A Candidate Giant Planet Imaged in the Habitable Zone of {\ensuremath{\alpha}} Centauri A. II. Binary Star Modeling, Planet and Exozodi Search, and Sensitivity Analysis}",
      journal = {\apjl},
         year = 2025,
        month = aug,
       volume = {989},
       number = {2},
          eid = {L23},
        pages = {L23},
          doi = {10.3847/2041-8213/adf53e},
archivePrefix = {arXiv},
       eprint = {2508.03812},
 primaryClass = {astro-ph.EP},
       adsurl = {https://ui.adsabs.harvard.edu/abs/2025ApJ...989L..23S}
}

@ARTICLE{Malin_2025,
       author = {{M{\^a}lin}, Mathilde and {Boccaletti}, Anthony and {Perrot}, Cl{\'e}ment and {Baudoz}, Pierre and {Rouan}, Daniel and {Lagage}, Pierre-Olivier and {Waters}, Rens and {G{\"u}del}, Manuel and {Henning}, Thomas and {Vandenbussche}, Bart and {Absil}, Olivier and {Barrado}, David and {Charnay}, Benjamin and {Choquet}, Elodie and {Cossou}, Christophe and {Danielski}, Camilla and {Decin}, Leen and {Glauser}, Adrian M. and {Pye}, John and {Olofsson}, Goran and {Glasse}, Alistair and {Patapis}, Polychronis and {Royer}, Pierre and {Scheithauer}, Silvia and {Serabyn}, Eugene and {Tremblin}, Pascal and {Whiteford}, Niall and {van Dishoeck}, Ewine F. and {Ostlin}, G{\"o}ran and {Ray}, Tom P. and {Wright}, Gillian},
        title = "{First unambiguous detection of ammonia in the atmosphere of a planetary mass companion with JWST/MIRI coronagraphs}",
      journal = {\aap},
         year = 2025,
        month = jan,
       volume = {693},
          eid = {A315},
        pages = {A315},
          doi = {10.1051/0004-6361/202452695},
archivePrefix = {arXiv},
       eprint = {2501.00104},
 primaryClass = {astro-ph.EP},
       adsurl = {https://ui.adsabs.harvard.edu/abs/2025A&A...693A.315M}
}

@ARTICLE{Wolff_2025,
       author = {{Wolff}, Schuyler G. and {G{\'a}sp{\'a}r}, Andr{\'a}s and {Rieke}, George and {Leisenring}, Jarron M. and {Sefilian}, Antranik A. and {Ygouf}, Marie and {Llop-Sayson}, Jorge},
        title = "{JWST/MIRI Imaging of the Warm Dust Component of the ϵ Eridani Debris Disk}",
      journal = {\aj},
         year = 2025,
        month = oct,
       volume = {170},
       number = {4},
          eid = {244},
        pages = {244},
          doi = {10.3847/1538-3881/adfcd6},
archivePrefix = {arXiv},
       eprint = {2509.24976},
 primaryClass = {astro-ph.EP},
       adsurl = {https://ui.adsabs.harvard.edu/abs/2025AJ....170..244W}
}

@ARTICLE{Bardalez-Gagliuffi_2025,
       author = {{Bardalez Gagliuffi}, Daniella C. and {Balmer}, William O. and {Pueyo}, Laurent and {Brandt}, Timothy D. and {Giovinazzi}, Mark R. and {Millholland}, Sarah and {Black}, Brennen and {Lu}, Tiger and {Rice}, Malena and {Mang}, James and {Morley}, Caroline and {Lacy}, Brianna and {Girard}, Julien H. and {Matthews}, Elisabeth C. and {Carter}, Aarynn L. and {Bowler}, Brendan P. and {Faherty}, Jacqueline K. and {Fontanive}, Clemence and {Rickman}, Emily},
        title = "{JWST Coronagraphic Images of 14 Her c: A Cold Giant Planet in a Dynamically Hot Multiplanet System}",
      journal = {\apjl},
         year = 2025,
        month = jul,
       volume = {988},
       number = {1},
          eid = {L18},
        pages = {L18},
          doi = {10.3847/2041-8213/ade30f},
archivePrefix = {arXiv},
       eprint = {2506.09201},
 primaryClass = {astro-ph.EP},
       adsurl = {https://ui.adsabs.harvard.edu/abs/2025ApJ...988L..18B}
}

@ARTICLE{tal-or_2019,
       author = {{Tal-Or}, Lev and {Trifonov}, Trifon and {Zucker}, Shay and {Mazeh}, Tsevi and {Zechmeister}, Mathias},
        title = "{Correcting HIRES/Keck radial velocities for small systematic errors}",
      journal = {\mnras},
         year = 2019,
        month = mar,
       volume = {484},
       number = {1},
        pages = {L8-L13},
          doi = {10.1093/mnrasl/sly227},
archivePrefix = {arXiv},
       eprint = {1810.02986},
 primaryClass = {astro-ph.EP},
       adsurl = {https://ui.adsabs.harvard.edu/abs/2019MNRAS.484L...8T}
}

@INPROCEEDINGS{Lajoie_2016,
       author = {{Lajoie}, Charles-Philippe and {Soummer}, R{\'e}mi and {Pueyo}, Laurent and {Hines}, Dean C. and {Nelan}, Edmund P. and {Perrin}, Marshall and {Clampin}, Mark and {Isaacs}, John C.},
        title = "{Small-grid dithers for the JWST coronagraphs}",
    booktitle = {Space Telescopes and Instrumentation 2016: Optical, Infrared, and Millimeter Wave},
         year = 2016,
       editor = {{MacEwen}, Howard A. and {Fazio}, Giovanni G. and {Lystrup}, Makenzie and {Batalha}, Natalie and {Siegler}, Nicholas and {Tong}, Edward C.},
       series = {Society of Photo-Optical Instrumentation Engineers (SPIE) Conference Series},
       volume = {9904},
        month = jul,
          eid = {99045K},
        pages = {99045K},
          doi = {10.1117/12.2233032},
       adsurl = {https://ui.adsabs.harvard.edu/abs/2016SPIE.9904E..5KL}
}

@ARTICLE{Ruane_2019,
       author = {{Ruane}, Garreth and {Ngo}, Henry and {Mawet}, Dimitri and {Absil}, Olivier and {Choquet}, {\'E}lodie and {Cook}, Therese and {Gomez Gonzalez}, Carlos and {Huby}, Elsa and {Matthews}, Keith and {Meshkat}, Tiffany and {Reggiani}, Maddalena and {Serabyn}, Eugene and {Wallack}, Nicole and {Xuan}, W. Jerry},
        title = "{Reference Star Differential Imaging of Close-in Companions and Circumstellar Disks with the NIRC2 Vortex Coronagraph at the W. M. Keck Observatory}",
      journal = {\aj},
         year = 2019,
        month = mar,
       volume = {157},
       number = {3},
          eid = {118},
        pages = {118},
          doi = {10.3847/1538-3881/aafee2},
archivePrefix = {arXiv},
       eprint = {1901.04090},
 primaryClass = {astro-ph.IM},
       adsurl = {https://ui.adsabs.harvard.edu/abs/2019AJ....157..118R}
}

@ARTICLE{Marois_2006,
       author = {{Marois}, Christian and {Lafreni{\`e}re}, David and {Doyon}, Ren{\'e} and {Macintosh}, Bruce and {Nadeau}, Daniel},
        title = "{Angular Differential Imaging: A Powerful High-Contrast Imaging Technique}",
      journal = {\apj},
         year = 2006,
        month = apr,
       volume = {641},
       number = {1},
        pages = {556-564},
          doi = {10.1086/500401},
archivePrefix = {arXiv},
       eprint = {astro-ph/0512335},
 primaryClass = {astro-ph},
       adsurl = {https://ui.adsabs.harvard.edu/abs/2006ApJ...641..556M}
}

@ARTICLE{rouan2000,
       author = {{Rouan}, D. and {Riaud}, P. and {Boccaletti}, A. and {Cl{\'e}net}, Y. and {Labeyrie}, A.},
        title = "{The Four-Quadrant Phase-Mask Coronagraph. I. Principle}",
      journal = {\pasp},
         year = 2000,
        month = nov,
       volume = {112},
       number = {777},
        pages = {1479-1486},
          doi = {10.1086/317707},
       adsurl = {https://ui.adsabs.harvard.edu/abs/2000PASP..112.1479R}
}

@ARTICLE{muller_1987,
       author = {{M{\"u}ller}, M. and {Weigelt}, G.},
        title = "{High-resolution astronomical imaging by roll deconvolution of Space Telescope data}",
      journal = {\aap},
         year = 1987,
        month = mar,
       volume = {175},
       number = {1-2},
        pages = {312-318},
       adsurl = {https://ui.adsabs.harvard.edu/abs/1987A&A...175..312M}
}

@ARTICLE{Boccaletti_2015,
       author = {{Boccaletti}, A. and {Lagage}, P.-O. and {Baudoz}, P. and {Beichman}, C. and {Bouchet}, P. and {Cavarroc}, C. and {Dubreuil}, D. and {Glasse}, Alistair and {Glauser}, A.~M. and {Hines}, D.~C. and {Lajoie}, C.-P. and {Lebreton}, J. and {Perrin}, M.~D. and {Pueyo}, L. and {Reess}, J.~M. and {Rieke}, G.~H. and {Ronayette}, S. and {Rouan}, D. and {Soummer}, R. and {Wright}, G.~S.},
        title = "{The Mid-Infrared Instrument for the James Webb Space Telescope, V: Predicted Performance of the MIRI Coronagraphs}",
      journal = {\pasp},
         year = 2015,
        month = jul,
       volume = {127},
       number = {953},
        pages = {633},
          doi = {10.1086/682256},
archivePrefix = {arXiv},
       eprint = {1508.02352},
 primaryClass = {astro-ph.IM},
       adsurl = {https://ui.adsabs.harvard.edu/abs/2015PASP..127..633B}
}

@ARTICLE{farhat_2023,
       author = {{Farhat}, Mohammad A. and {Sefilian}, Antranik A. and {Touma}, Jihad R.},
        title = "{The case of HD 106906 debris disc: a binary's revenge}",
      journal = {\mnras},
         year = 2023,
        month = may,
       volume = {521},
       number = {2},
        pages = {2067-2086},
          doi = {10.1093/mnras/stad316},
archivePrefix = {arXiv},
       eprint = {2210.07395},
 primaryClass = {astro-ph.EP},
       adsurl = {https://ui.adsabs.harvard.edu/abs/2023MNRAS.521.2067F}
}

@ARTICLE{nesvold_2017,
       author = {{Nesvold}, Erika R. and {Naoz}, Smadar and {Fitzgerald}, Michael P.},
        title = "{HD 106906: A Case Study for External Perturbations of a Debris Disk}",
      journal = {\apjl},
         year = 2017,
        month = mar,
       volume = {837},
       number = {1},
          eid = {L6},
        pages = {L6},
          doi = {10.3847/2041-8213/aa61a7},
archivePrefix = {arXiv},
       eprint = {1702.06578},
 primaryClass = {astro-ph.EP},
       adsurl = {https://ui.adsabs.harvard.edu/abs/2017ApJ...837L...6N}
}

@ARTICLE{fang_2013,
       author = {{Fang}, Julia and {Margot}, Jean-Luc},
        title = "{Are Planetary Systems Filled to Capacity? A Study Based on Kepler Results}",
      journal = {\apj},
         year = 2013,
        month = apr,
       volume = {767},
       number = {2},
          eid = {115},
        pages = {115},
          doi = {10.1088/0004-637X/767/2/115},
archivePrefix = {arXiv},
       eprint = {1302.7190},
 primaryClass = {astro-ph.EP},
       adsurl = {https://ui.adsabs.harvard.edu/abs/2013ApJ...767..115F}
}

@ARTICLE{Tabeshian_2017,
       author = {{Tabeshian}, Maryam and {Wiegert}, Paul A.},
        title = "{Detection and Characterization of Extrasolar Planets through Mean-motion Resonances. II. The Effect of the Planet{\textquoteright}s Orbital Eccentricity on Debris Disk Structures}",
      journal = {\apj},
         year = 2017,
        month = sep,
       volume = {847},
       number = {1},
          eid = {24},
        pages = {24},
          doi = {10.3847/1538-4357/aa831f},
archivePrefix = {arXiv},
       eprint = {1709.09978},
 primaryClass = {astro-ph.EP},
       adsurl = {https://ui.adsabs.harvard.edu/abs/2017ApJ...847...24T}
}

@ARTICLE{Bonsor2014,
       author = {{Bonsor}, Amy and {Raymond}, Sean N. and {Augereau}, Jean-Charles and {Ormel}, Chris W.},
        title = "{Planetesimal-driven migration as an explanation for observations of high levels of warm, exozodiacal dust}",
      journal = {\mnras},
         year = 2014,
        month = jul,
       volume = {441},
       number = {3},
        pages = {2380-2391},
          doi = {10.1093/mnras/stu721},
archivePrefix = {arXiv},
       eprint = {1404.2606},
 primaryClass = {astro-ph.EP},
       adsurl = {https://ui.adsabs.harvard.edu/abs/2014MNRAS.441.2380B}
}

@INPROCEEDINGS{Tremaine1993,
       author = {{Tremaine}, Scott},
        title = "{The distribution of comets around stars.}",
    booktitle = {Planets Around Pulsars},
         year = 1993,
       editor = {{Phillips}, J.~A. and {Thorsett}, Steve E. and {Kulkarni}, Shri R.},
       series = {Astronomical Society of the Pacific Conference Series},
       volume = {36},
        month = jan,
        pages = {335-344},
       adsurl = {https://ui.adsabs.harvard.edu/abs/1993ASPC...36..335T}
}

@ARTICLE{wyatt_2005,
       author = {{Wyatt}, M.~C.},
        title = "{Spiral structure when setting up pericentre glow: possible giant planets at hundreds of AU in the HD 141569 disk}",
      journal = {\aap},
         year = 2005,
        month = sep,
       volume = {440},
       number = {3},
        pages = {937-948},
          doi = {10.1051/0004-6361:20053391},
archivePrefix = {arXiv},
       eprint = {astro-ph/0506208},
 primaryClass = {astro-ph},
       adsurl = {https://ui.adsabs.harvard.edu/abs/2005A&A...440..937W}
}

@ARTICLE{faramaz_2014,
       author = {{Faramaz}, V. and {Beust}, H. and {Th{\'e}bault}, P. and {Augereau}, J. -C. and {Bonsor}, A. and {del Burgo}, C. and {Ertel}, S. and {Marshall}, J.~P. and {Milli}, J. and {Montesinos}, B. and {Mora}, A. and {Bryden}, G. and {Danchi}, W. and {Eiroa}, C. and {White}, G.~J. and {Wolf}, S.},
        title = "{Can eccentric debris disks be long-lived?. A first numerical investigation and application to {\ensuremath{\zeta}}$^{2}$ Reticuli}",
      journal = {\aap},
         year = 2014,
        month = mar,
       volume = {563},
          eid = {A72},
        pages = {A72},
          doi = {10.1051/0004-6361/201322469},
archivePrefix = {arXiv},
       eprint = {1312.5146},
 primaryClass = {astro-ph.EP},
       adsurl = {https://ui.adsabs.harvard.edu/abs/2014A&A...563A..72F}
}

@ARTICLE{boccaletti_2019,
       author = {{Boccaletti}, A. and {Th{\'e}bault}, P. and {Pawellek}, N. and {Lagrange}, A. -M. and {Galicher}, R. and {Desidera}, S. and {Milli}, J. and {Kral}, Q. and {Bonnefoy}, M. and {Augereau}, J. -C. and {Maire}, A.~L. and {Henning}, T. and {Beust}, H. and {Rodet}, L. and {Avenhaus}, H. and {Bhowmik}, T. and {Bonavita}, M. and {Chauvin}, G. and {Cheetham}, A. and {Cudel}, M. and {Feldt}, M. and {Gratton}, R. and {Hagelberg}, J. and {Janin-Potiron}, P. and {Langlois}, M. and {Menard}, F. and {Mesa}, D. and {Meyer}, M. and {Peretti}, S. and {Perrot}, C. and {Schmidt}, T. and {Sissa}, E. and {Vigan}, A. and {Rickman}, E. and {Magnard}, Y. and {Maurel}, D. and {Moeller-Nilsson}, O. and {Perret}, D. and {Sauvage}, J. -F.},
        title = "{Two cold belts in the debris disk around the G-type star NZ Lupi}",
      journal = {\aap},
         year = 2019,
        month = may,
       volume = {625},
          eid = {A21},
        pages = {A21},
          doi = {10.1051/0004-6361/201935135},
archivePrefix = {arXiv},
       eprint = {1904.02746},
 primaryClass = {astro-ph.EP},
       adsurl = {https://ui.adsabs.harvard.edu/abs/2019A&A...625A..21B}
}

@ARTICLE{feldt_2017,
       author = {{Feldt}, M. and {Olofsson}, J. and {Boccaletti}, A. and {Maire}, A.~L. and {Milli}, J. and {Vigan}, A. and {Langlois}, M. and {Henning}, Th. and {Moor}, A. and {Bonnefoy}, M. and {Wahhaj}, Z. and {Desidera}, S. and {Gratton}, R. and {K{\'o}sp{\'a}l}, {\'A}. and {Abraham}, P. and {Menard}, F. and {Chauvin}, G. and {Lagrange}, A.~M. and {Mesa}, D. and {Salter}, G. and {Buenzli}, E. and {Lannier}, J. and {Perrot}, C. and {Peretti}, S. and {Sissa}, E.},
        title = "{SPHERE/SHINE reveals concentric rings in the debris disk of HIP 73145}",
      journal = {\aap},
         year = 2017,
        month = may,
       volume = {601},
          eid = {A7},
        pages = {A7},
          doi = {10.1051/0004-6361/201629261},
archivePrefix = {arXiv},
       eprint = {1612.07621},
 primaryClass = {astro-ph.SR},
       adsurl = {https://ui.adsabs.harvard.edu/abs/2017A&A...601A...7F}
}

@ARTICLE{bonnefoy_2017,
       author = {{Bonnefoy}, M. and {Milli}, J. and {M{\'e}nard}, F. and {Vigan}, A. and {Lagrange}, A. -M. and {Delorme}, P. and {Boccaletti}, A. and {Lazzoni}, C. and {Galicher}, R. and {Desidera}, S. and {Chauvin}, G. and {Augereau}, J.~C. and {Mouillet}, D. and {Pinte}, C. and {van der Plas}, G. and {Gratton}, R. and {Beust}, H. and {Beuzit}, J.~L.},
        title = "{Belt(s) of debris resolved around the Sco-Cen star HIP 67497}",
      journal = {\aap},
         year = 2017,
        month = jan,
       volume = {597},
          eid = {L7},
        pages = {L7},
          doi = {10.1051/0004-6361/201628929},
archivePrefix = {arXiv},
       eprint = {1609.05638},
 primaryClass = {astro-ph.EP},
       adsurl = {https://ui.adsabs.harvard.edu/abs/2017A&A...597L...7B}
}

@ARTICLE{perrot_2016,
       author = {{Perrot}, C. and {Boccaletti}, A. and {Pantin}, E. and {Augereau}, J. -C. and {Lagrange}, A. -M. and {Galicher}, R. and {Maire}, A. -L. and {Mazoyer}, J. and {Milli}, J. and {Rousset}, G. and {Gratton}, R. and {Bonnefoy}, M. and {Brandner}, W. and {Buenzli}, E. and {Langlois}, M. and {Lannier}, J. and {Mesa}, D. and {Peretti}, S. and {Salter}, G. and {Sissa}, E. and {Chauvin}, G. and {Desidera}, S. and {Feldt}, M. and {Vigan}, A. and {Di Folco}, E. and {Dutrey}, A. and {P{\'e}ricaud}, J. and {Baudoz}, P. and {Benisty}, M. and {De Boer}, J. and {Garufi}, A. and {Girard}, J.~H. and {Menard}, F. and {Olofsson}, J. and {Quanz}, S.~P. and {Mouillet}, D. and {Christiaens}, V. and {Casassus}, S. and {Beuzit}, J. -L. and {Blanchard}, P. and {Carle}, M. and {Fusco}, T. and {Giro}, E. and {Hubin}, N. and {Maurel}, D. and {Moeller-Nilsson}, O. and {Sevin}, A. and {Weber}, L.},
        title = "{Discovery of concentric broken rings at sub-arcsec separations in the HD 141569A gas-rich, debris disk with VLT/SPHERE}",
      journal = {\aap},
         year = 2016,
        month = may,
       volume = {590},
          eid = {L7},
        pages = {L7},
          doi = {10.1051/0004-6361/201628396},
archivePrefix = {arXiv},
       eprint = {1605.00468},
 primaryClass = {astro-ph.EP},
       adsurl = {https://ui.adsabs.harvard.edu/abs/2016A&A...590L...7P}
}

@ARTICLE{trifonov_2020,
       author = {{Trifonov}, Trifon and {Tal-Or}, Lev and {Zechmeister}, Mathias and {Kaminski}, Adrian and {Zucker}, Shay and {Mazeh}, Tsevi},
        title = "{Public HARPS radial velocity database corrected for systematic errors}",
      journal = {\aap},
         year = 2020,
        month = apr,
       volume = {636},
          eid = {A74},
        pages = {A74},
          doi = {10.1051/0004-6361/201936686},
archivePrefix = {arXiv},
       eprint = {2001.05942},
 primaryClass = {astro-ph.EP},
       adsurl = {https://ui.adsabs.harvard.edu/abs/2020A&A...636A..74T}
}

@INPROCEEDINGS{Kammerer_2022,
       author = {{Kammerer}, Jens and {Girard}, Julien and {Carter}, Aarynn L. and {Perrin}, Marshall D. and {Cooper}, Rachel and {Thatte}, Deepashri and {Vandal}, Thomas and {Leisenring}, Jarron and {Wang}, Jason and {Balmer}, William O. and {Sivaramakrishnan}, Anand and {Pueyo}, Laurent and {Ward-Duong}, Kimberly and {Sunnquist}, Ben and {Adams Redai}, J{\'e}a.},
        title = "{Performance of near-infrared high-contrast imaging methods with JWST from commissioning}",
    booktitle = {Space Telescopes and Instrumentation 2022: Optical, Infrared, and Millimeter Wave},
         year = 2022,
       editor = {{Coyle}, Laura E. and {Matsuura}, Shuji and {Perrin}, Marshall D.},
       series = {Society of Photo-Optical Instrumentation Engineers (SPIE) Conference Series},
       volume = {12180},
        month = aug,
          eid = {121803N},
        pages = {121803N},
          doi = {10.1117/12.2628865},
archivePrefix = {arXiv},
       eprint = {2208.00996},
 primaryClass = {astro-ph.EP},
       adsurl = {https://ui.adsabs.harvard.edu/abs/2022SPIE12180E..3NK}
}

@ARTICLE{Morrison+2018,
       author = {{Morrison}, Sarah J. and {Kratter}, Kaitlin M.},
        title = "{Gap formation in planetesimal discs via divergently migrating planets}",
      journal = {\mnras},
         year = 2018,
        month = dec,
       volume = {481},
       number = {4},
        pages = {5180-5188},
          doi = {10.1093/mnras/sty2657},
archivePrefix = {arXiv},
       eprint = {1809.10209},
 primaryClass = {astro-ph.EP},
       adsurl = {https://ui.adsabs.harvard.edu/abs/2018MNRAS.481.5180M}
}

@INPROCEEDINGS{carter_pancake_2021,
       author = {{Carter}, Aarynn L. and {Skemer}, Andrew J.~I. and {Danielski}, Camilla and {Leisenring}, Jarron and {Wang}, Jason J. and {Van Gorkom}, Kyle and {York}, Brian and {Adams}, Jea and {Biller}, Beth and {Girard}, Julien H. and {Hinkley}, Sasha and {Nickson}, Bryony and {Perrin}, Marshall and {Pueyo}, Laurent},
        title = "{Simulating JWST high contrast observations with PanCAKE}",
    booktitle = {Techniques and Instrumentation for Detection of Exoplanets X},
         year = 2021,
       editor = {{Shaklan}, Stuart B. and {Ruane}, Garreth J.},
       series = {Society of Photo-Optical Instrumentation Engineers (SPIE) Conference Series},
       volume = {11823},
        month = sep,
          eid = {118230H},
        pages = {118230H},
          doi = {10.1117/12.2594501},
       adsurl = {https://ui.adsabs.harvard.edu/abs/2021SPIE11823E..0HC}
}

@ARTICLE{Sefilian2024,
       author = {{Sefilian}, Antranik A.},
        title = "{Massive Debris Disks May Hinder Secular Stirring by Planetary Companions: An Analytic Proof of Concept}",
      journal = {\apj},
         year = 2024,
        month = may,
       volume = {966},
       number = {1},
          eid = {140},
        pages = {140},
          doi = {10.3847/1538-4357/ad32d1},
archivePrefix = {arXiv},
       eprint = {2401.18020},
 primaryClass = {astro-ph.EP},
       adsurl = {https://ui.adsabs.harvard.edu/abs/2024ApJ...966..140S}
}

@ARTICLE{kervella+2019,
       author = {{Kervella}, Pierre and {Arenou}, Fr{\'e}d{\'e}ric and {Mignard}, Fran{\c{c}}ois and {Th{\'e}venin}, Fr{\'e}d{\'e}ric},
        title = "{Stellar and substellar companions of nearby stars from Gaia DR2. Binarity from proper motion anomaly}",
      journal = {\aap},
         year = 2019,
        month = mar,
       volume = {623},
          eid = {A72},
        pages = {A72},
          doi = {10.1051/0004-6361/201834371},
archivePrefix = {arXiv},
       eprint = {1811.08902},
 primaryClass = {astro-ph.SR},
       adsurl = {https://ui.adsabs.harvard.edu/abs/2019A&A...623A..72K}
}

@ARTICLE{Lagrange2009,
       author = {{Lagrange}, A. -M. and {Gratadour}, D. and {Chauvin}, G. and {Fusco}, T. and {Ehrenreich}, D. and {Mouillet}, D. and {Rousset}, G. and {Rouan}, D. and {Allard}, F. and {Gendron}, {\'E}. and {Charton}, J. and {Mugnier}, L. and {Rabou}, P. and {Montri}, J. and {Lacombe}, F.},
        title = "{A probable giant planet imaged in the {\ensuremath{\beta}} Pictoris disk. VLT/NaCo deep L'-band imaging}",
      journal = {\aap},
         year = 2009,
        month = jan,
       volume = {493},
       number = {2},
        pages = {L21-L25},
          doi = {10.1051/0004-6361:200811325},
archivePrefix = {arXiv},
       eprint = {0811.3583},
 primaryClass = {astro-ph},
       adsurl = {https://ui.adsabs.harvard.edu/abs/2009A&A...493L..21L}
}

@ARTICLE{Malhotra1995,
       author = {{Malhotra}, Renu},
        title = "{The Origin of Pluto's Orbit: Implications for the Solar System Beyond Neptune}",
      journal = {\aj},
         year = 1995,
        month = jul,
       volume = {110},
        pages = {420},
          doi = {10.1086/117532},
archivePrefix = {arXiv},
       eprint = {astro-ph/9504036},
 primaryClass = {astro-ph},
       adsurl = {https://ui.adsabs.harvard.edu/abs/1995AJ....110..420M}
}

@ARTICLE{boccaletti_2022,
       author = {{Boccaletti}, A. and {Cossou}, C. and {Baudoz}, P. and {Lagage}, P.~O. and {Dicken}, D. and {Glasse}, A. and {Hines}, D.~C. and {Aguilar}, J. and {Detre}, O. and {Nickson}, B. and {Noriega-Crespo}, A. and {G{\'a}sp{\'a}r}, A. and {Labiano}, A. and {Stark}, C. and {Rouan}, D. and {Reess}, J.~M. and {Wright}, G.~S. and {Rieke}, G. and {Garcia Marin}, M. and {Lajoie}, C. and {Girard}, J. and {Perrin}, M. and {Soummer}, R. and {Pueyo}, L.},
        title = "{JWST/MIRI coronagraphic performances as measured on-sky}",
      journal = {\aap},
         year = 2022,
        month = nov,
       volume = {667},
          eid = {A165},
        pages = {A165},
          doi = {10.1051/0004-6361/202244578},
archivePrefix = {arXiv},
       eprint = {2207.11080},
 primaryClass = {astro-ph.IM},
       adsurl = {https://ui.adsabs.harvard.edu/abs/2022A&A...667A.165B}
}

@INCOLLECTION{Morbidelli2020,
       author = {{Morbidelli}, Alessandro and {Nesvorn{\'y}}, David},
        title = "{Kuiper belt: formation and evolution}",
    booktitle = {The Trans-Neptunian Solar System},
         year = 2020,
       editor = {{Prialnik}, Dina and {Barucci}, Maria Antoinetta and {Young}, Leslie},
        pages = {25-59},
          doi = {10.1016/B978-0-12-816490-7.00002-3},
       adsurl = {https://ui.adsabs.harvard.edu/abs/2020tnss.book...25M}
}

@ARTICLE{hinkley_ers_2022a,
       author = {{Hinkley}, Sasha and {Carter}, Aarynn L. and {Ray}, Shrishmoy and {Skemer}, Andrew and {Biller}, Beth and {Choquet}, Elodie and {Millar-Blanchaer}, Maxwell A. and {Sallum}, Stephanie and {Miles}, Brittany and {Whiteford}, Niall and {Patapis}, Polychronis and {Perrin}, Marshall and {Pueyo}, Laurent and {Schneider}, Glenn and {Stapelfeldt}, Karl and {Wang}, Jason and {Ward-Duong}, Kimberly and {Bowler}, Brendan P. and {Boccaletti}, Anthony and {Girard}, Julien H. and {Hines}, Dean and {Kalas}, Paul and {Kammerer}, Jens and {Kervella}, Pierre and {Leisenring}, Jarron and {Pantin}, Eric and {Zhou}, Yifan and {Meyer}, Michael and {Liu}, Michael C. and {Bonnefoy}, Mickael and {Currie}, Thayne and {McElwain}, Michael and {Metchev}, Stanimir and {Wyatt}, Mark and {Absil}, Olivier and {Adams}, Jea and {Barman}, Travis and {Baraffe}, Isabelle and {Bonavita}, Mariangela and {Booth}, Mark and {Bryan}, Marta and {Chauvin}, Gael and {Chen}, Christine and {Danielski}, Camilla and {De Furio}, Matthew and {Factor}, Samuel M. and {Fitzgerald}, Michael P. and {Fortney}, Jonathan J. and {Grady}, Carol and {Greenbaum}, Alexandra and {Henning}, Thomas and {Hoch}, Kielan K.~W. and {Janson}, Markus and {Kennedy}, Grant and {Kenworthy}, Matthew and {Kraus}, Adam and {Kuzuhara}, Masayuki and {Lagage}, Pierre-Olivier and {Lagrange}, Anne-Marie and {Launhardt}, Ralf and {Lazzoni}, Cecilia and {Lloyd}, James and {Marino}, Sebastian and {Marley}, Mark and {Martinez}, Raquel and {Marois}, Christian and {Matthews}, Brenda and {Matthews}, Elisabeth C. and {Mawet}, Dimitri and {Mazoyer}, Johan and {Phillips}, Mark and {Petrus}, Simon and {Quanz}, Sascha P. and {Quirrenbach}, Andreas and {Rameau}, Julien and {Rebollido}, Isabel and {Rickman}, Emily and {Samland}, Matthias and {Sargent}, B. and {Schlieder}, Joshua E. and {Sivaramakrishnan}, Anand and {Stone}, Jordan M. and {Tamura}, Motohide and {Tremblin}, Pascal and {Uyama}, Taichi and {Vasist}, Malavika and {Vigan}, Arthur and {Wagner}, Kevin and {Ygouf}, Marie},
        title = "{The JWST Early Release Science Program for the Direct Imaging and Spectroscopy of Exoplanetary Systems}",
      journal = {\pasp},
         year = 2022,
        month = sep,
       volume = {134},
       number = {1039},
          eid = {095003},
        pages = {095003},
          doi = {10.1088/1538-3873/ac77bd},
archivePrefix = {arXiv},
       eprint = {2205.12972},
 primaryClass = {astro-ph.EP},
       adsurl = {https://ui.adsabs.harvard.edu/abs/2022PASP..134i5003H}
}

@misc{jwst_package_2022,
	title = {spacetelescope/jwst: {JWST} 1.6.2},
	url = {https://doi.org/10.5281/zenodo.6984366},
	publisher = {Zenodo},
	author = {Bushouse, Howard and Eisenhamer, Jonathan and Dencheva, Nadia and Davies, James and Greenfield, Perry and Morrison, Jane and Hodge, Phil and Simon, Bernie and Grumm, David and Droettboom, Michael and Slavich, Edward and Sosey, Megan and Pauly, Tyler and Miller, Todd and Jedrzejewski, Robert and Hack, Warren and Davis, David and Crawford, Steven and Law, David and Gordon, Karl and Regan, Michael and Cara, Mihai and MacDonald, Ken and Bradley, Larry and Shanahan, Clare and Jamieson, William},
	month = aug,
	year = {2022},
	doi = {10.5281/zenodo.6984366},
}

@software{wang_pyklip_2015,
       author = {{Wang}, Jason J. and {Ruffio}, Jean-Baptise and {De Rosa}, Robert J. and {Aguilar}, Jonathan and {Wolff}, Schuyler G. and {Pueyo}, Laurent},
        title = "{pyKLIP: PSF Subtraction for Exoplanets and Disks}",
 howpublished = {Astrophysics Source Code Library, record ascl:1506.001},
         year = 2015,
        month = jun,
          eid = {ascl:1506.001},
       adsurl = {https://ui.adsabs.harvard.edu/abs/2015ascl.soft06001W}
}

@article{argyriou_2023,
       author = {{Argyriou}, Ioannis and {Lage}, Craig and {Rieke}, George H. and {Gasman}, Danny and {Bouwman}, Jeroen and {Morrison}, Jane and {Libralato}, Mattia and {Dicken}, Daniel and {Brandl}, Bernhard R. and {{\'A}lvarez-M{\'a}rquez}, Javier and {Labiano}, Alvaro and {Regan}, Michael and {Ressler}, Michael E.},
        title = "{The brighter-fatter effect in the JWST MIRI Si:As IBC detectors. I. Observations, impact on science, and modeling}",
      journal = {\aap},
         year = 2023,
        month = dec,
       volume = {680},
          eid = {A96},
        pages = {A96},
          doi = {10.1051/0004-6361/202346490},
archivePrefix = {arXiv},
       eprint = {2303.13517},
 primaryClass = {astro-ph.IM},
       adsurl = {https://ui.adsabs.harvard.edu/abs/2023A&A...680A..96A}
}

@ARTICLE{lagrange+2025,
       author = {{Lagrange}, A. -M. and {Wilkinson}, C. and {M{\^a}lin}, M. and {Boccaletti}, A. and {Perrot}, C. and {Matr{\`a}}, L. and {Combes}, F. and {Rouan}, D. and {Beust}, H. and {Chomez}, A. and {Charnay}, B. and {Mazevet}, S. and {Flasseur}, O. and {Olofsson}, J. and {Bayo}, A. and {Kral}, Q. and {Chauvin}, G. and {Thebault}, P. and {Rubini}, P. and {Milli}, J. and {Kiefer}, F. and {Carter}, A. and {Crotts}, K. and {Radcliffe}, A. and {Mazoyer}, J. and {Bodrito}, T. and {Stasevic}, S. and {Delorme}, P. and {Langlois}, M.},
        title = "{Evidence for a sub-jovian planet in the young TWA7 disk}",
      journal = {arXiv e-prints},
         year = 2025,
        month = feb,
          eid = {arXiv:2502.15081},
        pages = {arXiv:2502.15081},
          doi = {10.48550/arXiv.2502.15081},
archivePrefix = {arXiv},
       eprint = {2502.15081},
 primaryClass = {astro-ph.EP},
       adsurl = {https://ui.adsabs.harvard.edu/abs/2025arXiv250215081L}
}

@ARTICLE{bowens-rubin_2025,
       author = {{Bowens-Rubin}, Rachel and {Mang}, James and {Limbach}, Mary Anne and {Carter}, Aarynn L. and {Stevenson}, Kevin B. and {Wagner}, Kevin and {Strampelli}, Giovanni and {Morley}, Caroline V. and {Kennedy}, Grant and {Matthews}, Elisabeth and {Vanderburg}, Andrew and {Salama}, Ma{\"\i}ssa},
        title = "{NIRCam yells at cloud: JWST MIRI imaging can directly detect exoplanets of the same temperature, mass, age, and orbital separation as Saturn and Jupiter}",
      journal = {arXiv e-prints},
         year = 2025,
        month = may,
          eid = {arXiv:2505.15995},
        pages = {arXiv:2505.15995},
          doi = {10.48550/arXiv.2505.15995},
archivePrefix = {arXiv},
       eprint = {2505.15995},
 primaryClass = {astro-ph.EP},
       adsurl = {https://ui.adsabs.harvard.edu/abs/2025arXiv250515995B}
}

@ARTICLE{brandt_2021_hr8799,
       author = {{Brandt}, G. Mirek and {Brandt}, Timothy D. and {Dupuy}, Trent J. and {Michalik}, Daniel and {Marleau}, Gabriel-Dominique},
        title = "{The First Dynamical Mass Measurement in the HR 8799 System}",
      journal = {\apjl},
         year = 2021,
        month = jul,
       volume = {915},
       number = {1},
          eid = {L16},
        pages = {L16},
          doi = {10.3847/2041-8213/ac0540},
archivePrefix = {arXiv},
       eprint = {2105.12820},
 primaryClass = {astro-ph.EP},
       adsurl = {https://ui.adsabs.harvard.edu/abs/2021ApJ...915L..16B}
}

@ARTICLE{wyatt_2008,
       author = {{Wyatt}, M.~C.},
        title = "{Evolution of debris disks.}",
      journal = {\araa},
         year = 2008,
        month = sep,
       volume = {46},
        pages = {339-383},
          doi = {10.1146/annurev.astro.45.051806.110525},
       adsurl = {https://ui.adsabs.harvard.edu/abs/2008ARA&A..46..339W}
}

@ARTICLE{gaspar_2023,
       author = {{G{\'a}sp{\'a}r}, Andr{\'a}s and {Wolff}, Schuyler Grace and {Rieke}, George H. and {Leisenring}, Jarron M. and {Morrison}, Jane and {Su}, Kate Y.~L. and {Ward-Duong}, Kimberly and {Aguilar}, Jonathan and {Ygouf}, Marie and {Beichman}, Charles and {Llop-Sayson}, Jorge and {Bryden}, Geoffrey},
        title = "{Spatially resolved imaging of the inner Fomalhaut disk using JWST/MIRI}",
      journal = {Nature Astronomy},
         year = 2023,
        month = jul,
       volume = {7},
        pages = {790-798},
          doi = {10.1038/s41550-023-01962-6},
archivePrefix = {arXiv},
       eprint = {2305.03789},
 primaryClass = {astro-ph.EP},
       adsurl = {https://ui.adsabs.harvard.edu/abs/2023NatAs...7..790G}
}

@ARTICLE{crotts_matthews_2024,
       author = {{Crotts}, Katie A. and {Matthews}, Brenda C.},
        title = "{The Search for Disk Perturbing Planets Around the Asymmetrical Debris Disk System HD 111520 Using REBOUND}",
      journal = {\apj},
         year = 2024,
        month = nov,
       volume = {975},
       number = {1},
          eid = {136},
        pages = {136},
          doi = {10.3847/1538-4357/ad7b28},
archivePrefix = {arXiv},
       eprint = {2410.03932},
 primaryClass = {astro-ph.EP},
       adsurl = {https://ui.adsabs.harvard.edu/abs/2024ApJ...975..136C}
}

@ARTICLE{regaly_2018,
       author = {{Reg{\'a}ly}, Zs. and {Dencs}, Z. and {Mo{\'o}r}, A. and {Kov{\'a}cs}, T.},
        title = "{On the cavity of a debris disc carved by a giant planet}",
      journal = {\mnras},
         year = 2018,
        month = jan,
       volume = {473},
       number = {3},
        pages = {3547-3558},
          doi = {10.1093/mnras/stx2604},
archivePrefix = {arXiv},
       eprint = {1710.01524},
 primaryClass = {astro-ph.EP},
       adsurl = {https://ui.adsabs.harvard.edu/abs/2018MNRAS.473.3547R}
}

@ARTICLE{lazzoni_2018,
       author = {{Lazzoni}, C. and {Desidera}, S. and {Marzari}, F. and {Boccaletti}, A. and {Langlois}, M. and {Mesa}, D. and {Gratton}, R. and {Kral}, Q. and {Pawellek}, N. and {Olofsson}, J. and {Bonnefoy}, M. and {Chauvin}, G. and {Lagrange}, A.~M. and {Vigan}, A. and {Sissa}, E. and {Antichi}, J. and {Avenhaus}, H. and {Baruffolo}, A. and {Baudino}, J.~L. and {Bazzon}, A. and {Beuzit}, J.~L. and {Biller}, B. and {Bonavita}, M. and {Brandner}, W. and {Bruno}, P. and {Buenzli}, E. and {Cantalloube}, F. and {Cascone}, E. and {Cheetham}, A. and {Claudi}, R.~U. and {Cudel}, M. and {Daemgen}, S. and {De Caprio}, V. and {Delorme}, P. and {Fantinel}, D. and {Farisato}, G. and {Feldt}, M. and {Galicher}, R. and {Ginski}, C. and {Girard}, J. and {Giro}, E. and {Janson}, M. and {Hagelberg}, J. and {Henning}, T. and {Incorvaia}, S. and {Kasper}, M. and {Kopytova}, T. and {LeCoroller}, H. and {Lessio}, L. and {Ligi}, R. and {Maire}, A.~L. and {M{\'e}nard}, F. and {Meyer}, M. and {Milli}, J. and {Mouillet}, D. and {Peretti}, S. and {Perrot}, C. and {Rouan}, D. and {Samland}, M. and {Salasnich}, B. and {Salter}, G. and {Schmidt}, T. and {Scuderi}, S. and {Sezestre}, E. and {Turatto}, M. and {Udry}, S. and {Wildi}, F. and {Zurlo}, A.},
        title = "{Dynamical models to explain observations with SPHERE in planetary systems with double debris belts}",
      journal = {\aap},
         year = 2018,
        month = mar,
       volume = {611},
          eid = {A43},
        pages = {A43},
          doi = {10.1051/0004-6361/201731426},
archivePrefix = {arXiv},
       eprint = {1710.03019},
 primaryClass = {astro-ph.EP},
       adsurl = {https://ui.adsabs.harvard.edu/abs/2018A&A...611A..43L}
}

@ARTICLE{mustill_2012,
       author = {{Mustill}, Alexander J. and {Wyatt}, Mark C.},
        title = "{Dependence of a planet's chaotic zone on particle eccentricity: the shape of debris disc inner edges}",
      journal = {\mnras},
         year = 2012,
        month = feb,
       volume = {419},
       number = {4},
        pages = {3074-3080},
          doi = {10.1111/j.1365-2966.2011.19948.x},
archivePrefix = {arXiv},
       eprint = {1110.1282},
 primaryClass = {astro-ph.EP},
       adsurl = {https://ui.adsabs.harvard.edu/abs/2012MNRAS.419.3074M}
}

@ARTICLE{booth_2023,
       author = {{Booth}, Mark and {Pearce}, Tim D. and {Krivov}, Alexander V. and {Wyatt}, Mark C. and {Dent}, William R.~F. and {Hales}, Antonio S. and {Lestrade}, Jean-Fran{\c{c}}ois and {Cruz-S{\'a}enz de Miera}, Fernando and {Faramaz}, Virginie C. and {L{\"o}hne}, Torsten and {Chavez-Dagostino}, Miguel},
        title = "{The clumpy structure of ϵ Eridani's debris disc revisited by ALMA}",
      journal = {\mnras},
         year = 2023,
        month = jun,
       volume = {521},
       number = {4},
        pages = {6180-6194},
          doi = {10.1093/mnras/stad938},
archivePrefix = {arXiv},
       eprint = {2303.13584},
 primaryClass = {astro-ph.EP},
       adsurl = {https://ui.adsabs.harvard.edu/abs/2023MNRAS.521.6180B}
}

@ARTICLE{telesco_2005,
       author = {{Telesco}, Charles M. and {Fisher}, R. Scott and {Wyatt}, Mark C. and {Dermott}, Stanley F. and {Kehoe}, Thomas J.~J. and {Novotny}, Steven and {Mari{\~n}as}, Naibi and {Radomski}, James T. and {Packham}, Christopher and {De Buizer}, James and {Hayward}, Thomas L.},
        title = "{Mid-infrared images of {\ensuremath{\beta}} Pictoris and the possible role of planetesimal collisions in the central disk}",
      journal = {\nat},
         year = 2005,
        month = jan,
       volume = {433},
       number = {7022},
        pages = {133-136},
          doi = {10.1038/nature03255},
       adsurl = {https://ui.adsabs.harvard.edu/abs/2005Natur.433..133T}
}

@ARTICLE{han_2023,
       author = {{Han}, Y. and {Wyatt}, M.~C. and {Dent}, W.~R.~F.},
        title = "{Has the dust clump in the debris disc of Beta Pictoris moved?}",
      journal = {\mnras},
         year = 2023,
        month = mar,
       volume = {519},
       number = {3},
        pages = {3257-3270},
          doi = {10.1093/mnras/stac3769},
archivePrefix = {arXiv},
       eprint = {2301.06891},
 primaryClass = {astro-ph.EP},
       adsurl = {https://ui.adsabs.harvard.edu/abs/2023MNRAS.519.3257H}
}

@ARTICLE{faramaz_2019,
       author = {{Faramaz}, Virginie and {Krist}, John and {Stapelfeldt}, Karl R. and {Bryden}, Geoffrey and {Mamajek}, Eric E. and {Matr{\`a}}, Luca and {Booth}, Mark and {Flaherty}, Kevin and {Hales}, Antonio S. and {Hughes}, A. Meredith and {Bayo}, Amelia and {Casassus}, Simon and {Cuadra}, Jorge and {Olofsson}, Johan and {Su}, Kate Y.~L. and {Wilner}, David J.},
        title = "{From Scattered-light to Millimeter Emission: A Comprehensive View of the Gigayear-old System of HD 202628 and its Eccentric Debris Ring}",
      journal = {\aj},
         year = 2019,
        month = oct,
       volume = {158},
       number = {4},
          eid = {162},
        pages = {162},
          doi = {10.3847/1538-3881/ab3ec1},
archivePrefix = {arXiv},
       eprint = {1909.04162},
 primaryClass = {astro-ph.EP},
       adsurl = {https://ui.adsabs.harvard.edu/abs/2019AJ....158..162F}
}

@ARTICLE{stasevic_2023,
       author = {{Stasevic}, S. and {Milli}, J. and {Mazoyer}, J. and {Lagrange}, A. -M. and {Bonnefoy}, M. and {Faramaz-Gorka}, V. and {M{\'e}nard}, F. and {Boccaletti}, A. and {Choquet}, E. and {Shuai}, L. and {Olofsson}, J. and {Chomez}, A. and {Ren}, B. and {Rubini}, P. and {Desgrange}, C. and {Gratton}, R. and {Chauvin}, G. and {Vigan}, A. and {Matthews}, E.},
        title = "{An inner warp discovered in the disk around HD 110058 using VLT/SPHERE and HST/STIS}",
      journal = {\aap},
         year = 2023,
        month = oct,
       volume = {678},
          eid = {A8},
        pages = {A8},
          doi = {10.1051/0004-6361/202346720},
archivePrefix = {arXiv},
       eprint = {2308.05613},
 primaryClass = {astro-ph.EP},
       adsurl = {https://ui.adsabs.harvard.edu/abs/2023A&A...678A...8S}
}

@ARTICLE{augereau_2001,
       author = {{Augereau}, J.~C. and {Nelson}, R.~P. and {Lagrange}, A.~M. and {Papaloizou}, J.~C.~B. and {Mouillet}, D.},
        title = "{Dynamical modeling of large scale asymmetries in the beta Pictoris dust disk}",
      journal = {\aap},
         year = 2001,
        month = may,
       volume = {370},
        pages = {447-455},
          doi = {10.1051/0004-6361:20010199},
archivePrefix = {arXiv},
       eprint = {astro-ph/0102069},
 primaryClass = {astro-ph},
       adsurl = {https://ui.adsabs.harvard.edu/abs/2001A&A...370..447A}
}

@ARTICLE{mouillet_1997,
       author = {{Mouillet}, D. and {Larwood}, J.~D. and {Papaloizou}, J.~C.~B. and {Lagrange}, A.~M.},
        title = "{A planet on an inclined orbit as an explanation of the warp in the beta Pictoris disc}",
      journal = {\mnras},
         year = 1997,
        month = dec,
       volume = {292},
       number = {4},
        pages = {896-904},
          doi = {10.1093/mnras/292.4.896},
archivePrefix = {arXiv},
       eprint = {astro-ph/9705100},
 primaryClass = {astro-ph},
       adsurl = {https://ui.adsabs.harvard.edu/abs/1997MNRAS.292..896M}
}

@ARTICLE{ren_2021,
       author = {{Ren}, Bin and {Choquet}, {\'E}lodie and {Perrin}, Marshall D. and {Mawet}, Dimitri and {Chen}, Christine H. and {Milli}, Julien and {Debes}, John H. and {Rebollido}, Isabel and {Stark}, Christopher C. and {Hagan}, J. Brendan and {Hines}, Dean C. and {Millar-Blanchaer}, Maxwell A. and {Pueyo}, Laurent and {Roberge}, Aki and {Schneider}, Glenn and {Serabyn}, Eugene and {Soummer}, R{\'e}mi and {Wolff}, Schuyler G.},
        title = "{A Layered Debris Disk around M Star TWA 7 in Scattered Light}",
      journal = {\apj},
         year = 2021,
        month = jun,
       volume = {914},
       number = {2},
          eid = {95},
        pages = {95},
          doi = {10.3847/1538-4357/ac03b9},
archivePrefix = {arXiv},
       eprint = {2105.09949},
 primaryClass = {astro-ph.EP},
       adsurl = {https://ui.adsabs.harvard.edu/abs/2021ApJ...914...95R}
}

@ARTICLE{milli_2017,
       author = {{Milli}, J. and {Vigan}, A. and {Mouillet}, D. and {Lagrange}, A. -M. and {Augereau}, J. -C. and {Pinte}, C. and {Mawet}, D. and {Schmid}, H.~M. and {Boccaletti}, A. and {Matr{\`a}}, L. and {Kral}, Q. and {Ertel}, S. and {Chauvin}, G. and {Bazzon}, A. and {M{\'e}nard}, F. and {Beuzit}, J. -L. and {Thalmann}, C. and {Dominik}, C. and {Feldt}, M. and {Henning}, T. and {Min}, M. and {Girard}, J.~H. and {Galicher}, R. and {Bonnefoy}, M. and {Fusco}, T. and {de Boer}, J. and {Janson}, M. and {Maire}, A. -L. and {Mesa}, D. and {Schlieder}, J.~E. and {SPHERE Consortium}},
        title = "{Near-infrared scattered light properties of the HR 4796 A dust ring. A measured scattering phase function from 13.6{\textdegree} to 166.6{\textdegree}}",
      journal = {\aap},
         year = 2017,
        month = mar,
       volume = {599},
          eid = {A108},
        pages = {A108},
          doi = {10.1051/0004-6361/201527838},
archivePrefix = {arXiv},
       eprint = {1701.00750},
 primaryClass = {astro-ph.EP},
       adsurl = {https://ui.adsabs.harvard.edu/abs/2017A&A...599A.108M}
}

@ARTICLE{crotts_2024,
       author = {{Crotts}, Katie A. and {Matthews}, Brenda C. and {Duch{\^e}ne}, Gaspard and {Esposito}, Thomas M. and {Dong}, Ruobing and {Hom}, Justin and {Oppenheimer}, Rebecca and {Rice}, Malena and {Wolff}, Schuyler G. and {Chen}, Christine H. and {Do {\'O}}, Clarissa R. and {Kalas}, Paul and {Lewis}, Briley L. and {Weinberger}, Alycia J. and {Wilner}, David J. and {Ammons}, Mark and {Arriaga}, Pauline and {De Rosa}, Robert J. and {Debes}, John H. and {Fitzgerald}, Michael P. and {Gonzales}, Eileen C. and {Hines}, Dean C. and {Hinkley}, Sasha and {Hughes}, A. Meredith and {Kolokolova}, Ludmilla and {Lee}, Eve J. and {L{\'o}pez}, Ronald A. and {Macintosh}, Bruce and {Mazoyer}, Johan and {Metchev}, Stanimir and {Millar-Blanchaer}, Maxwell A. and {Nielsen}, Eric L. and {Patience}, Jenny and {Perrin}, Marshall D. and {Pueyo}, Laurent and {Rantakyr{\"o}}, Fredrik T. and {Ren}, Bin B. and {Schneider}, Glenn and {Soummer}, Remi and {Stark}, Christopher C.},
        title = "{A Uniform Analysis of Debris Disks with the Gemini Planet Imager. I. An Empirical Search for Perturbations from Planetary Companions in Polarized Light Images}",
      journal = {\apj},
         year = 2024,
        month = feb,
       volume = {961},
       number = {2},
          eid = {245},
        pages = {245},
          doi = {10.3847/1538-4357/ad0e69},
archivePrefix = {arXiv},
       eprint = {2311.14599},
 primaryClass = {astro-ph.EP},
       adsurl = {https://ui.adsabs.harvard.edu/abs/2024ApJ...961..245C}
}

@ARTICLE{lawson_2024,
       author = {{Lawson}, Kellen and {Schlieder}, Joshua E. and {Leisenring}, Jarron M. and {Bogat}, Ell and {Beichman}, Charles A. and {Bryden}, Geoffrey and {G{\'a}sp{\'a}r}, Andr{\'a}s and {Groff}, Tyler D. and {McElwain}, Michael W. and {Meyer}, Michael R. and {Barclay}, Thomas and {Calissendorff}, Per and {De Furio}, Matthew and {Li}, Yiting and {Rieke}, Marcia J. and {Ygouf}, Marie and {Greene}, Thomas P. and {Girard}, Julien H. and {Gennaro}, Mario and {Kammerer}, Jens and {Rest}, Armin and {Roellig}, Thomas L. and {Sunnquist}, Ben},
        title = "{JWST/NIRCam Detection of the Fomalhaut C Debris Disk in Scattered Light}",
      journal = {\apjl},
         year = 2024,
        month = may,
       volume = {967},
       number = {1},
          eid = {L8},
        pages = {L8},
          doi = {10.3847/2041-8213/ad4496},
archivePrefix = {arXiv},
       eprint = {2405.00573},
 primaryClass = {astro-ph.EP},
       adsurl = {https://ui.adsabs.harvard.edu/abs/2024ApJ...967L...8L}
}

@ARTICLE{su_2024,
       author = {{Su}, Kate Y.~L. and {G{\'a}sp{\'a}r}, Andr{\'a}s and {Rieke}, George H. and {Malhotra}, Renu and {Matr{\'a}}, Luca and {Wolff}, Schuyler Grace and {Leisenring}, Jarron M. and {Beichman}, Charles and {Ygouf}, Marie},
        title = "{Imaging of the Vega Debris System Using JWST/MIRI}",
      journal = {\apj},
         year = 2024,
        month = dec,
       volume = {977},
       number = {2},
          eid = {277},
        pages = {277},
          doi = {10.3847/1538-4357/ad8cde},
archivePrefix = {arXiv},
       eprint = {2410.23636},
 primaryClass = {astro-ph.EP},
       adsurl = {https://ui.adsabs.harvard.edu/abs/2024ApJ...977..277S}
}

@ARTICLE{esposito_2020,
       author = {{Esposito}, Thomas M. and {Kalas}, Paul and {Fitzgerald}, Michael P. and {Millar-Blanchaer}, Maxwell A. and {Duch{\^e}ne}, Gaspard and {Patience}, Jennifer and {Hom}, Justin and {Perrin}, Marshall D. and {De Rosa}, Robert J. and {Chiang}, Eugene and {Czekala}, Ian and {Macintosh}, Bruce and {Graham}, James R. and {Ansdell}, Megan and {Arriaga}, Pauline and {Bruzzone}, Sebastian and {Bulger}, Joanna and {Chen}, Christine H. and {Cotten}, Tara and {Dong}, Ruobing and {Draper}, Zachary H. and {Follette}, Katherine B. and {Hung}, Li-Wei and {Lopez}, Ronald and {Matthews}, Brenda C. and {Mazoyer}, Johan and {Metchev}, Stan and {Rameau}, Julien and {Ren}, Bin and {Rice}, Malena and {Song}, Inseok and {Stahl}, Kevin and {Wang}, Jason and {Wolff}, Schuyler and {Zuckerman}, Ben and {Ammons}, S. Mark and {Bailey}, Vanessa P. and {Barman}, Travis and {Chilcote}, Jeffrey and {Doyon}, Rene and {Gerard}, Benjamin L. and {Goodsell}, Stephen J. and {Greenbaum}, Alexandra Z. and {Hibon}, Pascale and {Hinkley}, Sasha and {Ingraham}, Patrick and {Konopacky}, Quinn and {Maire}, J{\'e}r{\^o}me and {Marchis}, Franck and {Marley}, Mark S. and {Marois}, Christian and {Nielsen}, Eric L. and {Oppenheimer}, Rebecca and {Palmer}, David and {Poyneer}, Lisa and {Pueyo}, Laurent and {Rajan}, Abhijith and {Rantakyr{\"o}}, Fredrik T. and {Ruffio}, Jean-Baptiste and {Savransky}, Dmitry and {Schneider}, Adam C. and {Sivaramakrishnan}, Anand and {Soummer}, R{\'e}mi and {Thomas}, Sandrine and {Ward-Duong}, Kimberly},
        title = "{Debris Disk Results from the Gemini Planet Imager Exoplanet Survey's Polarimetric Imaging Campaign}",
      journal = {\aj},
         year = 2020,
        month = jul,
       volume = {160},
       number = {1},
          eid = {24},
        pages = {24},
          doi = {10.3847/1538-3881/ab9199},
archivePrefix = {arXiv},
       eprint = {2004.13722},
 primaryClass = {astro-ph.EP},
       adsurl = {https://ui.adsabs.harvard.edu/abs/2020AJ....160...24E}
}

@ARTICLE{matra_2025,
       author = {{Matr{\`a}}, L. and {Marino}, S. and {Wilner}, D.~J. and {Kennedy}, G.~M. and {Booth}, M. and {Krivov}, A.~V. and {Williams}, J.~P. and {Hughes}, A.~M. and {del Burgo}, C. and {Carpenter}, J. and {Davies}, C.~L. and {Ertel}, S. and {Kral}, Q. and {Lestrade}, J. -F. and {Marshall}, J.~P. and {Milli}, J. and {{\"O}berg}, K.~I. and {Pawellek}, N. and {Sepulveda}, A.~G. and {Wyatt}, M.~C. and {Matthews}, B.~C. and {MacGregor}, M.},
        title = "{REsolved ALMA and SMA Observations of Nearby Stars (REASONS): A population of 74 resolved planetesimal belts at millimetre wavelengths}",
      journal = {\aap},
         year = 2025,
        month = jan,
       volume = {693},
          eid = {A151},
        pages = {A151},
          doi = {10.1051/0004-6361/202451397},
archivePrefix = {arXiv},
       eprint = {2501.09058},
 primaryClass = {astro-ph.EP},
       adsurl = {https://ui.adsabs.harvard.edu/abs/2025A&A...693A.151M}
}

@ARTICLE{sibthorpe_2018,
       author = {{Sibthorpe}, B. and {Kennedy}, G.~M. and {Wyatt}, M.~C. and {Lestrade}, J. -F. and {Greaves}, J.~S. and {Matthews}, B.~C. and {Duch{\^e}ne}, G.},
        title = "{Analysis of the Herschel DEBRIS Sun-like star sample}",
      journal = {\mnras},
         year = 2018,
        month = apr,
       volume = {475},
       number = {3},
        pages = {3046-3064},
          doi = {10.1093/mnras/stx3188},
archivePrefix = {arXiv},
       eprint = {1803.00072},
 primaryClass = {astro-ph.EP},
       adsurl = {https://ui.adsabs.harvard.edu/abs/2018MNRAS.475.3046S}
}

@ARTICLE{beuzit_2019,
       author = {{Beuzit}, J. -L. and {Vigan}, A. and {Mouillet}, D. and {Dohlen}, K. and {Gratton}, R. and {Boccaletti}, A. and {Sauvage}, J. -F. and {Schmid}, H.~M. and {Langlois}, M. and {Petit}, C. and {Baruffolo}, A. and {Feldt}, M. and {Milli}, J. and {Wahhaj}, Z. and {Abe}, L. and {Anselmi}, U. and {Antichi}, J. and {Barette}, R. and {Baudrand}, J. and {Baudoz}, P. and {Bazzon}, A. and {Bernardi}, P. and {Blanchard}, P. and {Brast}, R. and {Bruno}, P. and {Buey}, T. and {Carbillet}, M. and {Carle}, M. and {Cascone}, E. and {Chapron}, F. and {Charton}, J. and {Chauvin}, G. and {Claudi}, R. and {Costille}, A. and {De Caprio}, V. and {de Boer}, J. and {Delboulb{\'e}}, A. and {Desidera}, S. and {Dominik}, C. and {Downing}, M. and {Dupuis}, O. and {Fabron}, C. and {Fantinel}, D. and {Farisato}, G. and {Feautrier}, P. and {Fedrigo}, E. and {Fusco}, T. and {Gigan}, P. and {Ginski}, C. and {Girard}, J. and {Giro}, E. and {Gisler}, D. and {Gluck}, L. and {Gry}, C. and {Henning}, T. and {Hubin}, N. and {Hugot}, E. and {Incorvaia}, S. and {Jaquet}, M. and {Kasper}, M. and {Lagadec}, E. and {Lagrange}, A. -M. and {Le Coroller}, H. and {Le Mignant}, D. and {Le Ruyet}, B. and {Lessio}, G. and {Lizon}, J. -L. and {Llored}, M. and {Lundin}, L. and {Madec}, F. and {Magnard}, Y. and {Marteaud}, M. and {Martinez}, P. and {Maurel}, D. and {M{\'e}nard}, F. and {Mesa}, D. and {M{\"o}ller-Nilsson}, O. and {Moulin}, T. and {Moutou}, C. and {Orign{\'e}}, A. and {Parisot}, J. and {Pavlov}, A. and {Perret}, D. and {Pragt}, J. and {Puget}, P. and {Rabou}, P. and {Ramos}, J. and {Reess}, J. -M. and {Rigal}, F. and {Rochat}, S. and {Roelfsema}, R. and {Rousset}, G. and {Roux}, A. and {Saisse}, M. and {Salasnich}, B. and {Santambrogio}, E. and {Scuderi}, S. and {Segransan}, D. and {Sevin}, A. and {Siebenmorgen}, R. and {Soenke}, C. and {Stadler}, E. and {Suarez}, M. and {Tiph{\`e}ne}, D. and {Turatto}, M. and {Udry}, S. and {Vakili}, F. and {Waters}, L.~B.~F.~M. and {Weber}, L. and {Wildi}, F. and {Zins}, G. and {Zurlo}, A.},
        title = "{SPHERE: the exoplanet imager for the Very Large Telescope}",
      journal = {\aap},
         year = 2019,
        month = nov,
       volume = {631},
          eid = {A155},
        pages = {A155},
          doi = {10.1051/0004-6361/201935251},
archivePrefix = {arXiv},
       eprint = {1902.04080},
 primaryClass = {astro-ph.IM},
       adsurl = {https://ui.adsabs.harvard.edu/abs/2019A&A...631A.155B}
}

@ARTICLE{macintosh_2014,
       author = {{Macintosh}, Bruce and {Graham}, James R. and {Ingraham}, Patrick and {Konopacky}, Quinn and {Marois}, Christian and {Perrin}, Marshall and {Poyneer}, Lisa and {Bauman}, Brian and {Barman}, Travis and {Burrows}, Adam S. and {Cardwell}, Andrew and {Chilcote}, Jeffrey and {De Rosa}, Robert J. and {Dillon}, Daren and {Doyon}, Rene and {Dunn}, Jennifer and {Erikson}, Darren and {Fitzgerald}, Michael P. and {Gavel}, Donald and {Goodsell}, Stephen and {Hartung}, Markus and {Hibon}, Pascale and {Kalas}, Paul and {Larkin}, James and {Maire}, Jerome and {Marchis}, Franck and {Marley}, Mark S. and {McBride}, James and {Millar-Blanchaer}, Max and {Morzinski}, Katie and {Norton}, Andrew and {Oppenheimer}, B.~R. and {Palmer}, David and {Patience}, Jennifer and {Pueyo}, Laurent and {Rantakyro}, Fredrik and {Sadakuni}, Naru and {Saddlemyer}, Leslie and {Savransky}, Dmitry and {Serio}, Andrew and {Soummer}, Remi and {Sivaramakrishnan}, Anand and {Song}, Inseok and {Thomas}, Sandrine and {Wallace}, J. Kent and {Wiktorowicz}, Sloane and {Wolff}, Schuyler},
        title = "{First light of the Gemini Planet Imager}",
      journal = {Proceedings of the National Academy of Science},
         year = 2014,
        month = sep,
       volume = {111},
       number = {35},
        pages = {12661-12666},
          doi = {10.1073/pnas.1304215111},
archivePrefix = {arXiv},
       eprint = {1403.7520},
 primaryClass = {astro-ph.EP},
       adsurl = {https://ui.adsabs.harvard.edu/abs/2014PNAS..11112661M}
}

@ARTICLE{morbidelli_2005,
       author = {{Morbidelli}, A. and {Levison}, H.~F. and {Tsiganis}, K. and {Gomes}, R.},
        title = "{Chaotic capture of Jupiter's Trojan asteroids in the early Solar System}",
      journal = {\nat},
         year = 2005,
        month = may,
       volume = {435},
       number = {7041},
        pages = {462-465},
          doi = {10.1038/nature03540},
       adsurl = {https://ui.adsabs.harvard.edu/abs/2005Natur.435..462M}
}

@ARTICLE{marino_2022,
       author = {{Marino}, Sebastian},
        title = "{Planetesimal/Debris discs}",
      journal = {arXiv e-prints},
         year = 2022,
        month = feb,
          eid = {arXiv:2202.03053},
        pages = {arXiv:2202.03053},
          doi = {10.48550/arXiv.2202.03053},
archivePrefix = {arXiv},
       eprint = {2202.03053},
 primaryClass = {astro-ph.EP},
       adsurl = {https://ui.adsabs.harvard.edu/abs/2022arXiv220203053M}
}

@ARTICLE{brandt_2021_betapic,
       author = {{Brandt}, G. Mirek and {Brandt}, Timothy D. and {Dupuy}, Trent J. and {Li}, Yiting and {Michalik}, Daniel},
        title = "{Precise Dynamical Masses and Orbital Fits for {\ensuremath{\beta}} Pic b and {\ensuremath{\beta}} Pic c}",
      journal = {\aj},
         year = 2021,
        month = apr,
       volume = {161},
       number = {4},
          eid = {179},
        pages = {179},
          doi = {10.3847/1538-3881/abdc2e},
archivePrefix = {arXiv},
       eprint = {2011.06215},
 primaryClass = {astro-ph.EP},
       adsurl = {https://ui.adsabs.harvard.edu/abs/2021AJ....161..179B}
}

@ARTICLE{Kiefer+2024,
       author = {{Kiefer}, Flavien and {Lagrange}, Anne-Marie and {Rubini}, Pascal and {Philipot}, Florian},
        title = "{Searching for substellar companion candidates with Gaia. I. Introducing the GaiaPMEX tool}",
      journal = {arXiv e-prints},
         year = 2024,
        month = sep,
          eid = {arXiv:2409.16992},
        pages = {arXiv:2409.16992},
          doi = {10.48550/arXiv.2409.16992},
archivePrefix = {arXiv},
       eprint = {2409.16992},
 primaryClass = {astro-ph.EP},
       adsurl = {https://ui.adsabs.harvard.edu/abs/2024arXiv240916992K}
}

@ARTICLE{limbach2024,
       author = {{Limbach}, Mary Anne and {Vanderburg}, Andrew and {Venner}, Alexander and {Blouin}, Simon and {Stevenson}, Kevin B. and {MacDonald}, Ryan J. and {Jenkins}, Sydney and {Bowens-Rubin}, Rachel and {Soares-Furtado}, Melinda and {Morley}, Caroline and {Janson}, Markus and {Debes}, John and {Xu}, Siyi and {Kleisioti}, Evangelia and {Kenworthy}, Matthew and {Butler}, Paul and {Crane}, Jeffrey D. and {Osip}, Dave and {Shectman}, Stephen and {Teske}, Johanna},
        title = "{The MIRI Exoplanets Orbiting White dwarfs (MEOW) Survey: Mid-infrared Excess Reveals a Giant Planet Candidate around a Nearby White Dwarf}",
      journal = {\apjl},
         year = 2024,
        month = sep,
       volume = {973},
       number = {1},
          eid = {L11},
        pages = {L11},
          doi = {10.3847/2041-8213/ad74ed},
archivePrefix = {arXiv},
       eprint = {2408.16813},
 primaryClass = {astro-ph.EP},
       adsurl = {https://ui.adsabs.harvard.edu/abs/2024ApJ...973L..11L}
}

@ARTICLE{lindegren_2021,
       author = {{Lindegren}, L. and {Klioner}, S.~A. and {Hern{\'a}ndez}, J. and {Bombrun}, A. and {Ramos-Lerate}, M. and {Steidelm{\"u}ller}, H. and {Bastian}, U. and {Biermann}, M. and {de Torres}, A. and {Gerlach}, E. and {Geyer}, R. and {Hilger}, T. and {Hobbs}, D. and {Lammers}, U. and {McMillan}, P.~J. and {Stephenson}, C.~A. and {Casta{\~n}eda}, J. and {Davidson}, M. and {Fabricius}, C. and {Gracia-Abril}, G. and {Portell}, J. and {Rowell}, N. and {Teyssier}, D. and {Torra}, F. and {Bartolom{\'e}}, S. and {Clotet}, M. and {Garralda}, N. and {Gonz{\'a}lez-Vidal}, J.~J. and {Torra}, J. and {Abbas}, U. and {Altmann}, M. and {Anglada Varela}, E. and {Balaguer-N{\'u}{\~n}ez}, L. and {Balog}, Z. and {Barache}, C. and {Becciani}, U. and {Bernet}, M. and {Bertone}, S. and {Bianchi}, L. and {Bouquillon}, S. and {Brown}, A.~G.~A. and {Bucciarelli}, B. and {Busonero}, D. and {Butkevich}, A.~G. and {Buzzi}, R. and {Cancelliere}, R. and {Carlucci}, T. and {Charlot}, P. and {Cioni}, M. -R.~L. and {Crosta}, M. and {Crowley}, C. and {del Peloso}, E.~F. and {del Pozo}, E. and {Drimmel}, R. and {Esquej}, P. and {Fienga}, A. and {Fraile}, E. and {Gai}, M. and {Garcia-Reinaldos}, M. and {Guerra}, R. and {Hambly}, N.~C. and {Hauser}, M. and {Jan{\ss}en}, K. and {Jordan}, S. and {Kostrzewa-Rutkowska}, Z. and {Lattanzi}, M.~G. and {Liao}, S. and {Licata}, E. and {Lister}, T.~A. and {L{\"o}ffler}, W. and {Marchant}, J.~M. and {Masip}, A. and {Mignard}, F. and {Mints}, A. and {Molina}, D. and {Mora}, A. and {Morbidelli}, R. and {Murphy}, C.~P. and {Pagani}, C. and {Panuzzo}, P. and {Pe{\~n}alosa Esteller}, X. and {Poggio}, E. and {Re Fiorentin}, P. and {Riva}, A. and {Sagrist{\`a} Sell{\'e}s}, A. and {Sanchez Gimenez}, V. and {Sarasso}, M. and {Sciacca}, E. and {Siddiqui}, H.~I. and {Smart}, R.~L. and {Souami}, D. and {Spagna}, A. and {Steele}, I.~A. and {Taris}, F. and {Utrilla}, E. and {van Reeven}, W. and {Vecchiato}, A.},
        title = "{Gaia Early Data Release 3. The astrometric solution}",
      journal = {\aap},
         year = 2021,
        month = may,
       volume = {649},
          eid = {A2},
        pages = {A2},
          doi = {10.1051/0004-6361/202039709},
archivePrefix = {arXiv},
       eprint = {2012.03380},
 primaryClass = {astro-ph.IM},
       adsurl = {https://ui.adsabs.harvard.edu/abs/2021A&A...649A...2L}
}

@ARTICLE{Andrew+2009,
       author = {{Smith}, Andrew W. and {Lissauer}, Jack J.},
        title = "{Orbital stability of systems of closely-spaced planets}",
      journal = {\icarus},
         year = 2009,
        month = may,
       volume = {201},
       number = {1},
        pages = {381-394},
          doi = {10.1016/j.icarus.2008.12.027},
       adsurl = {https://ui.adsabs.harvard.edu/abs/2009Icar..201..381S}
}

@ARTICLE{matthews+2024,
       author = {{Matthews}, E.~C. and {Carter}, A.~L. and {Pathak}, P. and {Morley}, C.~V. and {Phillips}, M.~W. and {P.~M.}, S. Krishanth and {Feng}, F. and {Bonse}, M.~J. and {Boogaard}, L.~A. and {Burt}, J.~A. and {Crossfield}, I.~J.~M. and {Douglas}, E.~S. and {Henning}, Th. and {Hom}, J. and {Ko}, C. -L. and {Kasper}, M. and {Lagrange}, A. -M. and {Petit dit de la Roche}, D. and {Philipot}, F.},
        title = "{A temperate super-Jupiter imaged with JWST in the mid-infrared}",
      journal = {\nat},
         year = 2024,
        month = sep,
       volume = {633},
       number = {8031},
        pages = {789-792},
          doi = {10.1038/s41586-024-07837-8},
archivePrefix = {arXiv},
       eprint = {2503.01599},
 primaryClass = {astro-ph.EP},
       adsurl = {https://ui.adsabs.harvard.edu/abs/2024Natur.633..789M}
}

@article{marino_2018,
	title = {A gap in the planetesimal disc around {HD} 107146 and asymmetric warm dust emission revealed by {ALMA}},
	volume = {479},
	issn = {0035-8711},
	url = {https://ui.adsabs.harvard.edu/abs/2018MNRAS.479.5423M},
	doi = {10.1093/mnras/sty1790},
	urldate = {2023-11-24},
	journal = {Monthly Notices of the Royal Astronomical Society},
	author = {Marino, S. and Carpenter, J. and Wyatt, M. C. and Booth, M. and Casassus, S. and Faramaz, V. and Guzman, V. and Hughes, A. M. and Isella, A. and Kennedy, G. M. and Matrà, L. and Ricci, L. and Corder, S.},
	month = oct,
	year = {2018},
	note = {ADS Bibcode: 2018MNRAS.479.5423M},
	pages = {5423--5439},
}

@article{marino_2019,
	title = {A gap in {HD} 92945's broad planetesimal disc revealed by {ALMA}},
	volume = {484},
	issn = {0035-8711},
	url = {https://ui.adsabs.harvard.edu/abs/2019MNRAS.484.1257M},
	doi = {10.1093/mnras/stz049},
	urldate = {2023-11-22},
	journal = {Monthly Notices of the Royal Astronomical Society},
	author = {Marino, S. and Yelverton, B. and Booth, M. and Faramaz, V. and Kennedy, G. M. and Matrà, L. and Wyatt, M. C.},
	month = mar,
	year = {2019},
	note = {ADS Bibcode: 2019MNRAS.484.1257M},
	pages = {1257--1269},
}

@article{marino_2020,
	title = {Insights into the planetary dynamics of {HD} 206893 with {ALMA}},
	volume = {498},
	issn = {0035-8711},
	url = {https://ui.adsabs.harvard.edu/abs/2020MNRAS.498.1319M},
	doi = {10.1093/mnras/staa2386},
	urldate = {2023-11-24},
	journal = {Monthly Notices of the Royal Astronomical Society},
	author = {Marino, S. and Zurlo, A. and Faramaz, V. and Milli, J. and Henning, Th and Kennedy, G. M. and Matrà, L. and Pérez, S. and Delorme, P. and Cieza, L. A. and Hughes, A. M.},
	month = oct,
	year = {2020},
	note = {ADS Bibcode: 2020MNRAS.498.1319M},
	pages = {1319--1334},
}

@article{macgregor_2019,
	title = {Multiple {Rings} of {Millimeter} {Dust} {Emission} in the {HD} 15115 {Debris} {Disk}},
	volume = {877},
	issn = {0004-637X},
	url = {https://ui.adsabs.harvard.edu/abs/2019ApJ...877L..32M},
	doi = {10.3847/2041-8213/ab21c2},
	urldate = {2024-02-07},
	journal = {The Astrophysical Journal},
	author = {MacGregor, Meredith A. and Weinberger, Alycia J. and Nesvold, Erika R. and Hughes, A. Meredith and Wilner, D. J. and Currie, Thayne and Debes, John H. and Donaldson, Jessica K. and Redfield, Seth and Roberge, Aki and Schneider, Glenn},
	month = jun,
	year = {2019},
	note = {ADS Bibcode: 2019ApJ...877L..32M},
	pages = {L32},
}

@ARTICLE{rafikov_2023,
       author = {{Rafikov}, Roman R.},
        title = "{Radial profiles of surface density in debris discs}",
      journal = {\mnras},
         year = 2023,
        month = mar,
       volume = {519},
       number = {4},
        pages = {5607-5622},
          doi = {10.1093/mnras/stac3411},
archivePrefix = {arXiv},
       eprint = {2207.07678},
 primaryClass = {astro-ph.EP},
       adsurl = {https://ui.adsabs.harvard.edu/abs/2023MNRAS.519.5607R}
}

@ARTICLE{rodet_2022,
       author = {{Rodet}, Laetitia and {Lai}, Dong},
        title = "{Eccentric debris belts reveal the dynamical history of the companion exoplanet}",
      journal = {\mnras},
         year = 2022,
        month = nov,
       volume = {516},
       number = {4},
        pages = {5544-5554},
          doi = {10.1093/mnras/stac2621},
archivePrefix = {arXiv},
       eprint = {2208.05041},
 primaryClass = {astro-ph.EP},
       adsurl = {https://ui.adsabs.harvard.edu/abs/2022MNRAS.516.5544R}
}

@ARTICLE{kennedy_2020,
       author = {{Kennedy}, Grant M.},
        title = "{The unexpected narrowness of eccentric debris rings: a sign of eccentricity during the protoplanetary disc phase}",
      journal = {Royal Society Open Science},
         year = 2020,
        month = jun,
       volume = {7},
       number = {6},
          eid = {200063},
        pages = {200063},
          doi = {10.1098/rsos.200063},
archivePrefix = {arXiv},
       eprint = {2005.14200},
 primaryClass = {astro-ph.EP},
       adsurl = {https://ui.adsabs.harvard.edu/abs/2020RSOS....700063K}
}

@ARTICLE{Pearce_2014,
       author = {{Pearce}, Tim D. and {Wyatt}, Mark C.},
        title = "{Dynamical evolution of an eccentric planet and a less massive debris disc}",
      journal = {\mnras},
         year = 2014,
        month = sep,
       volume = {443},
       number = {3},
        pages = {2541-2560},
          doi = {10.1093/mnras/stu1302},
archivePrefix = {arXiv},
       eprint = {1406.7294},
 primaryClass = {astro-ph.EP},
       adsurl = {https://ui.adsabs.harvard.edu/abs/2014MNRAS.443.2541P}
}

@ARTICLE{wyatt_1999,
       author = {{Wyatt}, M.~C. and {Dermott}, S.~F. and {Telesco}, C.~M. and {Fisher}, R.~S. and {Grogan}, K. and {Holmes}, E.~K. and {Pi{\~n}a}, R.~K.},
        title = "{How Observations of Circumstellar Disk Asymmetries Can Reveal Hidden Planets: Pericenter Glow and Its Application to the HR 4796 Disk}",
      journal = {\apj},
         year = 1999,
        month = dec,
       volume = {527},
       number = {2},
        pages = {918-944},
          doi = {10.1086/308093},
archivePrefix = {arXiv},
       eprint = {astro-ph/9908267},
 primaryClass = {astro-ph},
       adsurl = {https://ui.adsabs.harvard.edu/abs/1999ApJ...527..918W}
}

@ARTICLE{squicciarini_2025,
       author = {{Squicciarini}, V. and {Mazoyer}, J. and {Lagrange}, A. -M. and {Chomez}, A. and {Delorme}, P. and {Flasseur}, O. and {Kiefer}, F. and {Bergeon}, S. and {Albert}, D. and {Meunier}, N.},
        title = "{The COBREX archival survey: Improved constraints on the occurrence rate of wide-orbit substellar companions: I. A uniform re-analysis of 400 stars from the GPIES survey}",
      journal = {\aap},
         year = 2025,
        month = jan,
       volume = {693},
          eid = {A54},
        pages = {A54},
          doi = {10.1051/0004-6361/202452310},
archivePrefix = {arXiv},
       eprint = {2411.06157},
 primaryClass = {astro-ph.EP},
       adsurl = {https://ui.adsabs.harvard.edu/abs/2025A&A...693A..54S}
}

@ARTICLE{rebollido_2024,
       author = {{Rebollido}, Isabel and {Stark}, Christopher C. and {Kammerer}, Jens and {Perrin}, Marshall D. and {Lawson}, Kellen and {Pueyo}, Laurent and {Chen}, Christine and {Hines}, Dean and {Girard}, Julien H. and {Worthen}, Kadin and {Ingerbretsen}, Carl and {Betti}, Sarah and {Clampin}, Mark and {Golimowski}, David and {Hoch}, Kielan and {Lewis}, Nikole K. and {Lu}, Cicero X. and {van der Marel}, Roeland P. and {Rickman}, Emily and {Seager}, Sara and {Soummer}, R{\'e}mi and {Valenti}, Jeff A. and {Ward-Duong}, Kimberly and {Mountain}, C. Matt},
        title = "{JWST-TST High Contrast: Asymmetries, Dust Populations, and Hints of a Collision in the {\ensuremath{\beta}} Pictoris Disk with NIRCam and MIRI}",
      journal = {\aj},
         year = 2024,
        month = feb,
       volume = {167},
       number = {2},
          eid = {69},
        pages = {69},
          doi = {10.3847/1538-3881/ad1759},
archivePrefix = {arXiv},
       eprint = {2401.05271},
 primaryClass = {astro-ph.EP},
       adsurl = {https://ui.adsabs.harvard.edu/abs/2024AJ....167...69R}
}

@ARTICLE{xie_2025,
       author = {{Xie}, Chen and {Chen}, Christine H. and {Lisse}, Carey M. and {Hines}, Dean C. and {Beck}, Tracy and {Betti}, Sarah K. and {Pinilla-Alonso}, Noem{\'\i} and {Ingebretsen}, Carl and {Worthen}, Kadin and {G{\'a}sp{\'a}r}, Andr{\'a}s and {Wolff}, Schuyler G. and {Bolin}, Bryce T. and {Pueyo}, Laurent and {Perrin}, Marshall D. and {Stansberry}, John A. and {Leisenring}, Jarron M.},
        title = "{Water ice in the debris disk around HD 181327}",
      journal = {\nat},
         year = 2025,
        month = may,
       volume = {641},
       number = {8063},
        pages = {608-611},
          doi = {10.1038/s41586-025-08920-4},
archivePrefix = {arXiv},
       eprint = {2505.08863},
 primaryClass = {astro-ph.EP},
       adsurl = {https://ui.adsabs.harvard.edu/abs/2025Natur.641..608X}
}

@ARTICLE{cloutier_2024,
       author = {{Cloutier}, Ryan},
        title = "{Exoplanet Demographics: Physical and Orbital Properties}",
      journal = {arXiv e-prints},
         year = 2024,
        month = sep,
          eid = {arXiv:2409.13062},
        pages = {arXiv:2409.13062},
          doi = {10.48550/arXiv.2409.13062},
archivePrefix = {arXiv},
       eprint = {2409.13062},
 primaryClass = {astro-ph.EP},
       adsurl = {https://ui.adsabs.harvard.edu/abs/2024arXiv240913062C}
}

@article{daley_2019,
	title = {The {Mass} of {Stirring} {Bodies} in the {AU} {Mic} {Debris} {Disk} {Inferred} from {Resolved} {Vertical} {Structure}},
	volume = {875},
	issn = {0004-637X},
	url = {https://ui.adsabs.harvard.edu/abs/2019ApJ...875...87D},
	doi = {10.3847/1538-4357/ab1074},
	urldate = {2024-02-07},
	journal = {The Astrophysical Journal},
	author = {Daley, Cail and Hughes, A. Meredith and Carter, Evan S. and Flaherty, Kevin and Lambros, Zachary and Pan, Margaret and Schlichting, Hilke and Chiang, Eugene and Wyatt, Mark and Wilner, David and Andrews, Sean and Carpenter, John},
	month = apr,
	year = {2019},
	note = {ADS Bibcode: 2019ApJ...875...87D},
	pages = {87},
}

@article{harlan_1970,
	title = {Erratum: {MK} classifications for {F}- and {G}-type stars. {II} [{Astron}. {J}., {Vol}. 75, p. 165 - 166 (1970)].},
	volume = {75},
	issn = {0004-6256},
	shorttitle = {Erratum},
	url = {https://ui.adsabs.harvard.edu/abs/1970AJ.....75..507H},
	doi = {10.1086/110986},
	urldate = {2024-02-07},
	journal = {The Astronomical Journal},
	author = {Harlan, E. A. and Taylor, D. C.},
	month = may,
	year = {1970},
	note = {ADS Bibcode: 1970AJ.....75..507H},
	pages = {507--508},
}

@article{torres_2006,
	title = {Search for associations containing young stars ({SACY}). {I}. {Sample} and searching method},
	volume = {460},
	issn = {0004-6361},
	url = {https://ui.adsabs.harvard.edu/abs/2006A&A...460..695T},
	doi = {10.1051/0004-6361:20065602},
	urldate = {2024-02-07},
	journal = {Astronomy and Astrophysics},
	author = {Torres, C. A. O. and Quast, G. R. and da Silva, L. and de La Reza, R. and Melo, C. H. F. and Sterzik, M.},
	month = dec,
	year = {2006},
	note = {ADS Bibcode: 2006A\&A...460..695T},
	pages = {695--708},
}

@article{gray_2006,
	title = {Contributions to the {Nearby} {Stars} ({NStars}) {Project}: {Spectroscopy} of {Stars} {Earlier} than {M0} within 40 pc-{The} {Southern} {Sample}},
	volume = {132},
	issn = {0004-6256},
	shorttitle = {Contributions to the {Nearby} {Stars} ({NStars}) {Project}},
	url = {https://ui.adsabs.harvard.edu/abs/2006AJ....132..161G},
	doi = {10.1086/504637},
	urldate = {2024-02-07},
	journal = {The Astronomical Journal},
	author = {Gray, R. O. and Corbally, C. J. and Garrison, R. F. and McFadden, M. T. and Bubar, E. J. and McGahee, C. E. and O'Donoghue, A. A. and Knox, E. R.},
	month = jul,
	year = {2006},
	note = {ADS Bibcode: 2006AJ....132..161G},
	pages = {161--170},
}

@article{mesa_2021,
	title = {Limits on the presence of planets in systems with debris discs: {HD} 92945 and {HD} 107146},
	volume = {503},
	issn = {0035-8711},
	shorttitle = {Limits on the presence of planets in systems with debris discs},
	url = {https://ui.adsabs.harvard.edu/abs/2021MNRAS.503.1276M},
	doi = {10.1093/mnras/stab438},
	urldate = {2023-11-28},
	journal = {Monthly Notices of the Royal Astronomical Society},
	author = {Mesa, D. and Marino, S. and Bonavita, M. and Lazzoni, C. and Fontanive, C. and Pérez, S. and D'Orazi, V. and Desidera, S. and Gratton, R. and Engler, N. and Henning, T. and Janson, M. and Kral, Q. and Langlois, M. and Messina, S. and Milli, J. and Pawellek, N. and Perrot, C. and Rigliaco, E. and Rickman, E. and Squicciarini, V. and Vigan, A. and Wahhaj, Z. and Zurlo, A. and Boccaletti, A. and Bonnefoy, M. and Chauvin, G. and De Caprio, V. and Feldt, M. and Gluck, L. and Hagelberg, J. and Keppler, M. and Lagrange, A. -M. and Launhardt, R. and Maire, A. -L. and Meyer, M. and Moeller-Nilsson, O. and Pavlov, A. and Samland, M. and Schmidt, T. and Weber, L.},
	month = may,
	year = {2021},
	note = {ADS Bibcode: 2021MNRAS.503.1276M},
	pages = {1276--1289},
}

@article{hinkley_2023,
	title = {Direct discovery of the inner exoplanet in the {HD} 206893 system. {Evidence} for deuterium burning in a planetary-mass companion},
	volume = {671},
	issn = {0004-6361},
	url = {https://ui.adsabs.harvard.edu/abs/2023A&A...671L...5H},
	doi = {10.1051/0004-6361/202244727},
	urldate = {2023-11-22},
	journal = {Astronomy and Astrophysics},
	author = {Hinkley, S. and Lacour, S. and Marleau, G. -D. and Lagrange, A. -M. and Wang, J. J. and Kammerer, J. and Cumming, A. and Nowak, M. and Rodet, L. and Stolker, T. and Balmer, W. -O. and Ray, S. and Bonnefoy, M. and Mollière, P. and Lazzoni, C. and Kennedy, G. and Mordasini, C. and Abuter, R. and Aigrain, S. and Amorim, A. and Asensio-Torres, R. and Babusiaux, C. and Benisty, M. and Berger, J. -P. and Beust, H. and Blunt, S. and Boccaletti, A. and Bohn, A. and Bonnet, H. and Bourdarot, G. and Brandner, W. and Cantalloube, F. and Caselli, P. and Charnay, B. and Chauvin, G. and Chomez, A. and Choquet, E. and Christiaens, V. and Clénet, Y. and Coudé du Foresto, V. and Cridland, A. and Delorme, P. and Dembet, R. and Drescher, A. and Duvert, G. and Eckart, A. and Eisenhauer, F. and Feuchtgruber, H. and Galland, F. and Garcia, P. and Garcia Lopez, R. and Gardner, T. and Gendron, E. and Genzel, R. and Gillessen, S. and Girard, J. H. and Grandjean, A. and Haubois, X. and Heißel, G. and Henning, Th. and Hippler, S. and Horrobin, M. and Houllé, M. and Hubert, Z. and Jocou, L. and Keppler, M. and Kervella, P. and Kreidberg, L. and Lapeyrère, V. and Le Bouquin, J. -B. and Léna, P. and Lutz, D. and Maire, A. -L. and Mang, F. and Mérand, A. and Meunier, N. and Monnier, J. D. and Mouillet, D. and Nasedkin, E. and Ott, T. and Otten, G. P. P. L. and Paladini, C. and Paumard, T. and Perraut, K. and Perrin, G. and Philipot, F. and Pfuhl, O. and Pourré, N. and Pueyo, L. and Rameau, J. and Rickman, E. and Rubini, P. and Rustamkulov, Z. and Samland, M. and Shangguan, J. and Shimizu, T. and Sing, D. and Straubmeier, C. and Sturm, E. and Tacconi, L. J. and van Dishoeck, E. F. and Vigan, A. and Vincent, F. and Ward-Duong, K. and Widmann, F. and Wieprecht, E. and Wiezorrek, E. and Woillez, J. and Yazici, S. and Young, A. and Zicher, N.},
	month = mar,
	year = {2023},
	note = {ADS Bibcode: 2023A\&A...671L...5H},
	pages = {L5},
}

@article{marino_2021,
	title = {Constraining planetesimal stirring: how sharp are debris disc edges?},
	volume = {503},
	issn = {0035-8711},
	shorttitle = {Constraining planetesimal stirring},
	url = {https://ui.adsabs.harvard.edu/abs/2021MNRAS.503.5100M},
	doi = {10.1093/mnras/stab771},
	urldate = {2023-11-28},
	journal = {Monthly Notices of the Royal Astronomical Society},
	author = {Marino, Sebastian},
	month = may,
	year = {2021},
	note = {ADS Bibcode: 2021MNRAS.503.5100M},
	pages = {5100--5114},
}

@article{imaz_blanco_2023,
	title = {Inner edges of planetesimal belts: collisionally eroded or truncated?},
	volume = {522},
	issn = {0035-8711},
	shorttitle = {Inner edges of planetesimal belts},
	url = {https://ui.adsabs.harvard.edu/abs/2023MNRAS.522.6150I},
	doi = {10.1093/mnras/stad1221},
	urldate = {2023-11-28},
	journal = {Monthly Notices of the Royal Astronomical Society},
	author = {Imaz Blanco, Amaia and Marino, Sebastian and Matrà, Luca and Booth, Mark and Carpenter, John and Faramaz, Virginie and Henning, Thomas and Hughes, A. Meredith and Kennedy, Grant M. and Pérez, Sebastián and Ricci, Luca and Wyatt, Mark C.},
	month = jul,
	year = {2023},
	note = {ADS Bibcode: 2023MNRAS.522.6150I},
	pages = {6150--6169},
}

@ARTICLE{costa_2024,
       author = {{Costa}, Tyson and {Pearce}, Tim D. and {Krivov}, Alexander V.},
        title = "{Increasing planet-stirring efficiency of debris discs by 'projectile stirring' and 'resonant stirring'}",
      journal = {\mnras},
         year = 2024,
        month = jan,
       volume = {527},
       number = {3},
        pages = {7317-7336},
          doi = {10.1093/mnras/stad3582},
archivePrefix = {arXiv},
       eprint = {2311.10461},
 primaryClass = {astro-ph.EP},
       adsurl = {https://ui.adsabs.harvard.edu/abs/2024MNRAS.527.7317C}
}

@INPROCEEDINGS{delorme_2017,
       author = {{Delorme}, P. and {Meunier}, N. and {Albert}, D. and {Lagadec}, E. and {Le Coroller}, H. and {Galicher}, R. and {Mouillet}, D. and {Boccaletti}, A. and {Mesa}, D. and {Meunier}, J. -C. and {Beuzit}, J. -L. and {Lagrange}, A. -M. and {Chauvin}, G. and {Sapone}, A. and {Langlois}, M. and {Maire}, A. -L. and {Montarg{\`e}s}, M. and {Gratton}, R. and {Vigan}, A. and {Surace}, C.},
        title = "{The SPHERE Data Center: a reference for high contrast imaging processing}",
    booktitle = {SF2A-2017: Proceedings of the Annual meeting of the French Society of Astronomy and Astrophysics},
         year = 2017,
       editor = {{Reyl{\'e}}, C. and {Di Matteo}, P. and {Herpin}, F. and {Lagadec}, E. and {Lan{\c{c}}on}, A. and {Meliani}, Z. and {Royer}, F.},
        month = dec,
        pages = {Di},
          doi = {10.48550/arXiv.1712.06948},
archivePrefix = {arXiv},
       eprint = {1712.06948},
 primaryClass = {astro-ph.IM},
       adsurl = {https://ui.adsabs.harvard.edu/abs/2017sf2a.conf..347D}
}

@ARTICLE{kervella_2004,
       author = {{Kervella}, P. and {Th{\'e}venin}, F. and {Di Folco}, E. and {S{\'e}gransan}, D.},
        title = "{The angular sizes of dwarf stars and subgiants. Surface brightness relations calibrated by interferometry}",
      journal = {\aap},
         year = 2004,
        month = oct,
       volume = {426},
        pages = {297-307},
          doi = {10.1051/0004-6361:20035930},
archivePrefix = {arXiv},
       eprint = {astro-ph/0404180},
 primaryClass = {astro-ph},
       adsurl = {https://ui.adsabs.harvard.edu/abs/2004A&A...426..297K}
}

@ARTICLE{plavchan_2009,
       author = {{Plavchan}, Peter and {Werner}, M.~W. and {Chen}, C.~H. and {Stapelfeldt}, K.~R. and {Su}, K.~Y.~L. and {Stauffer}, J.~R. and {Song}, I.},
        title = "{New Debris Disks Around Young, Low-Mass Stars Discovered with the Spitzer Space Telescope}",
      journal = {\apj},
         year = 2009,
        month = jun,
       volume = {698},
       number = {2},
        pages = {1068-1094},
          doi = {10.1088/0004-637X/698/2/1068},
archivePrefix = {arXiv},
       eprint = {0904.0819},
 primaryClass = {astro-ph.SR},
       adsurl = {https://ui.adsabs.harvard.edu/abs/2009ApJ...698.1068P}
}

@ARTICLE{crotts_2025,
       author = {{Crotts}, Katie A. and {Carter}, Aarynn L. and {Lawson}, Kellen and {Mang}, James and {Biller}, Beth and {Booth}, Mark and {Ferrer-Chavez}, Rodrigo and {Girard}, Julien H. and {Lagrange}, Anne-Marie and {Liu}, Michael C. and {Marino}, Sebastian and {Millar-Blanchaer}, Maxwell A. and {Skemer}, Andy and {Strampelli}, Giovanni M. and {Wang}, Jason and {Absil}, Olivier and {Balmer}, William O. and {Bendahan-West}, Rapha{\"e}l and {Bogat}, Ellis and {Bowens-Rubin}, Rachel and {Chauvin}, Ga{\"e}l and {Fontanive}, Cl{\'e}mence and {Franson}, Kyle and {Kammerer}, Jens and {Leisenring}, Jarron and {Morley}, Caroline V. and {Rebollido}, Isabel and {Skaf}, Nour and {Sutlieff}, Ben J. and {Bruinsma}, Evelyn L. and {Hinkley}, Sasha and {Hoch}, Kielan and {James}, Andrew D. and {Kane}, Rohan and {Mawet}, Dimitri and {Meyer}, Michael R. and {Palatnick}, Skyler and {Perrin}, Marshall D. and {Ray}, Shrishmoy and {Rickman}, Emily and {Sanghi}, Aniket and {Subbotina Stephenson}, Klaus},
        title = "{Follow-Up Exploration of the TWA 7 Planet-Disk System with JWST NIRCam}",
      journal = {arXiv e-prints},
         year = 2025,
        month = jun,
          eid = {arXiv:2506.19932},
        pages = {arXiv:2506.19932},
archivePrefix = {arXiv},
       eprint = {2506.19932},
 primaryClass = {astro-ph.EP},
       adsurl = {https://ui.adsabs.harvard.edu/abs/2025arXiv250619932C}
}

@ARTICLE{Chen_2014,
       author = {{Chen}, Christine H. and {Mittal}, Tushar and {Kuchner}, Marc and {Forrest}, William J. and {Lisse}, Carey M. and {Manoj}, P. and {Sargent}, Benjamin A. and {Watson}, Dan M.},
        title = "{The Spitzer Infrared Spectrograph Debris Disk Catalog. I. Continuum Analysis of Unresolved Targets}",
      journal = {\apjs},
         year = 2014,
        month = apr,
       volume = {211},
       number = {2},
          eid = {25},
        pages = {25},
          doi = {10.1088/0067-0049/211/2/25},
       adsurl = {https://ui.adsabs.harvard.edu/abs/2014ApJS..211...25C}
}

@ARTICLE{holland_2017,
       author = {{Holland}, Wayne S. and {Matthews}, Brenda C. and {Kennedy}, Grant M. and {Greaves}, Jane S. and {Wyatt}, Mark C. and {Booth}, Mark and {Bastien}, Pierre and {Bryden}, Geoff and {Butner}, Harold and {Chen}, Christine H. and {Chrysostomou}, Antonio and {Davies}, Claire L. and {Dent}, William R.~F. and {Di Francesco}, James and {Duch{\^e}ne}, Gaspard and {Gibb}, Andy G. and {Friberg}, Per and {Ivison}, Rob J. and {Jenness}, Tim and {Kavelaars}, JJ and {Lawler}, Samantha and {Lestrade}, Jean-Fran{\c{c}}ois and {Marshall}, Jonathan P. and {Moro-Martin}, Amaya and {Pani{\'c}}, Olja and {Phillips}, Neil and {Serjeant}, Stephen and {Schieven}, Gerald H. and {Sibthorpe}, Bruce and {Vican}, Laura and {Ward-Thompson}, Derek and {van der Werf}, Paul and {White}, Glenn J. and {Wilner}, David and {Zuckerman}, Ben},
        title = "{SONS: The JCMT legacy survey of debris discs in the submillimetre}",
      journal = {\mnras},
         year = 2017,
        month = sep,
       volume = {470},
       number = {3},
        pages = {3606-3663},
          doi = {10.1093/mnras/stx1378},
archivePrefix = {arXiv},
       eprint = {1706.01218},
 primaryClass = {astro-ph.SR},
       adsurl = {https://ui.adsabs.harvard.edu/abs/2017MNRAS.470.3606H}
}

@ARTICLE{stanford-moore_2020,
       author = {{Stanford-Moore}, S. Adam and {Nielsen}, Eric L. and {De Rosa}, Robert J. and {Macintosh}, Bruce and {Czekala}, Ian},
        title = "{BAFFLES: Bayesian Ages for Field Lower-mass Stars}",
      journal = {\apj},
         year = 2020,
        month = jul,
       volume = {898},
       number = {1},
          eid = {27},
        pages = {27},
          doi = {10.3847/1538-4357/ab9a35},
archivePrefix = {arXiv},
       eprint = {2006.04811},
 primaryClass = {astro-ph.SR},
       adsurl = {https://ui.adsabs.harvard.edu/abs/2020ApJ...898...27S}
}

@ARTICLE{nederlander_2021,
       author = {{Nederlander}, Ava and {Hughes}, A. Meredith and {Fehr}, Anna J. and {Flaherty}, Kevin M. and {Su}, Kate Y.~L. and {Mo{\'o}r}, Attila and {Chiang}, Eugene and {Andrews}, Sean M. and {Wilner}, David J. and {Marino}, Sebastian},
        title = "{Resolving Structure in the Debris Disk around HD 206893 with ALMA}",
      journal = {\apj},
         year = 2021,
        month = aug,
       volume = {917},
       number = {1},
          eid = {5},
        pages = {5},
          doi = {10.3847/1538-4357/abdd32},
archivePrefix = {arXiv},
       eprint = {2101.08849},
 primaryClass = {astro-ph.EP},
       adsurl = {https://ui.adsabs.harvard.edu/abs/2021ApJ...917....5N}
}

@ARTICLE{morrison_2015,
       author = {{Morrison}, Sarah and {Malhotra}, Renu},
        title = "{Planetary Chaotic Zone Clearing: Destinations and Timescales}",
      journal = {\apj},
         year = 2015,
        month = jan,
       volume = {799},
       number = {1},
          eid = {41},
        pages = {41},
          doi = {10.1088/0004-637X/799/1/41},
archivePrefix = {arXiv},
       eprint = {1411.1378},
 primaryClass = {astro-ph.EP},
       adsurl = {https://ui.adsabs.harvard.edu/abs/2015ApJ...799...41M}
}

@ARTICLE{wyatt2003,
       author = {{Wyatt}, M.~C.},
        title = "{Resonant Trapping of Planetesimals by Planet Migration: Debris Disk Clumps and Vega's Similarity to the Solar System}",
      journal = {\apj},
         year = 2003,
        month = dec,
       volume = {598},
       number = {2},
        pages = {1321-1340},
          doi = {10.1086/379064},
archivePrefix = {arXiv},
       eprint = {astro-ph/0308253},
 primaryClass = {astro-ph},
       adsurl = {https://ui.adsabs.harvard.edu/abs/2003ApJ...598.1321W}
}

@article{milli_2017_hd206893,
	title = {Discovery of a low-mass companion inside the debris ring surrounding the {F5V} star {HD} 206893},
	volume = {597},
	issn = {0004-6361},
	url = {https://ui.adsabs.harvard.edu/abs/2017A&A...597L...2M},
	doi = {10.1051/0004-6361/201629908},
	urldate = {2024-02-07},
	journal = {Astronomy and Astrophysics},
	author = {Milli, J. and Hibon, P. and Christiaens, V. and Choquet, É. and Bonnefoy, M. and Kennedy, G. M. and Wyatt, M. C. and Absil, O. and Gómez González, C. A. and del Burgo, C. and Matrà, L. and Augereau, J. -C. and Boccaletti, A. and Delacroix, C. and Ertel, S. and Dent, W. R. F. and Forsberg, P. and Fusco, T. and Girard, J. H. and Habraken, S. and Huby, E. and Karlsson, M. and Lagrange, A. -M. and Mawet, D. and Mouillet, D. and Perrin, M. and Pinte, C. and Pueyo, L. and Reyes, C. and Soummer, R. and Surdej, J. and Tarricq, Y. and Wahhaj, Z.},
	month = jan,
	year = {2017},
	note = {ADS Bibcode: 2017A\&A...597L...2M},
	pages = {L2},
}

@article{kammerer_2021,
	title = {{GRAVITY} {K}-band spectroscopy of {HD} 206893 {B}. {Brown} dwarf or exoplanet},
	volume = {652},
	issn = {0004-6361},
	url = {https://ui.adsabs.harvard.edu/abs/2021A&A...652A..57K},
	doi = {10.1051/0004-6361/202140749},
	urldate = {2024-02-07},
	journal = {Astronomy and Astrophysics},
	author = {Kammerer, J. and Lacour, S. and Stolker, T. and Mollière, P. and Sing, D. K. and Nasedkin, E. and Kervella, P. and Wang, J. J. and Ward-Duong, K. and Nowak, M. and Abuter, R. and Amorim, A. and Asensio-Torres, R. and Bauböck, M. and Benisty, M. and Berger, J. -P. and Beust, H. and Blunt, S. and Boccaletti, A. and Bohn, A. and Bolzer, M. -L. and Bonnefoy, M. and Bonnet, H. and Brandner, W. and Cantalloube, F. and Caselli, P. and Charnay, B. and Chauvin, G. and Choquet, E. and Christiaens, V. and Clénet, Y. and Coudé du Foresto, V. and Cridland, A. and Dembet, R. and Dexter, J. and de Zeeuw, P. T. and Drescher, A. and Duvert, G. and Eckart, A. and Eisenhauer, F. and Gao, F. and Garcia, P. and Garcia Lopez, R. and Gendron, E. and Genzel, R. and Gillessen, S. and Girard, J. and Haubois, X. and Heißel, G. and Henning, T. and Hinkley, S. and Hippler, S. and Horrobin, M. and Houllé, M. and Hubert, Z. and Jocou, L. and Keppler, M. and Kreidberg, L. and Lagrange, A. -M. and Lapeyrère, V. and Le Bouquin, J. -B. and Léna, P. and Lutz, D. and Maire, A. -L. and Mérand, A. and Monnier, J. D. and Mouillet, D. and Müller, A. and Ott, T. and Otten, G. P. P. L. and Paladini, C. and Paumard, T. and Perraut, K. and Perrin, G. and Pfuhl, O. and Pueyo, L. and Rameau, J. and Rodet, L. and Rousset, G. and Rustamkulov, Z. and Shangguan, J. and Shimizu, T. and Stadler, J. and Straub, O. and Straubmeier, C. and Sturm, E. and Tacconi, L. J. and van Dishoeck, E. F. and Vigan, A. and Vincent, F. and von Fellenberg, S. D. and Widmann, F. and Wieprecht, E. and Wiezorrek, E. and Woillez, J. and Yazici, S.},
	month = aug,
	year = {2021},
	note = {ADS Bibcode: 2021A\&A...652A..57K},
	pages = {A57},
}

@article{nielsen_2019,
	title = {The {Gemini} {Planet} {Imager} {Exoplanet} {Survey}: {Giant} {Planet} and {Brown} {Dwarf} {Demographics} from 10 to 100 au},
	volume = {158},
	issn = {0004-6256},
	shorttitle = {The {Gemini} {Planet} {Imager} {Exoplanet} {Survey}},
	url = {https://ui.adsabs.harvard.edu/abs/2019AJ....158...13N},
	doi = {10.3847/1538-3881/ab16e9},
	urldate = {2024-03-30},
	journal = {The Astronomical Journal},
	author = {Nielsen, Eric L. and De Rosa, Robert J. and Macintosh, Bruce and Wang, Jason J. and Ruffio, Jean-Baptiste and Chiang, Eugene and Marley, Mark S. and Saumon, Didier and Savransky, Dmitry and Ammons, S. Mark and Bailey, Vanessa P. and Barman, Travis and Blain, Célia and Bulger, Joanna and Burrows, Adam and Chilcote, Jeffrey and Cotten, Tara and Czekala, Ian and Doyon, Rene and Duchêne, Gaspard and Esposito, Thomas M. and Fabrycky, Daniel and Fitzgerald, Michael P. and Follette, Katherine B. and Fortney, Jonathan J. and Gerard, Benjamin L. and Goodsell, Stephen J. and Graham, James R. and Greenbaum, Alexandra Z. and Hibon, Pascale and Hinkley, Sasha and Hirsch, Lea A. and Hom, Justin and Hung, Li-Wei and Dawson, Rebekah Ilene and Ingraham, Patrick and Kalas, Paul and Konopacky, Quinn and Larkin, James E. and Lee, Eve J. and Lin, Jonathan W. and Maire, Jérôme and Marchis, Franck and Marois, Christian and Metchev, Stanimir and Millar-Blanchaer, Maxwell A. and Morzinski, Katie M. and Oppenheimer, Rebecca and Palmer, David and Patience, Jennifer and Perrin, Marshall and Poyneer, Lisa and Pueyo, Laurent and Rafikov, Roman R. and Rajan, Abhijith and Rameau, Julien and Rantakyrö, Fredrik T. and Ren, Bin and Schneider, Adam C. and Sivaramakrishnan, Anand and Song, Inseok and Soummer, Remi and Tallis, Melisa and Thomas, Sandrine and Ward-Duong, Kimberly and Wolff, Schuyler},
	month = jul,
	year = {2019},
	note = {ADS Bibcode: 2019AJ....158...13N},
	pages = {13},
}

@article{vigan_2021,
	title = {The {SPHERE} infrared survey for exoplanets ({SHINE}). {III}. {The} demographics of young giant exoplanets below 300 au with {SPHERE}},
	volume = {651},
	issn = {0004-6361},
	url = {https://ui.adsabs.harvard.edu/abs/2021A&A...651A..72V},
	doi = {10.1051/0004-6361/202038107},
	urldate = {2024-03-30},
	journal = {Astronomy and Astrophysics},
	author = {Vigan, A. and Fontanive, C. and Meyer, M. and Biller, B. and Bonavita, M. and Feldt, M. and Desidera, S. and Marleau, G. -D. and Emsenhuber, A. and Galicher, R. and Rice, K. and Forgan, D. and Mordasini, C. and Gratton, R. and Le Coroller, H. and Maire, A. -L. and Cantalloube, F. and Chauvin, G. and Cheetham, A. and Hagelberg, J. and Lagrange, A. -M. and Langlois, M. and Bonnefoy, M. and Beuzit, J. -L. and Boccaletti, A. and D'Orazi, V. and Delorme, P. and Dominik, C. and Henning, Th. and Janson, M. and Lagadec, E. and Lazzoni, C. and Ligi, R. and Menard, F. and Mesa, D. and Messina, S. and Moutou, C. and Müller, A. and Perrot, C. and Samland, M. and Schmid, H. M. and Schmidt, T. and Sissa, E. and Turatto, M. and Udry, S. and Zurlo, A. and Abe, L. and Antichi, J. and Asensio-Torres, R. and Baruffolo, A. and Baudoz, P. and Baudrand, J. and Bazzon, A. and Blanchard, P. and Bohn, A. J. and Brown Sevilla, S. and Carbillet, M. and Carle, M. and Cascone, E. and Charton, J. and Claudi, R. and Costille, A. and De Caprio, V. and Delboulbé, A. and Dohlen, K. and Engler, N. and Fantinel, D. and Feautrier, P. and Fusco, T. and Gigan, P. and Girard, J. H. and Giro, E. and Gisler, D. and Gluck, L. and Gry, C. and Hubin, N. and Hugot, E. and Jaquet, M. and Kasper, M. and Le Mignant, D. and Llored, M. and Madec, F. and Magnard, Y. and Martinez, P. and Maurel, D. and Möller-Nilsson, O. and Mouillet, D. and Moulin, T. and Origné, A. and Pavlov, A. and Perret, D. and Petit, C. and Pragt, J. and Puget, P. and Rabou, P. and Ramos, J. and Rickman, E. L. and Rigal, F. and Rochat, S. and Roelfsema, R. and Rousset, G. and Roux, A. and Salasnich, B. and Sauvage, J. -F. and Sevin, A. and Soenke, C. and Stadler, E. and Suarez, M. and Wahhaj, Z. and Weber, L. and Wildi, F.},
	month = jul,
	year = {2021},
	note = {ADS Bibcode: 2021A\&A...651A..72V},
	pages = {A72},
}

@article{hughes_2018,
	title = {Debris {Disks}: {Structure}, {Composition}, and {Variability}},
	volume = {56},
	issn = {0066-4146, 1545-4282},
	shorttitle = {Debris {Disks}},
	url = {http://arxiv.org/abs/1802.04313},
	doi = {10.1146/annurev-astro-081817-052035},
	language = {en},
	number = {1},
	urldate = {2022-07-15},
	journal = {Annual Review of Astronomy and Astrophysics},
	author = {Hughes, A. Meredith and Duchene, Gaspard and Matthews, Brenda},
	month = sep,
	year = {2018},
	note = {arXiv:1802.04313 [astro-ph]},
	pages = {541--591},
}

@article{walsh_2011,
	title = {A low mass for {Mars} from {Jupiter}'s early gas-driven migration},
	volume = {475},
	issn = {0028-0836},
	url = {https://ui.adsabs.harvard.edu/abs/2011Natur.475..206W},
	doi = {10.1038/nature10201},
	urldate = {2024-03-30},
	journal = {Nature},
	author = {Walsh, Kevin J. and Morbidelli, Alessandro and Raymond, Sean N. and O'Brien, David P. and Mandell, Avi M.},
	month = jul,
	year = {2011},
	note = {ADS Bibcode: 2011Natur.475..206W},
	pages = {206--209},
}

@article{ida_2000,
	title = {Orbital {Migration} of {Neptune} and {Orbital} {Distribution} of {Trans}-{Neptunian} {Objects}},
	volume = {534},
	issn = {0004-637X},
	url = {https://ui.adsabs.harvard.edu/abs/2000ApJ...534..428I},
	doi = {10.1086/308720},
	urldate = {2024-03-30},
	journal = {The Astrophysical Journal},
	author = {Ida, Shigeru and Bryden, Geoffrey and Lin, D. N. C. and Tanaka, Hidekazu},
	month = may,
	year = {2000},
	note = {ADS Bibcode: 2000ApJ...534..428I},
	pages = {428--445},
}

@article{andrews_2018,
	title = {The {Disk} {Substructures} at {High} {Angular} {Resolution} {Project} ({DSHARP}). {I}. {Motivation}, {Sample}, {Calibration}, and {Overview}},
	volume = {869},
	issn = {0004-637X},
	url = {https://ui.adsabs.harvard.edu/abs/2018ApJ...869L..41A},
	doi = {10.3847/2041-8213/aaf741},
	urldate = {2024-03-30},
	journal = {The Astrophysical Journal},
	author = {Andrews, Sean M. and Huang, Jane and Pérez, Laura M. and Isella, Andrea and Dullemond, Cornelis P. and Kurtovic, Nicolás T. and Guzmán, Viviana V. and Carpenter, John M. and Wilner, David J. and Zhang, Shangjia and Zhu, Zhaohuan and Birnstiel, Tilman and Bai, Xue-Ning and Benisty, Myriam and Hughes, A. Meredith and Öberg, Karin I. and Ricci, Luca},
	month = dec,
	year = {2018},
	note = {ADS Bibcode: 2018ApJ...869L..41A},
	pages = {L41},
}

@article{macgregor_2017,
	title = {A {Complete} {ALMA} {Map} of the {Fomalhaut} {Debris} {Disk}},
	volume = {842},
	issn = {0004-637X},
	url = {https://ui.adsabs.harvard.edu/abs/2017ApJ...842....8M},
	doi = {10.3847/1538-4357/aa71ae},
	urldate = {2024-03-30},
	journal = {The Astrophysical Journal},
	author = {MacGregor, Meredith A. and Matrà, Luca and Kalas, Paul and Wilner, David J. and Pan, Margaret and Kennedy, Grant M. and Wyatt, Mark C. and Duchene, Gaspard and Hughes, A. Meredith and Rieke, George H. and Clampin, Mark and Fitzgerald, Michael P. and Graham, James R. and Holland, Wayne S. and Panić, Olja and Shannon, Andrew and Su, Kate},
	month = jun,
	year = {2017},
	note = {ADS Bibcode: 2017ApJ...842....8M},
	pages = {8},
}

@article{kennedy_2018,
	title = {{ALMA} observations of the narrow {HR} {4796A} debris ring},
	volume = {475},
	issn = {0035-8711},
	url = {https://ui.adsabs.harvard.edu/abs/2018MNRAS.475.4924K},
	doi = {10.1093/mnras/sty135},
	urldate = {2024-03-30},
	journal = {Monthly Notices of the Royal Astronomical Society},
	author = {Kennedy, Grant M. and Marino, Sebastian and Matrà, Luca and Panić, Olja and Wilner, David and Wyatt, Mark C. and Yelverton, Ben},
	month = apr,
	year = {2018},
	note = {ADS Bibcode: 2018MNRAS.475.4924K},
	pages = {4924--4938},
}

@article{kirsh_2009,
	title = {Simulations of planet migration driven by planetesimal scattering},
	volume = {199},
	issn = {0019-1035},
	url = {https://ui.adsabs.harvard.edu/abs/2009Icar..199..197K},
	doi = {10.1016/j.icarus.2008.05.028},
	urldate = {2024-03-30},
	journal = {Icarus},
	author = {Kirsh, David R. and Duncan, Martin and Brasser, Ramon and Levison, Harold F.},
	month = jan,
	year = {2009},
	note = {ADS Bibcode: 2009Icar..199..197K},
	pages = {197--209},
}

@article{pearce_2015,
	title = {Double-ringed debris discs could be the work of eccentric planets: explaining the strange morphology of {HD} 107146},
	volume = {453},
	issn = {0035-8711},
	shorttitle = {Double-ringed debris discs could be the work of eccentric planets},
	url = {https://ui.adsabs.harvard.edu/abs/2015MNRAS.453.3329P},
	doi = {10.1093/mnras/stv1847},
	urldate = {2023-11-22},
	journal = {Monthly Notices of the Royal Astronomical Society},
	author = {Pearce, Tim D. and Wyatt, Mark C.},
	month = nov,
	year = {2015},
	note = {ADS Bibcode: 2015MNRAS.453.3329P},
	pages = {3329--3340},
}

@article{sefilian_2021,
	title = {Formation of {Gaps} in {Self}-gravitating {Debris} {Disks} by {Secular} {Resonance} in a {Single}-planet {System}. {I}. {A} {Simplified} {Model}},
	volume = {910},
	issn = {0004-637X},
	url = {https://ui.adsabs.harvard.edu/abs/2021ApJ...910...13S},
	doi = {10.3847/1538-4357/abda46},
	urldate = {2023-11-28},
	journal = {The Astrophysical Journal},
	author = {Sefilian, Antranik A. and Rafikov, Roman R. and Wyatt, Mark C.},
	month = mar,
	year = {2021},
	note = {ADS Bibcode: 2021ApJ...910...13S},
	pages = {13},
}

@article{sefilian_2023,
	title = {Formation of {Gaps} in {Self}-gravitating {Debris} {Disks} by {Secular} {Resonance} in a {Single}-planet {System}. {II}. {Toward} a {Self}-consistent {Model}},
	volume = {954},
	issn = {0004-637X},
	url = {https://ui.adsabs.harvard.edu/abs/2023ApJ...954..100S},
	doi = {10.3847/1538-4357/ace68e},
	urldate = {2023-11-21},
	journal = {The Astrophysical Journal},
	author = {Sefilian, Antranik A. and Rafikov, Roman R. and Wyatt, Mark C.},
	month = sep,
	year = {2023},
	note = {ADS Bibcode: 2023ApJ...954..100S},
	pages = {100},
}

@article{yelverton_2018,
	title = {Empty gaps? {Depleting} annular regions in debris discs by secular resonance with a two-planet system},
	volume = {479},
	issn = {0035-8711},
	shorttitle = {Empty gaps?},
	url = {https://ui.adsabs.harvard.edu/abs/2018MNRAS.479.2673Y},
	doi = {10.1093/mnras/sty1678},
	urldate = {2023-11-22},
	journal = {Monthly Notices of the Royal Astronomical Society},
	author = {Yelverton, Ben and Kennedy, Grant M.},
	month = sep,
	year = {2018},
	note = {ADS Bibcode: 2018MNRAS.479.2673Y},
	pages = {2673--2691},
}

@article{carter_2021,
	title = {Direct imaging of sub-{Jupiter} mass exoplanets with {James} {Webb} {Space} {Telescope} coronagraphy},
	volume = {501},
	issn = {0035-8711},
	url = {https://ui.adsabs.harvard.edu/abs/2021MNRAS.501.1999C},
	doi = {10.1093/mnras/staa3579},
	urldate = {2024-03-30},
	journal = {Monthly Notices of the Royal Astronomical Society},
	author = {Carter, Aarynn L. and Hinkley, Sasha and Bonavita, Mariangela and Phillips, Mark W. and Girard, Julien H. and Perrin, Marshall and Pueyo, Laurent and Vigan, Arthur and Gagné, Jonathan and Skemer, Andrew J. I.},
	month = feb,
	year = {2021},
	note = {ADS Bibcode: 2021MNRAS.501.1999C},
	pages = {1999--2016},
}

@article{phillips_2020,
	title = {A new set of atmosphere and evolution models for cool {T}-{Y} brown dwarfs and giant exoplanets},
	volume = {637},
	issn = {0004-6361},
	url = {https://ui.adsabs.harvard.edu/abs/2020A&A...637A..38P},
	doi = {10.1051/0004-6361/201937381},
	urldate = {2024-03-30},
	journal = {Astronomy and Astrophysics},
	author = {Phillips, M. W. and Tremblin, P. and Baraffe, I. and Chabrier, G. and Allard, N. F. and Spiegelman, F. and Goyal, J. M. and Drummond, B. and Hébrard, E.},
	month = may,
	year = {2020},
	note = {ADS Bibcode: 2020A\&A...637A..38P},
	pages = {A38},
}

@article{linder_2019,
	title = {Evolutionary models of cold and low-mass planets: cooling curves, magnitudes, and detectability},
	volume = {623},
	issn = {0004-6361},
	shorttitle = {Evolutionary models of cold and low-mass planets},
	url = {https://ui.adsabs.harvard.edu/abs/2019A&A...623A..85L},
	doi = {10.1051/0004-6361/201833873},
	urldate = {2024-03-30},
	journal = {Astronomy and Astrophysics},
	author = {Linder, Esther F. and Mordasini, Christoph and Mollière, Paul and Marleau, Gabriel-Dominique and Malik, Matej and Quanz, Sascha P. and Meyer, Michael R.},
	month = mar,
	year = {2019},
	note = {ADS Bibcode: 2019A\&A...623A..85L},
	pages = {A85},
}

@article{kervella_2022,
	title = {Stellar and substellar companions from {Gaia} {EDR3}. {Proper}-motion anomaly and resolved common proper-motion pairs},
	volume = {657},
	issn = {0004-6361},
	url = {https://ui.adsabs.harvard.edu/abs/2022A&A...657A...7K},
	doi = {10.1051/0004-6361/202142146},
	urldate = {2023-11-22},
	journal = {Astronomy and Astrophysics},
	author = {Kervella, Pierre and Arenou, Frédéric and Thévenin, Frédéric},
	month = jan,
	year = {2022},
	note = {ADS Bibcode: 2022A\&A...657A...7K},
	pages = {A7},
}

@article{friebe_2022,
	title = {Gap carving by a migrating planet embedded in a massive debris disc},
	volume = {512},
	issn = {0035-8711},
	url = {https://ui.adsabs.harvard.edu/abs/2022MNRAS.512.4441F},
	doi = {10.1093/mnras/stac664},
	urldate = {2023-11-22},
	journal = {Monthly Notices of the Royal Astronomical Society},
	author = {Friebe, Marc F. and Pearce, Tim D. and Löhne, Torsten},
	month = may,
	year = {2022},
	note = {ADS Bibcode: 2022MNRAS.512.4441F},
	pages = {4441--4454},
}

@article{gaiadr3_collaboration_2023,
	title = {Gaia {Data} {Release} 3. {Summary} of the content and survey properties},
	volume = {674},
	issn = {0004-6361},
	url = {https://ui.adsabs.harvard.edu/abs/2023A&A...674A...1G},
	doi = {10.1051/0004-6361/202243940},
    urldate = {2024-07-10},
	journal = {Astronomy and Astrophysics},
	author = {{Gaia Collaboration} and Vallenari, A. and Brown, A. G. A. and Prusti, T. and de Bruijne, J. H. J. and Arenou, F. and Babusiaux, C. and Biermann, M. and Creevey, O. L. and Ducourant, C. and Evans, D. W. and Eyer, L. and Guerra, R. and Hutton, A. and Jordi, C. and Klioner, S. A. and Lammers, U. L. and Lindegren, L. and Luri, X. and Mignard, F. and Panem, C. and Pourbaix, D. and Randich, S. and Sartoretti, P. and Soubiran, C. and Tanga, P. and Walton, N. A. and Bailer-Jones, C. A. L. and Bastian, U. and Drimmel, R. and Jansen, F. and Katz, D. and Lattanzi, M. G. and van Leeuwen, F. and Bakker, J. and Cacciari, C. and Castañeda, J. and De Angeli, F. and Fabricius, C. and Fouesneau, M. and Frémat, Y. and Galluccio, L. and Guerrier, A. and Heiter, U. and Masana, E. and Messineo, R. and Mowlavi, N. and Nicolas, C. and Nienartowicz, K. and Pailler, F. and Panuzzo, P. and Riclet, F. and Roux, W. and Seabroke, G. M. and Sordo, R. and Thévenin, F. and Gracia-Abril, G. and Portell, J. and Teyssier, D. and Altmann, M. and Andrae, R. and Audard, M. and Bellas-Velidis, I. and Benson, K. and Berthier, J. and Blomme, R. and Burgess, P. W. and Busonero, D. and Busso, G. and Cánovas, H. and Carry, B. and Cellino, A. and Cheek, N. and Clementini, G. and Damerdji, Y. and Davidson, M. and de Teodoro, P. and Nuñez Campos, M. and Delchambre, L. and Dell'Oro, A. and Esquej, P. and Fernández-Hernández, J. and Fraile, E. and Garabato, D. and García-Lario, P. and Gosset, E. and Haigron, R. and Halbwachs, J. -L. and Hambly, N. C. and Harrison, D. L. and Hernández, J. and Hestroffer, D. and Hodgkin, S. T. and Holl, B. and Janßen, K. and Jevardat de Fombelle, G. and Jordan, S. and Krone-Martins, A. and Lanzafame, A. C. and Löffler, W. and Marchal, O. and Marrese, P. M. and Moitinho, A. and Muinonen, K. and Osborne, P. and Pancino, E. and Pauwels, T. and Recio-Blanco, A. and Reylé, C. and Riello, M. and Rimoldini, L. and Roegiers, T. and Rybizki, J. and Sarro, L. M. and Siopis, C. and Smith, M. and Sozzetti, A. and Utrilla, E. and van Leeuwen, M. and Abbas, U. and Ábrahám, P. and Abreu Aramburu, A. and Aerts, C. and Aguado, J. J. and Ajaj, M. and Aldea-Montero, F. and Altavilla, G. and Álvarez, M. A. and Alves, J. and Anders, F. and Anderson, R. I. and Anglada Varela, E. and Antoja, T. and Baines, D. and Baker, S. G. and Balaguer-Núñez, L. and Balbinot, E. and Balog, Z. and Barache, C. and Barbato, D. and Barros, M. and Barstow, M. A. and Bartolomé, S. and Bassilana, J. -L. and Bauchet, N. and Becciani, U. and Bellazzini, M. and Berihuete, A. and Bernet, M. and Bertone, S. and Bianchi, L. and Binnenfeld, A. and Blanco-Cuaresma, S. and Blazere, A. and Boch, T. and Bombrun, A. and Bossini, D. and Bouquillon, S. and Bragaglia, A. and Bramante, L. and Breedt, E. and Bressan, A. and Brouillet, N. and Brugaletta, E. and Bucciarelli, B. and Burlacu, A. and Butkevich, A. G. and Buzzi, R. and Caffau, E. and Cancelliere, R. and Cantat-Gaudin, T. and Carballo, R. and Carlucci, T. and Carnerero, M. I. and Carrasco, J. M. and Casamiquela, L. and Castellani, M. and Castro-Ginard, A. and Chaoul, L. and Charlot, P. and Chemin, L. and Chiaramida, V. and Chiavassa, A. and Chornay, N. and Comoretto, G. and Contursi, G. and Cooper, W. J. and Cornez, T. and Cowell, S. and Crifo, F. and Cropper, M. and Crosta, M. and Crowley, C. and Dafonte, C. and Dapergolas, A. and David, M. and David, P. and de Laverny, P. and De Luise, F. and De March, R. and De Ridder, J. and de Souza, R. and de Torres, A. and del Peloso, E. F. and del Pozo, E. and Delbo, M. and Delgado, A. and Delisle, J. -B. and Demouchy, C. and Dharmawardena, T. E. and Di Matteo, P. and Diakite, S. and Diener, C. and Distefano, E. and Dolding, C. and Edvardsson, B. and Enke, H. and Fabre, C. and Fabrizio, M. and Faigler, S. and Fedorets, G. and Fernique, P. and Fienga, A. and Figueras, F. and Fournier, Y. and Fouron, C. and Fragkoudi, F. and Gai, M. and Garcia-Gutierrez, A. and Garcia-Reinaldos, M. and García-Torres, M. and Garofalo, A. and Gavel, A. and Gavras, P. and Gerlach, E. and Geyer, R. and Giacobbe, P. and Gilmore, G. and Girona, S. and Giuffrida, G. and Gomel, R. and Gomez, A. and González-Núñez, J. and González-Santamaría, I. and González-Vidal, J. J. and Granvik, M. and Guillout, P. and Guiraud, J. and Gutiérrez-Sánchez, R. and Guy, L. P. and Hatzidimitriou, D. and Hauser, M. and Haywood, M. and Helmer, A. and Helmi, A. and Sarmiento, M. H. and Hidalgo, S. L. and Hilger, T. and Hładczuk, N. and Hobbs, D. and Holland, G. and Huckle, H. E. and Jardine, K. and Jasniewicz, G. and Jean-Antoine Piccolo, A. and Jiménez-Arranz, Ó. and Jorissen, A. and Juaristi Campillo, J. and Julbe, F. and Karbevska, L. and Kervella, P. and Khanna, S. and Kontizas, M. and Kordopatis, G. and Korn, A. J. and Kóspál, Á. and Kostrzewa-Rutkowska, Z. and Kruszyńska, K. and Kun, M. and Laizeau, P. and Lambert, S. and Lanza, A. F. and Lasne, Y. and Le Campion, J. -F. and Lebreton, Y. and Lebzelter, T. and Leccia, S. and Leclerc, N. and Lecoeur-Taibi, I. and Liao, S. and Licata, E. L. and Lindstrøm, H. E. P. and Lister, T. A. and Livanou, E. and Lobel, A. and Lorca, A. and Loup, C. and Madrero Pardo, P. and Magdaleno Romeo, A. and Managau, S. and Mann, R. G. and Manteiga, M. and Marchant, J. M. and Marconi, M. and Marcos, J. and Marcos Santos, M. M. S. and Marín Pina, D. and Marinoni, S. and Marocco, F. and Marshall, D. J. and Martin Polo, L. and Martín-Fleitas, J. M. and Marton, G. and Mary, N. and Masip, A. and Massari, D. and Mastrobuono-Battisti, A. and Mazeh, T. and McMillan, P. J. and Messina, S. and Michalik, D. and Millar, N. R. and Mints, A. and Molina, D. and Molinaro, R. and Molnár, L. and Monari, G. and Monguió, M. and Montegriffo, P. and Montero, A. and Mor, R. and Mora, A. and Morbidelli, R. and Morel, T. and Morris, D. and Muraveva, T. and Murphy, C. P. and Musella, I. and Nagy, Z. and Noval, L. and Ocaña, F. and Ogden, A. and Ordenovic, C. and Osinde, J. O. and Pagani, C. and Pagano, I. and Palaversa, L. and Palicio, P. A. and Pallas-Quintela, L. and Panahi, A. and Payne-Wardenaar, S. and Peñalosa Esteller, X. and Penttilä, A. and Pichon, B. and Piersimoni, A. M. and Pineau, F. -X. and Plachy, E. and Plum, G. and Poggio, E. and Prša, A. and Pulone, L. and Racero, E. and Ragaini, S. and Rainer, M. and Raiteri, C. M. and Rambaux, N. and Ramos, P. and Ramos-Lerate, M. and Re Fiorentin, P. and Regibo, S. and Richards, P. J. and Rios Diaz, C. and Ripepi, V. and Riva, A. and Rix, H. -W. and Rixon, G. and Robichon, N. and Robin, A. C. and Robin, C. and Roelens, M. and Rogues, H. R. O. and Rohrbasser, L. and Romero-Gómez, M. and Rowell, N. and Royer, F. and Ruz Mieres, D. and Rybicki, K. A. and Sadowski, G. and Sáez Núñez, A. and Sagristà Sellés, A. and Sahlmann, J. and Salguero, E. and Samaras, N. and Sanchez Gimenez, V. and Sanna, N. and Santoveña, R. and Sarasso, M. and Schultheis, M. and Sciacca, E. and Segol, M. and Segovia, J. C. and Ségransan, D. and Semeux, D. and Shahaf, S. and Siddiqui, H. I. and Siebert, A. and Siltala, L. and Silvelo, A. and Slezak, E. and Slezak, I. and Smart, R. L. and Snaith, O. N. and Solano, E. and Solitro, F. and Souami, D. and Souchay, J. and Spagna, A. and Spina, L. and Spoto, F. and Steele, I. A. and Steidelmüller, H. and Stephenson, C. A. and Süveges, M. and Surdej, J. and Szabados, L. and Szegedi-Elek, E. and Taris, F. and Taylor, M. B. and Teixeira, R. and Tolomei, L. and Tonello, N. and Torra, F. and Torra, J. and Torralba Elipe, G. and Trabucchi, M. and Tsounis, A. T. and Turon, C. and Ulla, A. and Unger, N. and Vaillant, M. V. and van Dillen, E. and van Reeven, W. and Vanel, O. and Vecchiato, A. and Viala, Y. and Vicente, D. and Voutsinas, S. and Weiler, M. and Wevers, T. and Wyrzykowski, Ł. and Yoldas, A. and Yvard, P. and Zhao, H. and Zorec, J. and Zucker, S. and Zwitter, T.},
	month = jun,
	year = {2023},
	note = {ADS Bibcode: 2023A\&A...674A...1G},
	pages = {A1},
}

@article{brandt2024b,
	title = {Likelihood-based {Jump} {Detection} and {Cosmic} {Ray} {Rejection} for {Detectors} {Read} {Out} {Up}-the-ramp},
	volume = {136},
	issn = {0004-6280},
	url = {https://ui.adsabs.harvard.edu/abs/2024PASP..136d5005B},
	doi = {10.1088/1538-3873/ad38da},
	urldate = {2024-12-20},
	journal = {Publications of the Astronomical Society of the Pacific},
	author = {Brandt, Timothy D.},
	month = apr,
	year = {2024},
	note = {Publisher: IOP
ADS Bibcode: 2024PASP..136d5005B},
	pages = {045005},
}

@article{brandt2024a,
	title = {Optimal {Fitting} and {Debiasing} for {Detectors} {Read} {Out} {Up}-the-{Ramp}},
	volume = {136},
	issn = {0004-6280},
	url = {https://ui.adsabs.harvard.edu/abs/2024PASP..136d5004B},
	doi = {10.1088/1538-3873/ad38d9},
	urldate = {2024-12-20},
	journal = {Publications of the Astronomical Society of the Pacific},
	author = {Brandt, Timothy D.},
	month = apr,
	year = {2024},
	note = {Publisher: IOP
ADS Bibcode: 2024PASP..136d5004B},
	pages = {045004},
}

@article{godoy2024,
	title = {A new atmospheric characterization of the sub-stellar companion {HR} 2562 {B} with {JWST}/{MIRI} observations},
	volume = {689},
	issn = {0004-6361},
	url = {https://ui.adsabs.harvard.edu/abs/2024A&A...689A.185G},
	doi = {10.1051/0004-6361/202449951},
	urldate = {2024-12-20},
	journal = {Astronomy and Astrophysics},
	author = {Godoy, N. and Choquet, E. and Serabyn, E. and Danielski, C. and Stolker, T. and Charnay, B. and Hinkley, S. and Lagage, P. O. and Ressler, M. E. and Tremblin, P. and Vigan, A.},
	month = sep,
	year = {2024},
	note = {Publisher: EDP
ADS Bibcode: 2024A\&A...689A.185G},
	pages = {A185},
}

@article{pearce2024,
	title = {The effect of sculpting planets on the steepness of debris-disc inner edges},
	volume = {527},
	issn = {0035-8711},
	url = {https://ui.adsabs.harvard.edu/abs/2024MNRAS.527.3876P},
	doi = {10.1093/mnras/stad3462},
	urldate = {2024-08-30},
	journal = {Monthly Notices of the Royal Astronomical Society},
	author = {Pearce, Tim D. and Krivov, Alexander V. and Sefilian, Antranik A. and Jankovic, Marija R. and Löhne, Torsten and Morgner, Tobias and Wyatt, Mark C. and Booth, Mark and Marino, Sebastian},
	month = jan,
	year = {2024},
	note = {Publisher: OUP
ADS Bibcode: 2024MNRAS.527.3876P},
	pages = {3876--3899},
}

@article{pearce2022,
	title = {Planet populations inferred from debris discs. {Insights} from 178 debris systems in the {ISPY}, {LEECH}, and {LIStEN} planet-hunting surveys},
	volume = {659},
	issn = {0004-6361},
	url = {https://ui.adsabs.harvard.edu/abs/2022A&A...659A.135P},
	doi = {10.1051/0004-6361/202142720},
	urldate = {2023-11-28},
	journal = {Astronomy and Astrophysics},
	author = {Pearce, Tim D. and Launhardt, Ralf and Ostermann, Robert and Kennedy, Grant M. and Gennaro, Mario and Booth, Mark and Krivov, Alexander V. and Cugno, Gabriele and Henning, Thomas K. and Quirrenbach, Andreas and Barcucci, Arianna Musso and Matthews, Elisabeth C. and Ruh, Henrik L. and Stone, Jordan M.},
	month = mar,
	year = {2022},
	note = {ADS Bibcode: 2022A\&A...659A.135P},
	pages = {A135},
}

@book{murray_1999,
	title = {Solar {System} {Dynamics}},
	url = {https://ui.adsabs.harvard.edu/abs/1999ssd..book.....M},
	urldate = {2025-02-04},
	author = {Murray, Carl D. and Dermott, Stanley F.},
	month = feb,
	year = {1999},
	doi = {10.1017/CBO9781139174817},
	note = {Publication Title: Solar System Dynamics
ADS Bibcode: 1999ssd..book.....M},
}

@article{Pearce_wyatt_2014,
  title = {Dynamical Evolution of an Eccentric Planet and a Less Massive Debris Disc},
  author = {Pearce, Tim D. and Wyatt, Mark C.},
  year = {2014},
  month = sep,
  journal = {Monthly Notices of the Royal Astronomical Society},
  volume = {443},
  pages = {2541--2560},
  issn = {0035-8711},
  doi = {10.1093/mnras/stu1302}
}

@article{Wisdom1980,
  title = {The Resonance Overlap Criterion and the Onset of Stochastic Behavior in the Restricted Three-Body Problem},
  author = {Wisdom, J.},
  year = {1980},
  month = aug,
  journal = {The Astronomical Journal},
  volume = {85},
  pages = {1122--1133},
  publisher = {IOP},
  issn = {0004-6256},
  doi = {10.1086/112778}
}

@article{Gladman1993,
  title = {Dynamics of {{Systems}} of {{Two Close Planets}}},
  author = {Gladman, Brett},
  year = {1993},
  month = nov,
  journal = {Icarus},
  volume = {106},
  number = {1},
  pages = {247--263},
  issn = {0019-1035},
  doi = {10.1006/icar.1993.1169}
}

@article{Shannon+2016,
  title = {The Unseen Planets of Double Belt Debris Disc Systems},
  author = {Shannon, Andrew and Bonsor, Amy and Kral, Quentin and Matthews, Elisabeth},
  year = {2016},
  month = oct,
  journal = {Monthly Notices of the Royal Astronomical Society},
  volume = {462},
  pages = {L116-L120},
  publisher = {OUP},
  issn = {0035-8711},
  doi = {10.1093/mnrasl/slw143}
}

@article{Tabeshian+2016,
  title = {Detection and {{Characterization}} of {{Extrasolar Planets}} through {{Mean-motion Resonances}}. {{I}}. {{Simulations}} of {{Hypothetical Debris Disks}}},
  author = {Tabeshian, Maryam and Wiegert, Paul A.},
  year = {2016},
  month = feb,
  journal = {The Astrophysical Journal},
  volume = {818},
  pages = {159},
  publisher = {IOP},
  issn = {0004-637X},
  doi = {10.3847/0004-637X/818/2/159}
}

@article{golimowski2011,
  title = {Hubble and {{Spitzer Space Telescope Observations}} of the {{Debris Disk}} around the Nearby {{K Dwarf HD}} 92945},
  author = {Golimowski, D. A. and Krist, J. E. and Stapelfeldt, K. R. and Chen, C. H. and Ardila, D. R. and Bryden, G. and Clampin, M. and Ford, H. C. and Illingworth, G. D. and Plavchan, P. and Rieke, G. H. and Su, K. Y. L.},
  year = {2011},
  month = jul,
  journal = {The Astronomical Journal},
  volume = {142},
  pages = {30},
  issn = {0004-6256},
  doi = {10.1088/0004-6256/142/1/30},
  urldate = {2023-11-22}
}

@article{Ardila2004,
  title = {A {{Resolved Debris Disk}} around the {{G2 V Star HD}} 107146},
  author = {Ardila, D. R. and Golimowski, D. A. and Krist, J. E. and Clampin, M. and Williams, J. P. and Blakeslee, J. P. and Ford, H. C. and Hartig, G. F. and Illingworth, G. D.},
  year = {2004},
  month = dec,
  journal = {The Astrophysical Journal},
  volume = {617},
  pages = {L147-L150},
  publisher = {IOP},
  issn = {0004-637X},
  doi = {10.1086/427434},
  urldate = {2025-03-13}
}

@article{Schneider2014,
  title = {Probing for {{Exoplanets Hiding}} in {{Dusty Debris Disks}}: {{Disk Imaging}}, {{Characterization}}, and {{Exploration}} with {{HST}}/{{STIS Multi-roll Coronagraphy}}},
  shorttitle = {Probing for {{Exoplanets Hiding}} in {{Dusty Debris Disks}}},
  author = {Schneider, Glenn and Grady, Carol A. and Hines, Dean C. and Stark, Christopher C. and Debes, John H. and Carson, Joe and Kuchner, Marc J. and Perrin, Marshall D. and Weinberger, Alycia J. and Wisniewski, John P. and Silverstone, Murray D. and {Jang-Condell}, Hannah and Henning, Thomas and Woodgate, Bruce E. and Serabyn, Eugene and {Moro-Martin}, Amaya and Tamura, Motohide and Hinz, Phillip M. and Rodigas, Timothy J.},
  year = {2014},
  month = oct,
  journal = {The Astronomical Journal},
  volume = {148},
  pages = {59},
  publisher = {IOP},
  issn = {0004-6256},
  doi = {10.1088/0004-6256/148/4/59},
  urldate = {2025-03-13}
}

@article{chambers1996,
  title = {The {{Stability}} of {{Multi-Planet Systems}}},
  author = {Chambers, J.E. and Wetherill, G.W. and Boss, A.P.},
  year = {1996},
  month = feb,
  journal = {Icarus},
  volume = {119},
  number = {2},
  pages = {261--268},
  issn = {00191035},
  doi = {10.1006/icar.1996.0019},
  urldate = {2023-11-22},
  langid = {english}
}

@software{2020ExoDMC,
       author = {{Bonavita}, Mariangela},
        title = "{Exo-DMC: Exoplanet Detection Map Calculator}",
 howpublished = {Astrophysics Source Code Library, record ascl:2010.008},
         year = 2020,
        month = oct,
          eid = {ascl:2010.008},
       adsurl = {https://ui.adsabs.harvard.edu/abs/2020ascl.soft10008B}
}

@ARTICLE{Squicciarini2022,
       author = {{Squicciarini}, V. and {Bonavita}, M.},
        title = "{MADYS: the Manifold Age Determination for Young Stars. I. Isochronal age estimates and model comparison}",
      journal = {\aap},
         year = 2022,
        month = oct,
       volume = {666},
          eid = {A15},
        pages = {A15},
          doi = {10.1051/0004-6361/202244193},
archivePrefix = {arXiv},
       eprint = {2206.02446},
 primaryClass = {astro-ph.SR},
       adsurl = {https://ui.adsabs.harvard.edu/abs/2022A&A...666A..15S}
}

@article{Mesa2023,
  title = {{{AF Lep}} b: {{The}} Lowest-Mass Planet Detected by Coupling Astrometric and Direct Imaging Data},
  shorttitle = {{{AF Lep}} b},
  author = {Mesa, D. and Gratton, R. and Kervella, P. and Bonavita, M. and Desidera, S. and D'Orazi, V. and Marino, S. and Zurlo, A. and Rigliaco, E.},
  year = {2023},
  month = apr,
  journal = {Astronomy and Astrophysics},
  volume = {672},
  pages = {A93},
  publisher = {EDP},
  issn = {0004-6361},
  doi = {10.1051/0004-6361/202345865},
  urldate = {2025-03-24}
}

@article{Franson2023,
  title = {Astrometric {{Accelerations}} as {{Dynamical Beacons}}: {{A Giant Planet Imaged}} inside the {{Debris Disk}} of the {{Young Star AF Lep}}},
  shorttitle = {Astrometric {{Accelerations}} as {{Dynamical Beacons}}},
  author = {Franson, Kyle and Bowler, Brendan P. and Zhou, Yifan and Pearce, Tim D. and Bardalez Gagliuffi, Daniella C. and Biddle, Lauren I. and Brandt, Timothy D. and Crepp, Justin R. and Dupuy, Trent J. and Faherty, Jacqueline and {Jensen-Clem}, Rebecca and Morgan, Marvin and Sanghi, Aniket and Theissen, Christopher A. and Tran, Quang H. and Wolf, Trevor N.},
  year = {2023},
  month = jun,
  journal = {The Astrophysical Journal},
  volume = {950},
  pages = {L19},
  publisher = {IOP},
  issn = {0004-637X},
  doi = {10.3847/2041-8213/acd6f6},
  urldate = {2025-03-24}
}

@INPROCEEDINGS{lajoie_2014,
       author = {{Lajoie}, Charles-Philippe and {Soummer}, R{\'e}mi and {Hines}, Dean C. and {Rieke}, George H.},
        title = "{Simulations of JWST MIRI 4QPM coronagraphs operations and performances}",
    booktitle = {Space Telescopes and Instrumentation 2014: Optical, Infrared, and Millimeter Wave},
         year = 2014,
       editor = {{Oschmann}, Jr., Jacobus M. and {Clampin}, Mark and {Fazio}, Giovanni G. and {MacEwen}, Howard A.},
       series = {Society of Photo-Optical Instrumentation Engineers (SPIE) Conference Series},
       volume = {9143},
        month = aug,
          eid = {91433R},
        pages = {91433R},
          doi = {10.1117/12.2056284},
       adsurl = {https://ui.adsabs.harvard.edu/abs/2014SPIE.9143E..3RL}
}

@ARTICLE{dicken_2024,
       author = {{Dicken}, Dan and {Mar{\'\i}n}, Macarena Garc{\'\i}a and {Shivaei}, Irene and {Guillard}, Pierre and {Libralato}, Mattia and {Glasse}, Alistair and {Gordon}, Karl D. and {Cossou}, Christophe and {Kavanagh}, Patrick and {Temim}, Tea and {Flagey}, Nicolas and {Klaassen}, Pamela and {Rieke}, George H. and {Wright}, Gillian and {Alberts}, Stacey and {Azzollini}, Ruyman and {{\'A}lvarez-M{\'a}rquez}, Javier and {Bouchet}, Patrice and {Bright}, Stacey and {Cracraft}, Misty and {Coulais}, Alain and {Detre}, Ors Hunor and {Engesser}, Mike and {Fox}, Ori D. and {Gaspar}, Andras and {Gastaud}, Ren{\'e} and {Glauser}, Adrian M. and {Hines}, Dean C. and {Kendrew}, Sarah and {Labiano}, Alvaro and {Lagage}, Pierre-Oliver and {Lee}, David and {Law}, David R. and {Morrison}, Jane E. and {Noriega-Crespo}, Alberto and {Jones}, Olivia and {Patapis}, Polychronis and {Scheithauer}, Silvia and {Sloan}, G.~C. and {Tamas}, Laszlo},
        title = "{JWST MIRI flight performance: Imaging}",
      journal = {\aap},
         year = 2024,
        month = sep,
       volume = {689},
          eid = {A5},
        pages = {A5},
          doi = {10.1051/0004-6361/202449451},
archivePrefix = {arXiv},
       eprint = {2403.16686},
 primaryClass = {astro-ph.IM},
       adsurl = {https://ui.adsabs.harvard.edu/abs/2024A&A...689A...5D}
}

@ARTICLE{kennedy_2010,
       author = {{Kennedy}, G.~M. and {Wyatt}, M.~C.},
        title = "{Are debris discs self-stirred?}",
      journal = {\mnras},
         year = 2010,
        month = jun,
       volume = {405},
       number = {2},
        pages = {1253-1270},
          doi = {10.1111/j.1365-2966.2010.16528.x},
archivePrefix = {arXiv},
       eprint = {1002.3469},
 primaryClass = {astro-ph.EP},
       adsurl = {https://ui.adsabs.harvard.edu/abs/2010MNRAS.405.1253K}
}

@article{DeRosa2023,
  title = {Direct Imaging Discovery of a Super-{{Jovian}} around the Young {{Sun-like}} Star {{AF Leporis}}},
  author = {De Rosa, Robert J. and Nielsen, Eric L. and Wahhaj, Zahed and Ruffio, Jean-Baptiste and Kalas, Paul G. and Peck, Anne E. and Hirsch, Lea A. and Roberson, William},
  year = {2023},
  month = apr,
  journal = {Astronomy and Astrophysics},
  volume = {672},
  pages = {A94},
  publisher = {EDP},
  issn = {0004-6361},
  doi = {10.1051/0004-6361/202345877},
  urldate = {2025-03-24}
}

@ARTICLE{Tremaine1998,
       author = {{Tremaine}, Scott},
        title = "{Resonant Relaxation in Protoplanetary Disks}",
      journal = {\aj},
         year = 1998,
        month = oct,
       volume = {116},
       number = {4},
        pages = {2015-2022},
          doi = {10.1086/300567},
archivePrefix = {arXiv},
       eprint = {astro-ph/9805334},
 primaryClass = {astro-ph},
       adsurl = {https://ui.adsabs.harvard.edu/abs/1998AJ....116.2015T}
}

@BOOK{depater_2001,
       author = {{de Pater}, Imke and {Lissauer}, Jack J.},
        title = "{Planetary Sciences}",
         year = 2001,
       adsurl = {https://ui.adsabs.harvard.edu/abs/2001plsc.book.....D}
}




\appendix

\section{LIKELY ramp fitting}
\begin{figure}
    \centering
    \includegraphics[trim=0.25cm 0.25cm 0.25cm 0.25cm, clip=true,width=\linewidth]{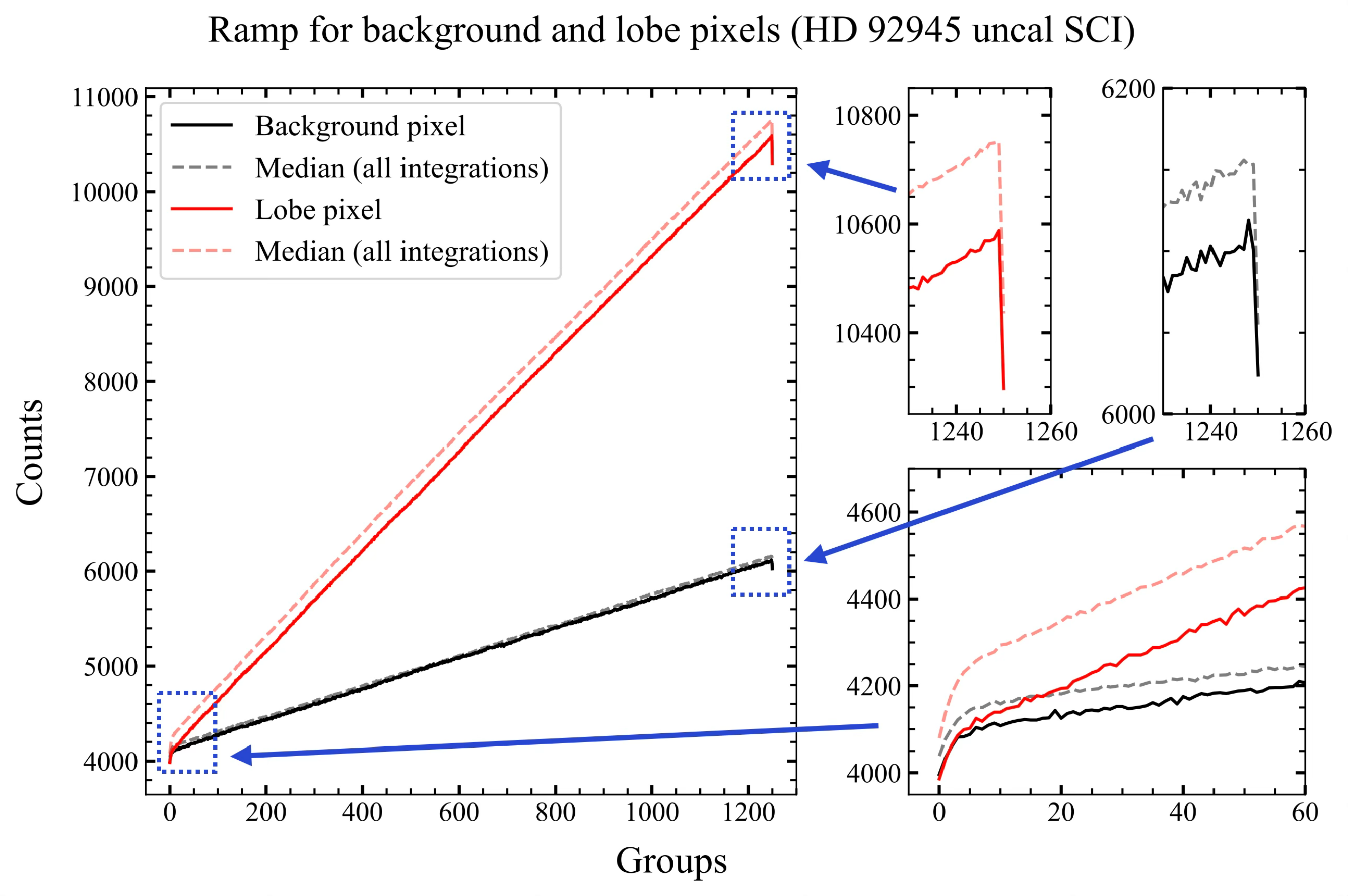}
    \caption{Counts per group for two representative pixels during the first integration of the HD\,92945 uncalibrated science observation: one in a bright PSF lobe (red) and one in a background region (black). Solid lines show the first integration, while dashed lines indicate the median over all integrations. The right-hand panels zoom in on the start and end of the ramp, with blue arrows and boxes marking the regions shown. The y-axis ticks in these zoomed panels correspond to 50 counts.}
    \label{fig: ramp_fitting_visualisation}
\end{figure}

During the ramp-fitting step described in \S\ref{sec: data-reduction}.a, we excluded the first 10 and last 2 groups of each integration. The initial groups are affected by detector settling effects following the reset, while the end of the ramp shows a systematic drop in counts. Both effects lead to deviations from the linear ramp expected for well-behaved groups. Figure~\ref{fig: ramp_fitting_visualisation} illustrates this for two representative pixels: one located within a bright PSF lobe (red lines) and another in the background region (black lines), both taken from the first integration of the HD\,92945 uncalibrated science observation. For comparison, we also plot the median group values across all integrations for the same pixels (dashed lines), which show the same behaviour. This effect is present in all MIRI coronagraphy observations of this program. The right-hand panels zoom in on the start and end of the group sequence, highlighting the slope changes that motivate the exclusion of these groups from the fit.

\section{Brighter-Fatter Effect (BFE) correction}
\label{app: bfe}
\begin{figure*}
    \centering
    \includegraphics[trim=0.3cm 0.2cm 0.5cm 0.5cm, clip=true,width=0.7\linewidth]{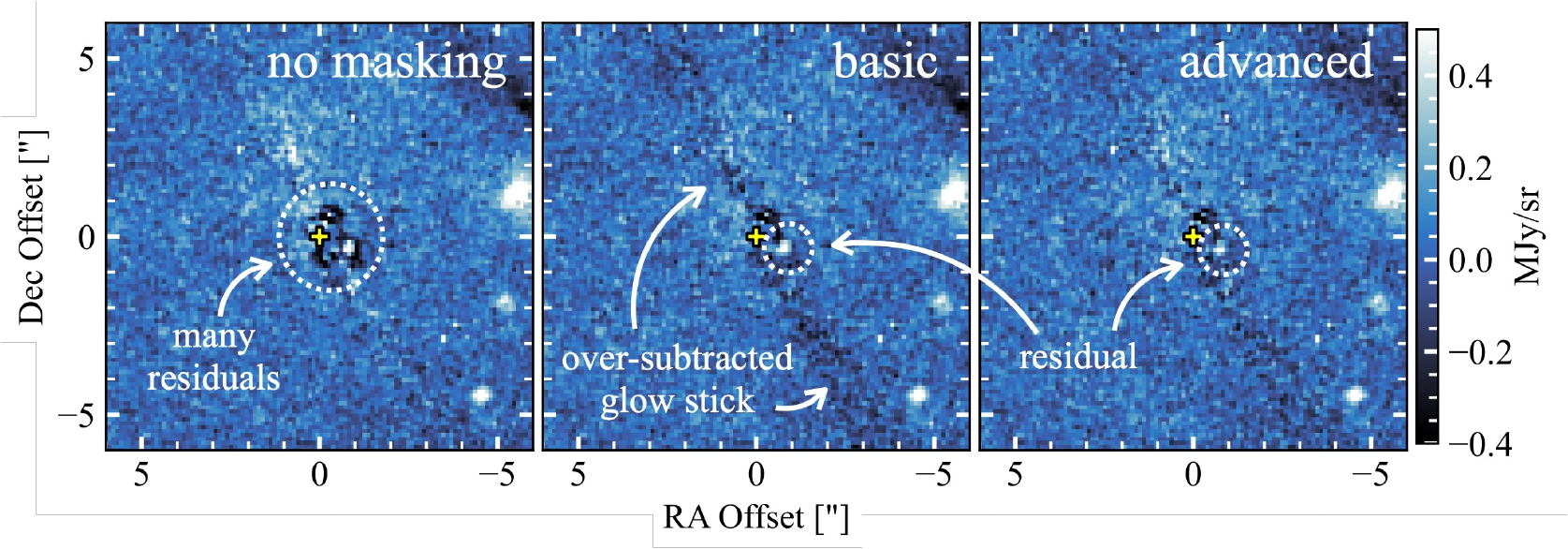}
    \vspace{-2mm}
    \caption{HD\,92945 PSF-subtracted images using ADI+RDI with different BFE correction methods: no group masking, `basic' method, and `advanced' method. Different residual patterns are visible across methods, with the `advanced' approach producing the cleanest images.}
    \label{fig: BFE correction}    
    \vspace{-3mm}
\end{figure*}

The BFE correction can remove significant residuals that otherwise impact the quality of the reductions (Figure~X). Three methods are available: `custom', `basic' and `advanced'. Figure~\ref{fig: BFE correction} shows the differences between no BFE correction, the `basic' and `advanced' methods. Without correction, significant residuals remain in the inner regions of the image. The `basic' method reduces residuals near the PSF lobes, but the number of groups trimmed causes the "glow stick" to be over-subtracted. The best reduction is obtained with the `advanced' method.

The `basic' and `advanced' methods differ in the region of optimisation but follow the same principle. More specifically, the `advanced' group-masking method operates as follows. First, we calculate the median of the last groups across all science integrations, and repeat this across all reference integrations. We then independently median subtract each of the median group images to remove any uniform background, and identify and mask all pixels below an 85\% percentile threshold to isolate the routine to the brightest pixels likely to be affected by BFE.

Each median group image is then split into a "central" box region of $20{\times}20$ pixels, centred on the 4QPM centre, and an "external" region of pixels outside of this box. Using the central and external regions, we independently calculate the absolute differences between each median reference group image and the median science group image, and determine the corresponding reference groups at which the difference for each region is minimized. This defines a maximum and minimum number of groups to be masked.

Next, for each reference independently, we calculate the median of the final reference group across all integrations, perform a median subtraction to match the earlier steps, and determine the peak pixel count in the central region, and the minimum pixel count in the external region. A linear interpolation between the pixel count values, and the previously identified maximum and minimum number of groups to be masked, is then used to identify the number of groups that must be masked during the ramp fitting for each pixel based on the counts in the median subtracted final group image. The number of groups to mask for pixels with counts outside the defined minimum or maximum, are clamped to the minimum of maximum number of groups to mask, respectively.

We emphasize that the success of this BFE correction technique is only possible because the reference observations experience a more significant BFE due to their higher detector counts compared to the science observations. Groups can be readily removed as a function of pixel position from the references without significantly affecting the overall signal-to-noise of the subtracted image. If the situation were reversed, and the science observations experienced higher detector counts, then to mimic this technique groups would need to be trimmed from the science observations, directly impacting the achievable SNR. A more general approach to a BFE correction, for example using a deconvolution kernel \citep{argyriou_2023}, may be more appropriate in these situations.

\section{Background observation cleaning}
\label{app: bg_cleaning}
\begin{figure}
    \centering
    \includegraphics[trim=0.25cm 0.25cm 0.25cm 0.25cm, clip=true, width=\linewidth]{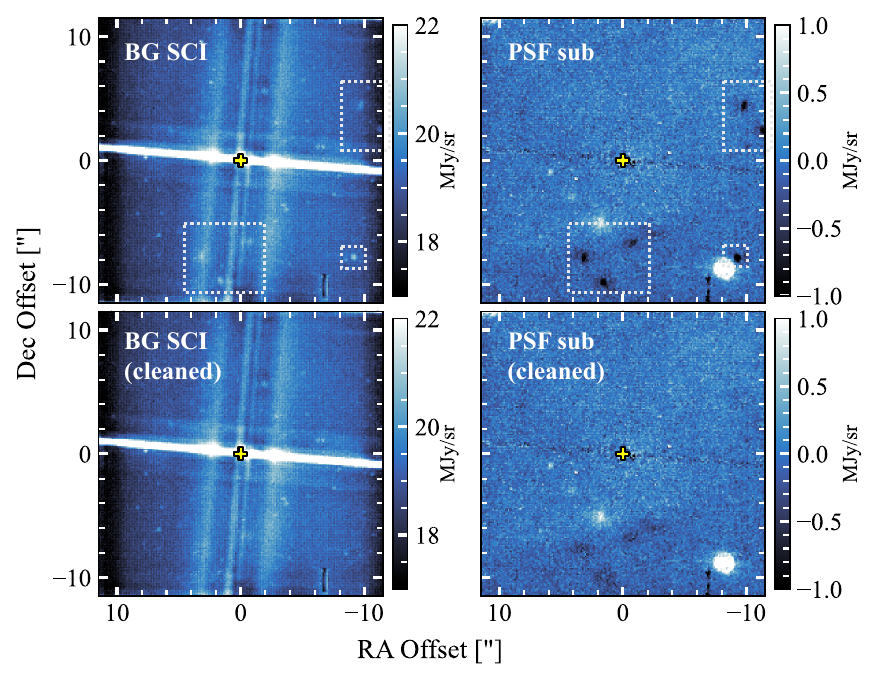}
    \caption{Effect of the background cleaning process. The left column shows the median of the HD\,92945 science background observations, while the right column presents the resulting RDI PSF-subtracted images. All the panels are in the detectors reference frame and not rotated North-East. The top row corresponds to the case without background cleaning, where white boxes highlight background contaminants (left) and their associated negative residual (right). The bottom row shows the results after applying the cleaning procedure.}
    \label{fig:bg_cleaning}
\end{figure}

To automatically identify and remove the contaminants observed in the HD\,92945 science background observations, we computed the difference between the two available science background images and divided it by the noise maps added in quadrature. Each noise map was defined as the standard deviation of each pixel across all integrations for the corresponding science background. The resulting signal-to-noise ratio (SNR) map is given by:
\begin{equation}
    \mathrm{SNR} = \frac{\mathrm{bkg}_1 - \mathrm{bkg}_2}{\sqrt{\sigma_{\mathrm{bkg}_1}^2 + \sigma_{\mathrm{bkg}_2}^2}},
\end{equation}
where bkg$_1$ and bkg$_2$ are the two science background observations, and $\sigma_{\mathrm{bkg}_1}$ and $\sigma_{\mathrm{bkg}_2}$ their respective noise maps. Pixels with SNR~$>3\sigma$ represent sources observed in bkg$_1$ to be masked, while SNR~$<-3\sigma$ identifies contaminants in bkg$_2$. When multiple background files of the same type are available (science or reference backgrounds), \texttt{spaceKLIP} takes their median for the background subtraction step, yielding a single, cleaned background image. Figure~\ref{fig:bg_cleaning} illustrates the effect of the background cleaning process which removes strong negative residual sources, resulting in cleaner PSF-subtracted images.

\begin{figure}
    \centering
    \includegraphics[trim=0.25cm 0.25cm 0.25cm 0.25cm, clip=true, width=\linewidth]{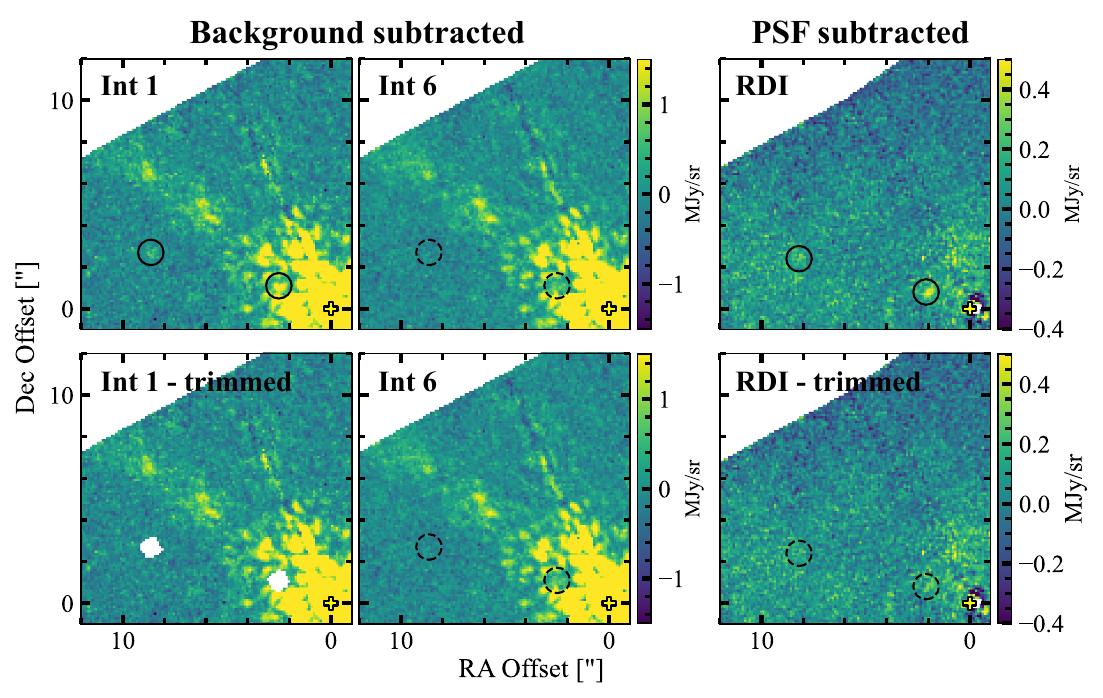}
    \caption{Effect of persistence trimming. The first two panels show background-subtracted science images zoomed in on the location affected by persistence, for the first (left) and last (centre) integrations. The right panels show the corresponding RDI PSF-subtracted images. Examples are for HD\,92945 and are rotated North-East. The top row shows data without persistence trimming, while the bottom row includes the trimming process. Solid black circles mark regions affected by persistence, dashed black circles mark regions where no persistence is observed, and the white areas in the bottom left panel show the masked pixels during the persistence trimming process.}
    \label{fig:persistence}
\end{figure}

\section{Persistence trimming}
\label{app: persistence_trimming}
To locate the persistence, we use the position of the star in the two TA images. We observe that persistence produces significant residuals throughout the data reduction process (Figure~\ref{fig:persistence}), most prominently in the first integration and to a lesser extent in the second, before becoming negligible. We therefore mask all pixels within a radius of 5 pixels from the persistence location in the first two integrations. This step is applied only to the science observations, which have 6 integrations in total, as the reference observations only have 2 integrations and the effect cannot be clearly identified. The impact of this trimming can be seen in the PSF-subtracted images in the right panels of Figure~\ref{fig:persistence}.

\section{PSF-fitting}
\begin{figure*}
    \centering
    \begin{subfigure}{\linewidth}
        \centering
        \includegraphics[trim=1.5cm 0.5cm 2.3cm 1.cm, clip=true, width=0.49\textwidth]{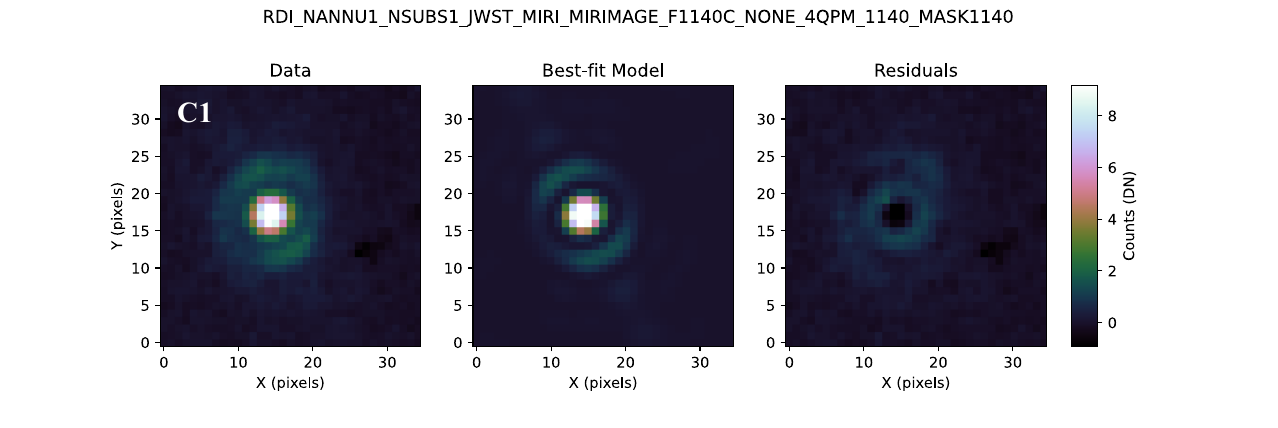}
        \includegraphics[trim=1.5cm 0.5cm 2.3cm 1.cm, clip=true, width=0.49\textwidth]{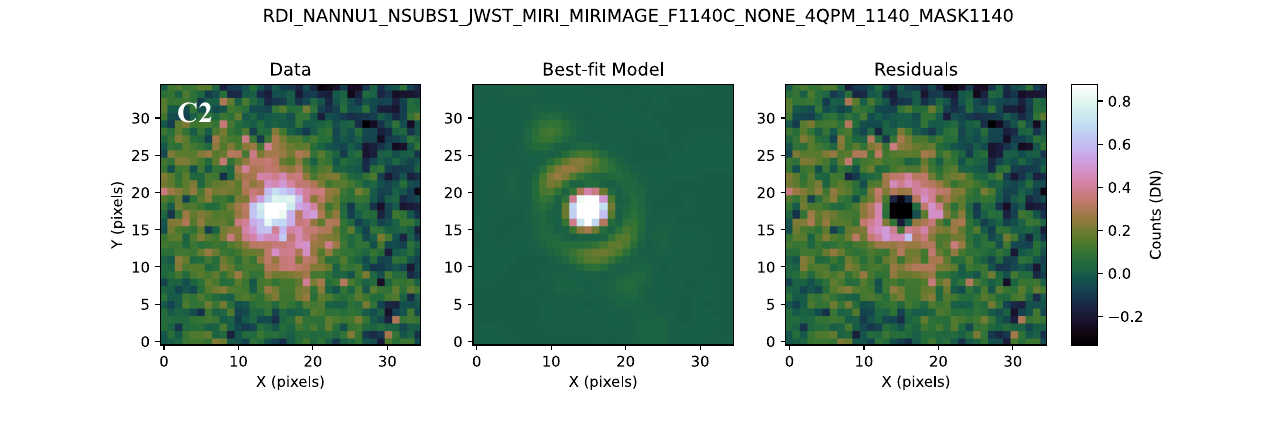}
        \includegraphics[trim=1.5cm 0.5cm 2.3cm 1.cm, clip=true, width=0.49\textwidth]{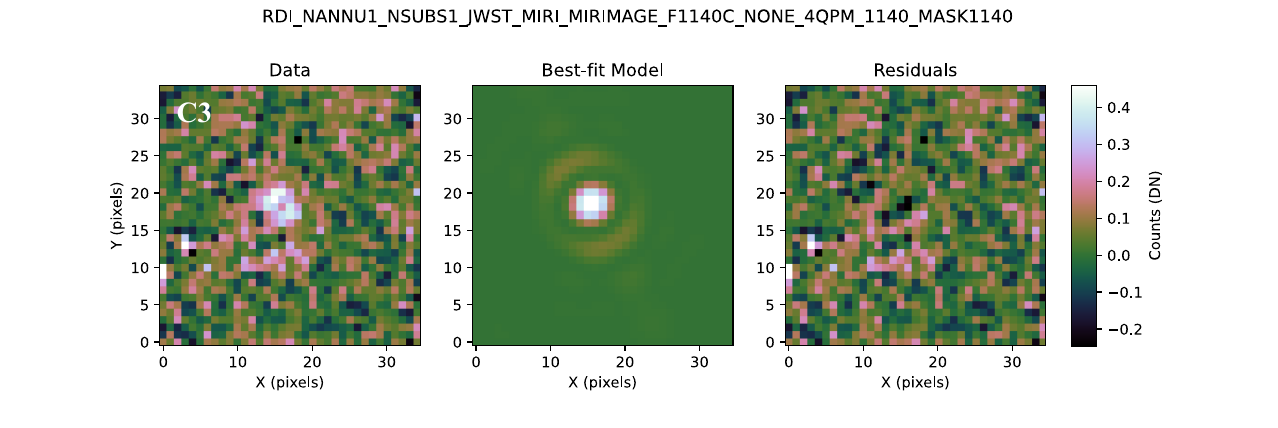}
        \includegraphics[trim=1.5cm 0.5cm 2.3cm 1.cm, clip=true, width=0.49\textwidth]{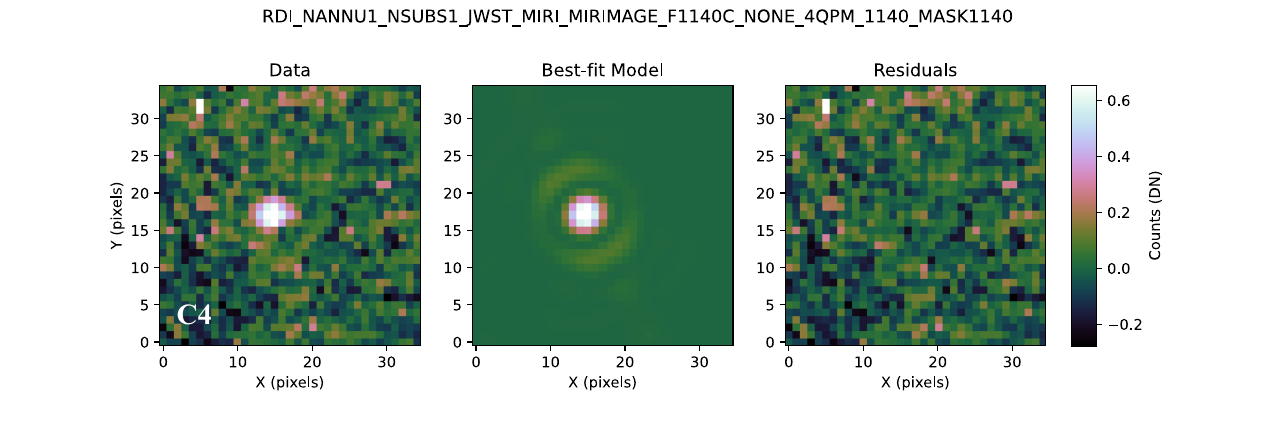}
        \subcaption{Four candidates observed in the field-of-view of HD\,92945.}
        \label{PSF-fitting-hd92945}
    \end{subfigure}
    
    \begin{subfigure}{\linewidth}
        \includegraphics[trim=1.5cm 0.5cm 2.3cm 1.cm, clip=true, width=0.49\textwidth]{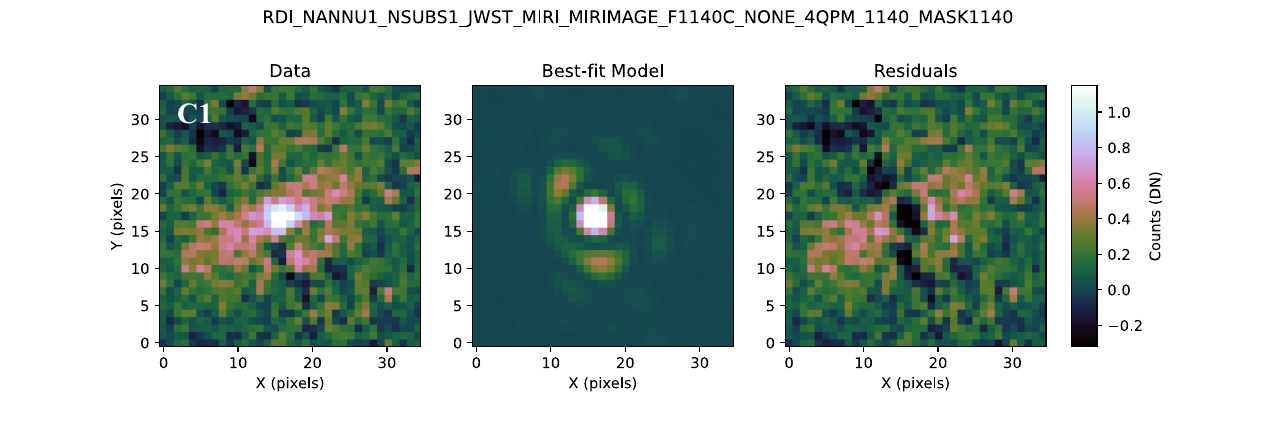}
        \subcaption{Single candidate observed in the field-of-view of HD\,107146.}
        \label{PSF-fitting-hd107146}
    \end{subfigure}
    
    \begin{subfigure}{\linewidth}
        \includegraphics[trim=1.5cm 0.5cm 2.2cm 1.cm, clip=true, width=0.49\textwidth]{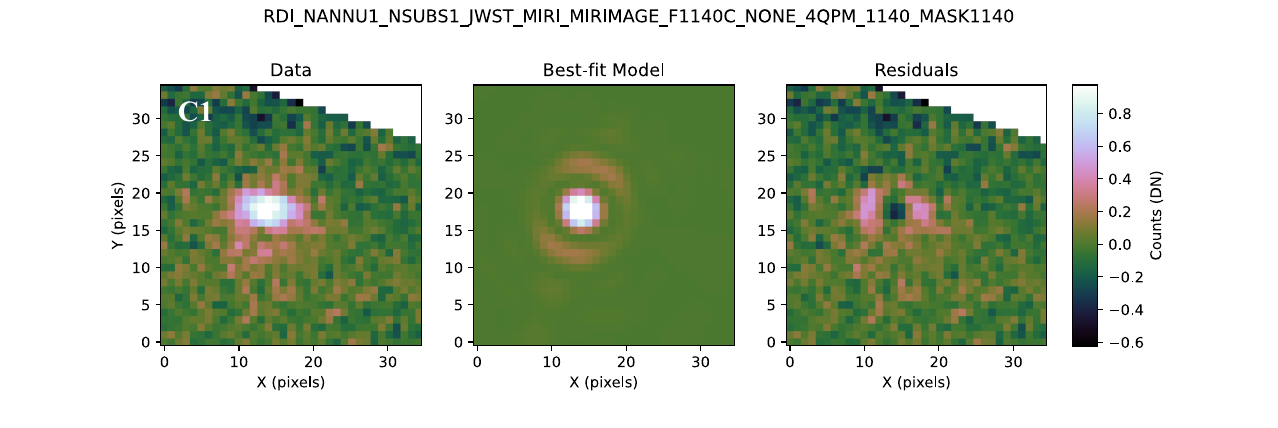}
        \includegraphics[trim=1.5cm 0.5cm 2.3cm 1.cm, clip=true, width=0.49\textwidth]{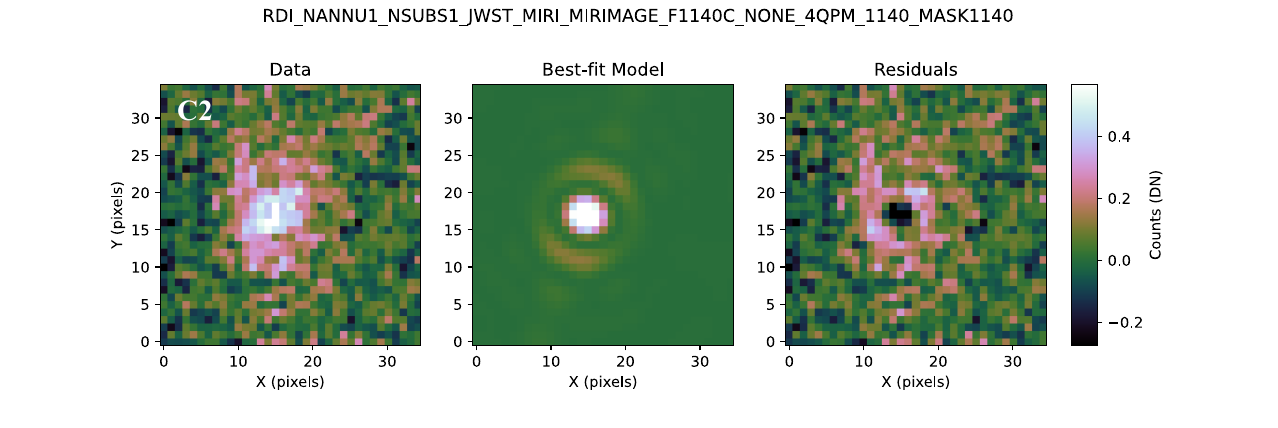}
        \includegraphics[trim=1.5cm 0.5cm 2.4cm 1.cm, clip=true, width=0.49\textwidth]{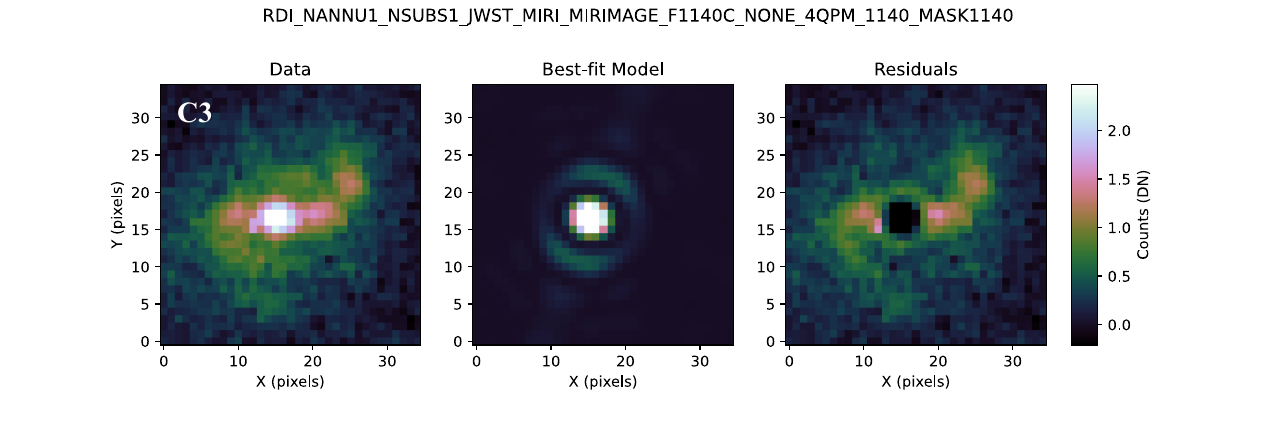}
        \subcaption{Three candidates observed in the field-of-view of HD\,206893.}
        \label{PSF-fitting-hd206893}
    \end{subfigure}
    \caption{Point-source fitting to the candidates observed in the MIRI coronagraphic observations. For each candidate, the left panel shows the source as observed in the data, the middle panel shows the PSF model generated from \texttt{spaceKLIP}, and the residuals from the PSF-fitting are seen in the right panel.}
    \label{fig: psf-fitting}
\end{figure*}

Figure~\ref{fig: psf-fitting} illustrates the PSF-fitting process for all the sources observed in the field-of-view of the three targets. The sources are each fitted with a point source, and residuals to the fit indicate whether a source is more consistent with a star or planet (point-like) or a background galaxy (extended).

\bsp	
\label{lastpage}
\end{document}